\documentclass[11pt,a4paper]{article}

\usepackage[truedimen,margin=30mm]{geometry} 
\usepackage{mathrsfs}
\usepackage{amssymb}
\usepackage{amsmath}
\usepackage{ascmac}
\usepackage{amsthm}
\usepackage{amsmath}
\usepackage[dvipdfmx]{graphicx}
\usepackage[]{color}
\usepackage{booktabs}
\usepackage{setspace}
\usepackage{natbib}
\usepackage{float}
\usepackage{mathtools}
\usepackage{url}
\usepackage{comment}
\usepackage[symbol]{footmisc}
\def\diag{\text{diag}}

\def\ep{{\varepsilon}}

\def\zero{{\text{\boldmath$0$}}}

\def\a{{\text{\boldmath $a$}}}
\def\e{{\text{\boldmath $e$}}}

\def\f{{\text{\boldmath $f$}}}

\def\h{{\text{\boldmath $h$}}}

\def\m{{\text{\boldmath $m$}}}

\def\u{{\text{\boldmath $u$}}}
\def\v{{\text{\boldmath $v$}}}
\def\w{{\text{\boldmath $w$}}}
\def\x{{\text{\boldmath $x$}}}
\def\y{{\text{\boldmath $y$}}}

\def\A{{\text{\boldmath $A$}}}

\def\C{{\text{\boldmath $C$}}}

\def\F{{\text{\boldmath $F$}}}

\def\I{{\text{\boldmath $I$}}}

\def\F{{\text{\boldmath $F$}}}
\def\H{{\text{\boldmath $H$}}}
\def\Q{{\text{\boldmath $Q$}}}

\def\V{{\text{\boldmath $V$}}}

\def\R{{\text{\boldmath $R$}}}

\def\u{{\text{\boldmath $u$}}}
\def\W{{\text{\boldmath $W$}}}

\def\bbe{{\text{\boldmath $\beta$}}}

\def\beps{{\text{\boldmath $\epsilon$}}}

\def\blam{{\text{\boldmath $\lambda$}}}

\def\bthe{{\text{\boldmath $\theta$}}}
\def\bPhi{{\text{\boldmath $\Phi$}}}

\def\bLam{{\text{\boldmath $\Lambda$}}}
\def\bSig{{\text{\boldmath $\Sigma$}}}

\def\LTE{\text{LTE}}
\def\EMP{\text{EMPGAP}}
\def\IMP{\text{ImpINF}}

\def\PIT{\text{PIT}}
\def\CRPS{\text{CRPS}}
\def\RCS{\text{RCS}}
\def\RTCS{\text{RTCS}}
\def\DRQS{\text{DRQS}}

\def\DQLM{\text{DQLM}}
\def\FQBART{\text{FQBART}}

\def\Dp{\text{Dp}}
\def\ep{\text{ep}}
\def\eq{\text{eq}}
\def\rr{\text{r}}
\def\lr{\text{lr}}

\title{\textbf{Dynamic Bayesian regression quantile synthesis for forecasting outlook-at-risk}
}
\date{March 10, 2026}
\begin{document}

\maketitle
\begin{center}
Genya Kobayashi$^{1}$\footnote{Author of correspondence: \url{gkobayashi@meiji.ac.jp}},
Shonosuke Sugasawa$^{2}$,
Yuta Yamauchi$^{3}$, and
Dongu Han$^{1}$
\end{center}

\noindent
$^1$School of Commerce, Meiji University\\
$^2$Department of Economics, Keio University\\
$^3$Department of Economics, Nagoya University

\begin{abstract}
This paper proposes dynamic Bayesian regression quantile synthesis (DRQS), a novel method for quantile forecasting within the Bayesian predictive synthesis (BPS) framework designed to combine quantile-specific information from multiple agent models. 
While existing BPS approaches primarily focus on mean forecasting, our method directly targets the conditional quantiles of the response variable by utilizing the asymmetric Laplace distribution for the synthesis function. 
The resulting framework can be interpreted as a dynamic quantile linear model with latent predictors.
We extend the univariate DRQS to a multivariate setting—factor DRQS (FDRQS)—by introducing a time-varying latent factor structure for the synthesis weights. 
This allows the model to leverage cross-sectional dependencies and shared information across multiple time series simultaneously. 
We develop an efficient Markov chain Monte Carlo (MCMC) algorithm for posterior inference, utilizing data augmentation and forward-filtering backward-sampling.
Empirical applications to US inflation and global GDP growth  demonstrate the improved performance of the proposed methods for quantile forecasting.  
In particular, FDRQS exhibits superior resilience during periods of extreme economic stress, such as the COVID-19 pandemic, by adaptively rebalancing agent contributions and capturing emergent global dependencies.

\end{abstract}

\textbf{keywords:}
asymmetric Laplace distribution;
Bayesian quantile regression;
dynamic latent factor model;
multiplicative gamma process;
macro economic forecasting;

\section{Introduction}

Accurately forecasting the tails of a distribution remains a fundamental challenge in econometrics and finance, particularly when underlying risks exhibit pronounced asymmetry. 
Traditional mean-based models often fail to capture the full spectrum of outcomes during periods of economic instability or extreme shocks. Consequently, quantile regression \citep[QR;][]{koenker1978regression, Koenker2005} has emerged as the standard tool for modeling specific parts of the distribution of a target variable. 
This is evidenced by its widespread applications to macroeconomic variables through frameworks such as Growth-at-Risk \citep[GaR;][]{GIGLIO16,Nicolo17} and Inflation-at-Risk \citep[IaR;][]{wang2024inflation}.
Recent empirical works has emphasized the importance of forecasting risks associated with macroeconomic variables, such as the probability that future inflation deviates substantially from a target range. 
In these contexts, the accurate forecasting of the conditional distribution, rather than the conditional mean alone, is indispensable for robust policy analysis and risk management. 

Nearly five decades since its inception by \cite{koenker1978regression}, the Quantile Regression (QR) framework has evolved into a cornerstone of macroeconomic forecasting, spawning a burgeoning literature dedicated to capturing complex distributional dynamics. 
Recent extensions have introduced time-varying parameters \citep{PFARRHOFER22,DQLM}, quantile factor models to account for pervasive common variation \citep{Chen21,Korobilis24,Clark24}, and Mixed Data Sampling (MIDAS) structures for handling disparate frequencies \citep{Mitchell22,Ferrara}. 
Flexible multivariate forecasting models have also been proposed for macroeconomic variables through, for example, the multivariate asymmetric Laplace distribution for quantile vector autoregressive models \citep{Iacopini} or by combining nonlinear components such as BART with vector autoregressive structures \citep{Clark23}. 

Another line of innovative works reconstructs full predictive distributions by mapping quantile forecasts onto parametric distributions \citep{Adrian}, a technique further refined by Bayesian and nonparametric extensions \citep{Mitchell24}. 
Collectively, these methodologies underscore the versatility of quantile-based approaches in characterizing the sophisticated tail risks inherent in modern macroeconomic landscapes.

Parallel to these modeling advancements, a robust consensus has emerged that forecast combination frequently outperforms individual specifications, yielding superior predictive accuracy and resilience against model misspecification. In the context of macroeconomic density forecasting, this approach has gained significant traction; for instance, \cite{Chernis23} utilize regression tree-based methods to synthesize disparate density forecasts. Consequently, forecast combination has evolved from a supplementary technique into a cornerstone of macroeconomic forecasting, providing a systematic framework for navigating the inherent uncertainties of complex economic environments.

In particular, Bayesian predictive synthesis \citep[BPS;][]{MW19} offers a theoretically coherent probabilistic framework for synthesizing predictive distributions from multiple agent models. 
While BPS has been successfully applied across  various forecasting contexts \citep[e.g.,][]{McAlinn20,kobayashi2024predicting,tallman2024bayesian}, including scalable approaches for large-scale forecasting problems developed \citep{Chernis24}, extant BPS methodologies remain predominantly centered on combining predictive means or full predictive densities. 
Consequently, there remains a critical methodological gap; there approaches do not directly target the specific quantiles of the predictive distribution, which a re essential for precise risk characterization.

This paper proposes a dynamic Bayesian regression quantile synthesis (DRQS), a novel approach designed specifically to bridge these methodological gaps. 
Our approach synthesizes quantile-specific predictive information from multiple agent models by embedding a dynamic latent factor quantile linear structure directly into the syntehsis function. 
Consequently, the DRQS model can be viewed as an extension of the dynamic quantile linear model \citep{DQLM} and dynamic BPS \citep{MW19}, wherein the latent predictors represent the quantile forecasts of the agents. 

Furthermore, we extend the univariate DRQS specifications to a multivariate setting by incorporating a time-varying latent factor structure for the synthesis weights, which captures cross-series dependencies and permits the model to adapt to shifting economic environments. 
The model is called the factor DRQS (FDRQS). 
For instance, while parsimonious models may suffice during periods of economic stability, out framework dynamically reallocates wieght toward more flexible specifications during structural breaks or episodes of extreme market stress. 
By synthesizing quantile forecasts within a dynamic and probabilistically coherent manner, the proposed DRQS approach yields superior predictive distributions for characterizing tail-risk dynamics. 

Regarding combining quantile forecasts, \cite{QBMA} proposed Bayesian model averaging for dynamic quantile regression, where forecasts are combined through convex combinations of quantile forecasts. 
However, such approaches typically do not provide a unified probabilistic framework for synthesizing quantile-specific predictive information across multiple models. 
Their particle-based computational scheme facilitates online weight updates, its implementation and hyperparameter tuning can be cumbersome. 

In contrast, the proposed DRQS framework offers a more streamlined and robust alternative. 
The posterior computation for DRQS consists of the simple forward filtering and backward sampling algorithm within a conditionally Gaussian DLM, after augmenting the latent variables. 
Moreover, recognizing that macroeconomic risks are frequently interconnected across borders or sectors, our approach is specifically designed to handle multivariate settings, capturing the latent dependencies inherent in globalized economic data. 

This paper is organized as follows.
Section~\ref{sec:DRQS} introduces DRQS the framework with the model specification for univariate time series and illustrates its utility through the application to US inflation-at-risk. 
Section~\ref{sec:fdrqs} extends the model to a multivariate setting by incorporating a latent factor structure for the synthesis weights, enabling the model to account for cross-series dependencies. 
In Section~\ref{sec:gar}, we evaluate the performance of quantile forecasting of FDRQS through a comprehensie analysis of global global growth-at-risk.
The concluding remarks are given in Section~\ref{sec:conc}.

\section{Dynamic regression quantile synthesis}
\label{sec:DRQS}

\subsection{Bayesian predictive synthesis}
Consider a multivariate time series data denoted by $\y_t=(y_{1t},\dots,y_{Nt})'$ for $t=1,2,\dots,T$. 
An analyst elicits $J$ agent models with the predictive density denoted by $h_{tj}(\y_{t})$ for $j=1,\dots,J$. 
Then to predict $\y_t$ at the time $t-1$, the collection of $J$ predictive densities forms an information set denoted by $\mathcal{H}_t=\left\{h_{t1}(\y_{t}),\dots,h_{tJ}(\y_{t})\right\}$. 
The available information is $\left\{\y_{1:t-1}, \mathcal{H}_{1:t}\right\}$. 
The BPS considers the following synthesised predictive distribution \citep{MW19, masuda2024proofs}:
\begin{equation}\label{eqn:bps}
    p(\y_t|\bPhi_t,\y_{1:t-1},\mathcal{H}_{1:t})\equiv p(\y_t|\bPhi_t,\mathcal{H}_t)=\int\alpha(\y_t|\f_t,\bPhi_t)\left[\prod_{j=1}^J h_{tj}(\f_{tj})\right]d\f_t, 
\end{equation}
where $\alpha(\y_t|\f_t,\bPhi_t)$ denotes the synthesis function which controls how to combine $J$ predictions from the agent models,  $\bPhi_t$ is the collection of parameters of the synthesis function, including the synthesis weights for the $J$ predictions, and $\f_t$ is the collection of $\f_{tj}=\left\{f_{1t1},\dots,f_{Ntj}\right\}_{j=1}^J$ which denotes the draws from the predictive densities and are considered latent factors or latent predictors.  
The design of the synthesis function and prior structure of the synthesis weights are left to the analyst's discretion. 
Thus, the model can vary depending on the purpose of the analysis and the type and potential dependence structure of the data. 

For example, to synthesize forecasts on some multiple macro economic variables, \cite{McAlinn20} employed the multivariate dynamic linear model (DLM) with latent predictors, using the multivariate normal density for the synthesis function and the random walk for the synthesis weights. 
The model is specified in the form given by
\begin{equation}\label{eqn:dlm}
\begin{split}
\y_t &= \F_t\bthe_t + \v_t,\quad \v_t\sim N(\zero, \V_t)\\
\bthe_t &= \bthe_{t-1} + \w_t,\quad \w_t\sim N(\zero, \W_t), 
\end{split}
\end{equation}
where $F_t$ is the $N\times N(J+1)$ matrix that includes $1$ and $\f_{tj}$, 
$\bthe_t$ is the $N(J+1)$ vector of synthesis weights (see Section \ref{sec:fdrqs} for more details), $\V_t$ and $\W_t$ are the covariance matrices of the observation and state equations, respectively. 
The latent predictors $\f_{tj}$ follow the multivariate normal distribution $N(\a_{tj}, \A_{tj})$ with $\a_{tj}$ and $\A_{tj}$ equal to the mean and variance, respectively, of the corresponding  predictive distribution of $j$th agent model, whose density corresponds to $h_{tj}(\f_{tj})$.

\subsection{DRQS for single time series} 
We first develop DRQS for a single time series denoted by $y_t$. 
For forecasting quantiles of the target variable, it is natural to follow the Bayesian quantile regression approach of \cite{YU01} and employ the density of the asymmetric Laplace distribution for the synthesis function. 
The density of the asymmetric Laplace distribution, denoted by $AL(\tau,\sigma)$, is given by
\begin{equation}\label{eqn:AL}
\alpha_\tau(\epsilon|\sigma)=\frac{\tau(1-\tau)}{\sigma}\exp\left\{-\rho_\tau\left(\frac{\epsilon}{\sigma}\right)\right\},\quad \epsilon\in\mathbb{R},
\end{equation}
for $t=1,\dots,T$, where $\sigma>0$ is the scale parameter, $\tau\in(0,1)$ is the shape parameter and $\rho_\tau(u)=u(\tau-I(u<0))$ is the so-called check function for quantile regression \citep{Koenker2005}. 
The asymmetric Laplace distribution is frequently used as an error term in a Bayesian quantile regression model, since $\tau$th quantile of the asymmetric Laplace distribution is equal to zero: $\int_{-\infty}^0\alpha_\tau(\epsilon|\sigma)d\epsilon =\tau$. 

Then, similar to \eqref{eqn:dlm}, we consider DRQS in the form of the latent factor dynamic quantile linear model for $\tau$th quantile given by
\begin{equation}\label{eqn:drqs}
\begin{split}
y_t &= \F_{\tau,t}'\bthe_{\tau,t} + \epsilon_{\tau,t},\quad \epsilon_{\tau,t}\sim AL(\tau, \sigma_{\tau,t})\\
\bthe_{\tau, t} &= \bthe_{\tau,t-1} + \w_{\tau,t},\quad \w_{\tau,t}\sim N(\zero, \sigma_{\tau,t}\W_{\tau,t}), 
\end{split}
\end{equation}
where $\F_{\tau,t}=(1,\f_{\tau,t}')'$, $\f_{\tau,t}=(f_{\tau,t1},\dots,f_{\tau,tJ})'$ and $\bPhi_{\tau,t}=\left\{\bthe_{\tau,t},\sigma_{\tau,t}\right\}$. 
Since $\epsilon_{\tau,t}$ follows the asymmetric Laplace, the conditional quantile at time $t$ is  $Q_t(\tau|\f_{\tau,t},\bthe_{\tau,t})=\F_{\tau,t}'\bthe_{\tau,t}$. 
This model can be seen as a quantile extension of the DLM used in the mean synthesis \citep{MW19}, which used the normal density for the synthesis function, and an extension of DQLM \citep{DQLM} with the latent predictors $\F_{\tau,t}$ (see also Section~\ref{sec:dqlm} of the Supplementary Materials).

The DRQS model is completed with the processes for the time varying $\W_t$ and $\sigma_t$, and the prior settings for $\f_{\tau,t}$, $\theta_{\tau,0}$  and $\sigma_{\tau,0}$. 
For $f_{\tau,tj}$, we assume $f_{\tau,tj}\sim N(a_{\tau,tj},A_{\tau,tj})$ where $a_{\tau,tj}$ and $A_{\tau,tj}$ denote the predictive mean and variance of $\tau$th quantile. 
We consider the standard specification in DLM for $\W_t$ with the discount factor denoted by $\delta_\tau\in(0,1]$ \citep{WH97,MW19}. 
For $\sigma_t$, it is assumed to follow the gamma-beta random walk \citep[GBRW;][]{WH97,Prado} model, often used in the context of DLM, given by
\begin{equation}\label{eqn:gbrw}
\sigma_{\tau,t}=\frac{\beta_\tau}{\gamma_{\tau,t}}\sigma_{\tau,t-1},\quad \gamma_{\tau,t}\sim Beta\left(\frac{\beta_\tau n_{\tau,t-1}}{2},\frac{(1-\beta_\tau)n_{\tau,t-1}}{2}\right), 
\end{equation}
where $\beta_\tau\in(0,1]$ denotes the discount factor. 
For $\bthe_{\tau,0}$ and $\sigma_{\tau,0}$, as in \cite{MW19}, we assume the conjugate prior $\bthe_{\tau,0}|\sigma_{\tau,0}\sim N(\m_0,(\sigma_{\tau,0}/s_0)\C_0)$ and $\sigma_{\tau,0}^{-1}\sim Ga(n_0,s_0)$. 
While we have used the subscript $\tau$ to denote that the quantities depend on the targeted quantile level $\tau$ so far, in what follows it is suppressed for notational simplicity unless it is ambiguous.  

The model is estimated using the Markov chain Monte Carlo method.  
We employ a mixture representation for the asymmetric Laplace distribution, allowing the observation equation in \eqref{eqn:drqs} to be rewritten as
$
y_t|\f_t,\bthe_t,\sigma_t,v_t\sim N(\F_{t}'\bthe_{t} + \kappa_{1} v_{t},\sigma_{t}\kappa_{2}v_{t})$
where $v_{t}\sim Exp(\sigma_{t})$, 
$\kappa_{1}=\frac{1-2\tau}{\tau(1-\tau)}$ and $\kappa_{2}=\frac{2}{\tau(1-\tau)}$, as often done in the Bayesian quantile regression literature \citep{kozumi11}. 
Therefore, conditional on $\f_t$ and $v_t$, \eqref{eqn:drqs} is regarded as the dynamic linear model, to which the forward filtering backward sampling (FFBS) algorithm \citep{fruhwirth1994data, DQLM} is applied. 
The detail of the Gibbs sampler is provided in Section~\ref{sec:gibbs_s} of the Supplementary Materials. 

\subsection{Forecasting quantiles with DRQS}\label{sec:pred}
Here we describe how to obtain the 1-step ahead forecast of $\tau$th quantile for time $t=T+1$ at $t=T$, denoted by $Q_{{T+1}}(\tau)$, targeting the posterior predictive distribution of $Q_{T+1}(\tau)$  given by:
\begin{equation}\label{eqn:pred}
\begin{split}
    p\left(Q_{{T+1}}(\tau)|\mathcal{H}_{T+1}\right)
    =\iint& p\left(Q_{{T+1}}(\tau|\f_{T+1},\bthe_{T+1})|\f_{T+1},\bPhi_{T+1},\mathcal{H}_{T+1}\right)\prod_{j=1}^Jh_{T+1,j}(f_{T+1,j})\\
    &\quad\quad \times p(\bPhi_{T+1}|\bPhi_{1:T},\mathcal{H}_T)p(\bPhi_{1:T}|\mathcal{H}_T)d\bPhi_{1:T+1}d\f_{T+1}, 
\end{split}
\end{equation}
where $p\left(Q_{{T+1}}(\tau|\f_{T+1},\bthe_{T+1})|\f_{T+1},\bPhi_{T+1},\mathcal{H}_{T+1}\right)$ is the predictive distribution of the conditional quantile, which has a point mass at $Q_{T+1}(\tau|\f_{T+1},\bthe_{T+1})=\F_{T+1}'\bthe_{T+1}$,  $p(\bPhi_{T+1}|\bPhi_{1:T},\mathcal{H}_T)$ is the predictive distribution of the state $\bPhi_{T+1}$ and $p(\bPhi_{1:T}|\mathcal{H}_T)$ is the posterior distribution of $\bPhi_{1:T}$. 

To sample from this predictive distribution, we follow the standard procedure of BPS based on DLM  as follows \citep[see also][]{MW19, McAlinn20}. 
Given an MCMC draw from the posterior $p(\bPhi_{1:T}|\mathcal{H}_T)$, we first draw $\bPhi_{T+1}$ from $p(\bPhi_{T+1}|\bPhi_{1:T},\mathcal{H}_T)$ based on the state equations described in \eqref{eqn:drqs} and \eqref{eqn:gbrw}. 
Then, we draw $\f_{T+1,j}$ from the predictive distribution of the agent with the density $h_{T+1,j}(f_{T+1,j})$ for $j=1\dots,J$ to obtain the forecast of the conditional quantile at $t=T+1$ given by $Q_{T+1}(\tau|\f_{T+1},\bthe_{T+1})=\F_{T+1}'\bthe_{T+1}$. 
Repeating this to each posterior draw, we obtain the draws from \eqref{eqn:pred}, based on which we get the point forecast and prediction interval.

\subsection{Forecasting US inflation-at-risk}\label{sec:iar}
We first provide a small demonstration of the proposed DRQS  using the US inflation rate data. 
The inflation rate $y_t$ is defined by $y_t=400 \log (Y_t/Y_{t-h})/h$, where $Y_t$ denotes the quarterly consumer price index (CPI). 
We consider $h=1$ and $4$, which correspond to the annualised quarterly inflation rate and average annual inflation rate. 
Figure~\ref{fig:supp_iar_data} of the Supplementary Material provides the times series plots of the data. 

The agent models used in this analysis are $J=4$ DQLMs \citep[][see \ref{sec:dqlm} of the Supplementary Material]{DQLM}, which have different predictor specifications on the combinations of long-term inflation expectation, $\LTE_t$, unemployment gap, $\EMP_t$, inflation of the import price index $\IMP_t$, and their lags, similar to \cite{wang2024inflation}. 
The models are denoted by DQLM1, DQLM2, DQLM3 and DQLM4. 
Let us denote the predictor vector of DQLM by $\x_t$. 
The specifications of $\x_t$ are as follows: 
$\x_t=(1,y_{t-1})'$ for the simplest simplest DQLM1, 
$\x_t=(1,y_{t-1},\LTE_{t-1})'$ for DQLM2,
$\x_t=(1,y_{t-1},\allowbreak \LTE_{t-1},\allowbreak \EMP_{t-1}, \allowbreak \IMP_{t-1})'$ for DQLM3 and 
$\x_t=(1,y_{t-1}, y_{t-2},\allowbreak \LTE_{t-1}, \LTE_{t-2}, \allowbreak \EMP_{t-1}, \allowbreak \IMP_{t-1})'$ for DQLM4.

We first fitted DQLMs using the data between 1983Q1 and 1997Q4 to obtain forecasts for 1998Q1 for nineteen quantiles $\tau=0.05,0.1,\dots,0.9,0.95$. 
Then we expanded the forecast window up to 2019Q4. 
For all DQLMs, we used the discount factor $0.95$ for the covariance matrix of the state equation (see Section~\ref{sec:dqlm} of the Supplementary Materials). 
Then, for each $\tau$, we fitted the DRQS model starting with the data between 1998Q1 and 2013Q1, and then expanded the forecast window up to 2019Q4. 
The hyper parameters for DRQS are set as follows: $\m_0=(0,1/4,1/4,1/4,1/4)'$ so that the agent models initially receive the equal weights, $\C_0=\diag(1000,1,1,1,1)$ to allow the initial intercept term to vary across different $\tau$, and $n_0=s_0=0.01$ as we do not have information on the initial state of the scale parameter. 
For the discount factors, we set $\delta=\beta=0.9$ to ensure the synthesis weights adapt quickly to new information.  
For each model and forecast window, the MCMC algorithms were run to obtain 3000 posterior draws after discarding the first 1000 as burn-in. 

The predictive performance of the models for the quantiles is evaluated based on the quantile-weighted continuous ranked probability score (CRPS) proposed by \cite{Gneiting01072011}. 
CRPS for one-step ahead quantile forecast for time $t$ is defined as
\begin{equation}\label{eqn:crps}
    \CRPS_{t}^{(m)}=\int_0^1 2\left(I\left\{y_{t}<\hat{Q}^{(m)}_{t}(\tau)\right\}-\tau\right)\left(\hat{Q}^{(m)}_{t}(\tau)-y_{t}\right)\nu(\tau)d\tau,
\end{equation}
where $y_{t}$ is the observed value to be predicted, $\hat{Q}^{(m)}_{t}(\tau)$ is the $\tau$th quantile forecast of the model $m$ and $\nu(\tau)$ is the weighting function. 
We use the posterior predictive mean of $Q_{t}(\tau)$ for $\hat{Q}_{t}(\tau)$. 
Three choices of $\nu(\tau)$ are considered: $\nu_\texttt{none}(\tau)=1$, $\nu_\texttt{right}(\tau)=\tau^2$ and $\nu_\texttt{left}(\tau)=(1-\tau)^2$. 
As the subscripts suggest, while the first one does not put specific emphasis on any part of the distribution (unweighted), 
the latter two focus on the forecast performance in the respective tails. 
The integral with respective $\tau$ in \eqref{eqn:crps} is approximated by the trapezoidal rule over the grid $0.05,0.1,\dots,0.95$. 
For interpretability of the results, below we often use the 
cumulative CRPS for DRQS, DQLM2, DQLM3 and DQLM4 relative to that for DQLM1 (RCS; relative cumulative score) up to time $t^*$: 
\[
\RCS_{t^*}^{(m)}=\frac{\sum_{t=t_s}^{t^*}\CRPS_t^{(m)}}{\sum_{t=t_s}^{t^*}\CRPS_t^{(\DQLM1)}}, 
\] 
where $m\in\left\{\DRQS,\DQLM2,\DQLM3, \DQLM4\right\}$, $t_s=\text{2014Q2}$ and $t^*$ ranges from 2014Q2 to 2019Q4. 
The RCS below $1$ indicates that the quantile forecasting under the corresponding model performs better than under DQLM1. 

\begin{figure}[H]
    \centering
    \includegraphics[width=\textwidth]{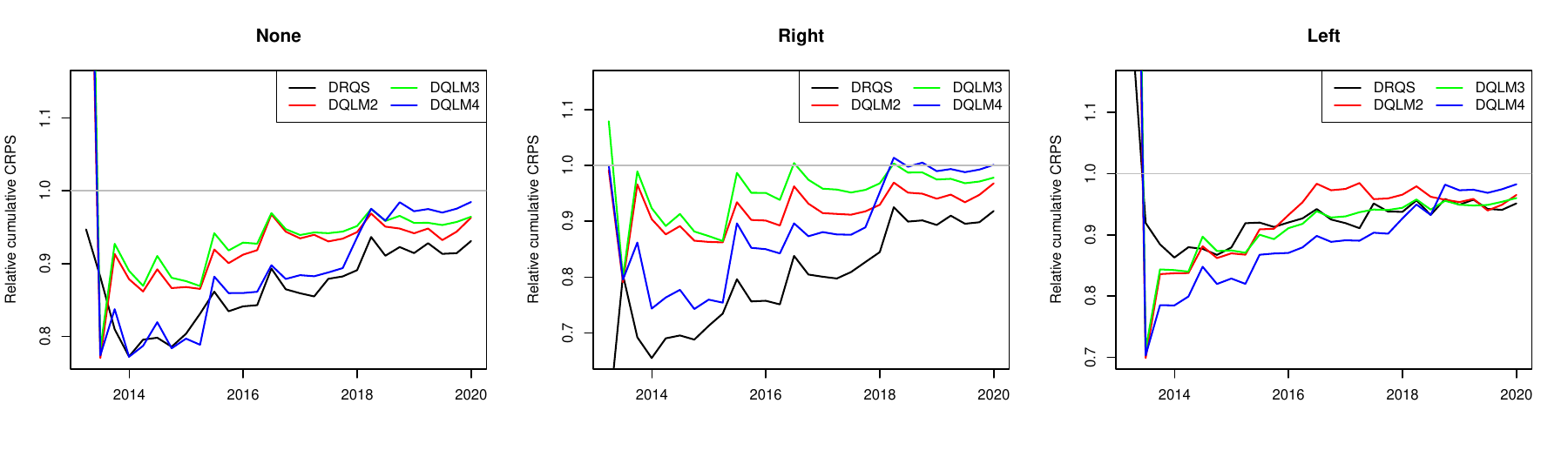}
    \includegraphics[width=\textwidth]{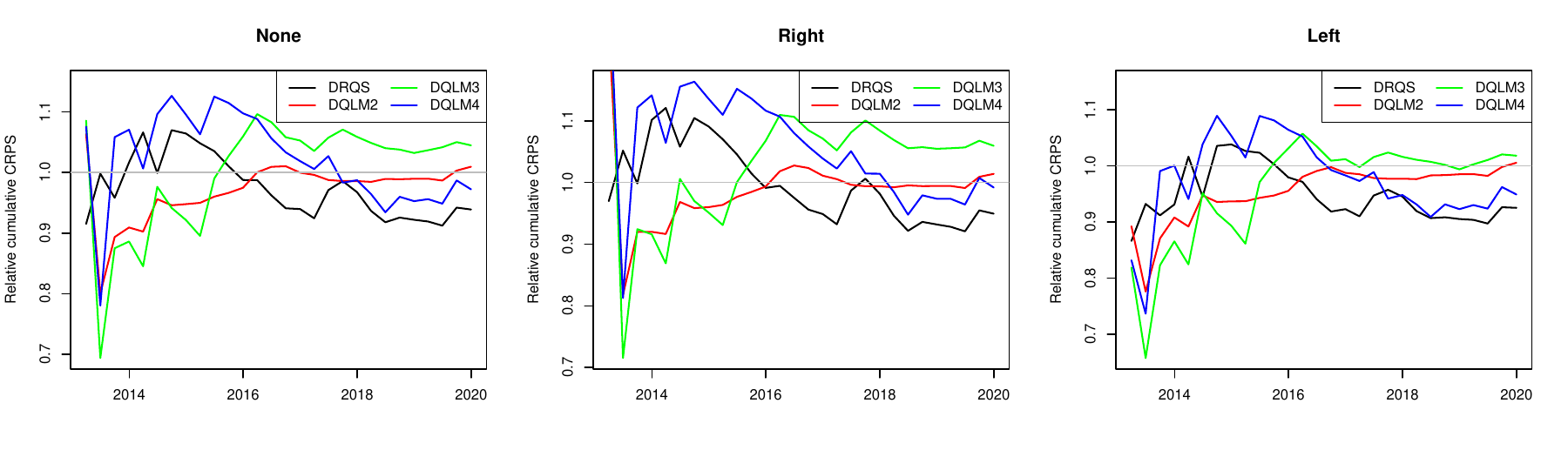}
    \caption{Cumulative CRPS relative to DQLM1 for $h=1$ (top panels) and $h=4$ (bottom panels).}
    \label{fig:crps_iar}
\end{figure}

Figure~\ref{fig:crps_iar} presents the times series plot of RCS. 
The figure shows that the proposed DRQS performs reasonably well for $h=1$, achieving the smallest RCS for the \texttt{none} and \texttt{right} cases throughout most of the sample. 
While DQLM4 initially shows the smallest RCS for the \texttt{left} case, those for DQLM2, DQLM3, DQLM4 and DRQS become comparable towards the end of the data period. 
Similarly, for $h=4$, DRQS maintains the smallest RCS for the second half of the data period for all three cases.  

\begin{figure}[H]
    \centering
    \includegraphics[width=\textwidth]{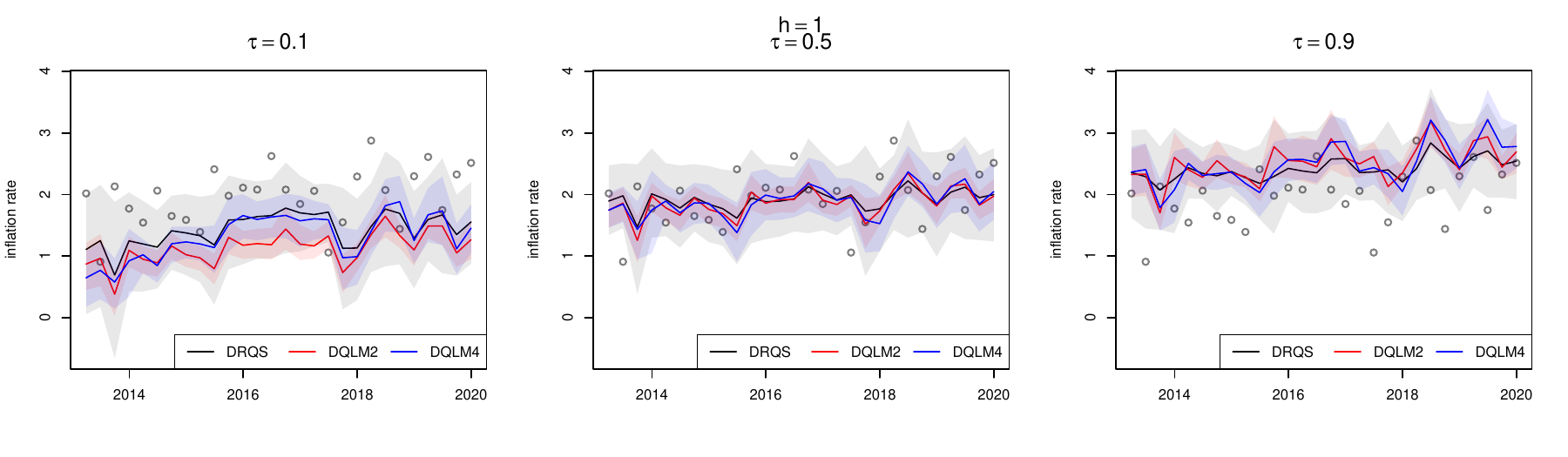}
    \includegraphics[width=\textwidth]{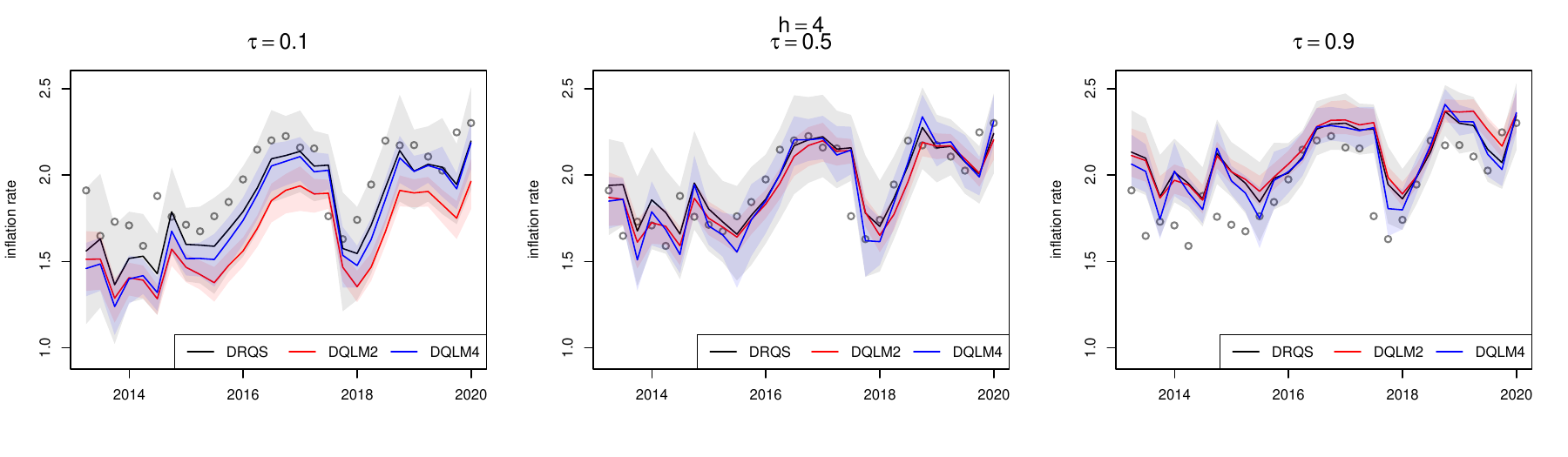}
    \caption{One-step ahead predictive means (solid lines) and 95\% intervals (shaded areas) of $0.1$th, $0.5$th and $0.9$th quantiles under DRQS, DQLM2 and DQLM4 for $h=1$ (top panels) and $h=4$ (bottom panels). Points indicate the observed inflation rates. }
    \label{fig:pred_iar}
\end{figure}

Figure~\ref{fig:pred_iar} presents the one-step ahead predictive means and 95\% intervals under subset of the models, DRQS, DQLM2 and DQLM4, for $\tau=0.1$, $0.5$ and $0.9$ for $h=1$ and $4$. 
The points in the figure indicate the observed inflation rates. 
These models are selected, for visibility, based on CRPS in Figure~\ref{fig:crps_iar} and also on the synthesis weights shown in Figures~\ref{fig:supp_iar_theta_1} and \ref{fig:supp_iar_theta_4} of the Supplementary Materials. 
The latter figures show that the posterior means of the weights for DQLM2 and DQLM4 are positive for most or all of the data period, depending on $\tau$ and $h$, while those for DQLM1 and DQLM3 are negative or close to zero. 
In Figure~\ref{fig:pred_iar}, DRQS and DQLM4 generally produced similar patterns, especially for $h=4$, for which the synthesis weight for DQLM4 is dominant. 
However, the 95\% intervals for DRQS are much wider than the agent models, reflecting the additional layer of uncertainty inherent in the synthesis process.  
Furthermore, it is observed that DQLM2 produced much lower $0.1$th quantile forecasts compared to both DRQS and DQLM4.

\section{Factor DRQS for multiple time series}\label{sec:fdrqs}
This section extends DRQS to deal with multiple time series $\y_t=(y_{1t},\dots,y_{Nt})'$ simultaneously. 
Also, let us denote the $N$ vector of error terms as $\beps_{t}=(\epsilon_{1t},\dots,\epsilon_{Nt})'$. 
Then we begin with the following latent factor DLM, which is similar to that of \eqref{eqn:dlm}: 
\begin{equation}\label{eqn:mdrqs}
\begin{split}
\y_t &= \F_{t}\bthe_{t} + \beps_{t},\quad \beps_{it}\sim AL(\tau,\sigma_{it}),\\
\bthe_{t}&=\bthe_{t-1} + \tilde{\w}_t, \quad \tilde{\w}_t\sim N(\zero, \tilde{\W}_t), 
\end{split}
\end{equation}
where 
$\F_{t}=\begin{bmatrix}
 \I_{N}& \F_{t1}\dots\F_{tJ}
\end{bmatrix}
$ is the $N\times N(J+1)$ matrix with the $N\times N$ identity matrix $\I_N$ and $N\times N$ matrix $
\F_{tj}=\diag(f_{1tj},\dots,f_{Ntj})$ for $ j=1,\dots,J$, $f_{itj}$ denotes the latent predictor corresponding to $j$th agent in $i$th series at $t$th period, 
$\bthe_{t}=(\bthe_{t0}', \bthe_{t1}',\dots,\bthe_{tJ})'$ is the $N(J+1)$ vector of synthesis weights with $\bthe_{tj}=(\theta_{1tj},\dots,\theta_{Ntj})'$ for $j=0,1,\dots,J$, and $\tilde{\W}_t$ is the $N(J+1)\times N(J+1)$ covariance matrix of the state equation. 
Note that the error terms $\epsilon_{it}$ independently follow the asymmetric Laplace distribution. 
The time varying scale parameter $\sigma_{it}$ follows GBRW with the discount factor $\beta_i\in(0,1]$ independently for $i=1,\dots,N$ with the initial prior $\sigma_{i0}^{-1}\sim Ga(n_{i0},s_{i0})$ as in the previous section. 

Using this specification \eqref{eqn:mdrqs} as it is would face the following potential limitations. 
Firstly, when the number of time series $N$ and/or the number of agents $J$ is large, the cost of the posterior computation for the synthesis weights $\bthe_{t}$ increases significantly, since it involves many matrix operations, especially matrix inversions. 
Even in our real data application in Section~\ref{sec:gar} with relatively small $N=18$ and $J=4$, the dimension of $\bthe_t$ is ninety. 
Secondly, the observation equation of \eqref{eqn:mdrqs} just stacks $N$ observation equations of the form of \eqref{eqn:drqs} which independently follow the univariate asymmetric Laplace distributions. 
Therefore, the cross-sectional information would not be sufficiently leveraged. 
While it is possible to explicitly introduce dependence by adopting, such as, the multivariate asymmetric Laplace distribution as considered by \cite{Iacopini}, the existence of the correlation matrix renders the posterior computation cumbersome. 

To overcome these limitations, we introduce the factor model structure into the synthesis weights, following the recent stream of quantile regression modelling for multiple outcomes \citep[e.g.,][]{Chen21, Clark24, Korobilis24}. 
Specifically, we consider introducing the factor structure over the multiple series, then the synthesis weight of $j$th agent model for $i$th series at $t$th period is expressed as
\begin{equation}\label{eqn:factor}
    \theta_{itj}=\blam_{ij}'\u_{tj},\quad i=1,\dots,N, \quad t=1,\dots,T,\quad j=0,1,\dots,J,
\end{equation}
where $\blam_{ij}=(\lambda_{i1j},\dots,\lambda_{iLj})'$ is the $L<N$ vector of factor loadings of $j$th agent model for $i$th series and $\u_{tj}=(u_{t1j},\dots,u_{tLj})'$ is the corresponding vector of latent factors. 
Instead of choosing the number of factors $L$ based on model selection, we adopt the multiplicative gamma process (MGP) of \cite{MGP} for the factor loadings, which gradually shrinks the factor loadings through diminishing variances. 
The model of MGP for the factor loadings is given by
\begin{equation}\label{eqn:mgp}
\begin{split}
    \lambda_{i\ell j}&\sim N(0,\phi_{i\ell j}^{-1}\omega_{\ell j}^{-1}),\quad \phi_{i\ell j}\sim Ga(\nu_j/2,\nu_j/2),\\
    \delta_{1j}&\sim Ga(a_{1j},1),\quad \delta_{\ell j}\sim Ga(a_{2j},1),\quad \ell\geq 2,
\end{split}
\end{equation}
for $j=0,1,\dots,J$, where $\omega_{\ell j}=\prod_{h=1}^\ell \delta_{hj}$. 
For an appropriate combination of values $a_{1j}$ and $a_{2j}$, the prior variance of $\lambda_{i\ell j}$ monotonically decreases with $\ell$ \citep{DURANTE2017198}. 
Therefore, we fit the model with moderate but computationally reasonable numbers of factors $L<N$ and let the MGP shrink the excess factor loadings. 
To complete the factor structure, we assume the random walk process for the time varying latent factors, $\u_t\sim N(\u_{t-1},\W_t)$ for $t=1,\dots,T$, where $\u_t=(\u_{t0}',\u_{t1}',\dots,\u_{tJ}')'$ is the $L(J+1)$ vector of the collection of the latent factors and $\W_t$ is the $L(J+1)\times L(J+1)$ full covariance matrix with the single discount factor $\delta$. 
The initial state is specified as $\u_0\sim N(\m_0,\C_0)$. 

While \cite{Chernis24} also utilized a factor structure within (univariate) BPS, their approach focuses on dimensionality reduction of the agent space $\f_t$ when $J$ is large. 
In contrast, out motivation is the contemporaneous leveraging of information across $N$ times series. By imposing a factor structure on the synthesis weights $\bthe_t$, we capture cross-sectional dependencies and achieve computational efficiency without the complexity of a fully-specified multivariate asymmetric Laplace distribution.

As with the univariate DRQS, we have developed a Gibbs sampler for the posterior computation of FDRQS. 
By leveraging the mixture representation of the asymmetric Laplace distribution, each step of the sampler involves draws from standard conditional distributions. 
This includes an FFBS algorithm for the latent factors $\u_t$ and independent GBRW updates for the scale parameters $\sigma_{it}$. 
Full details of the Gibbs sampler are provided in Section~\ref{sec:gibbs_f} of the Supplementary Materials.

\section{Forecasting global growth-at-risk}\label{sec:gar}
\subsection{Data and setting}
\begin{figure}[H]
    \centering
    \includegraphics[width=\linewidth]{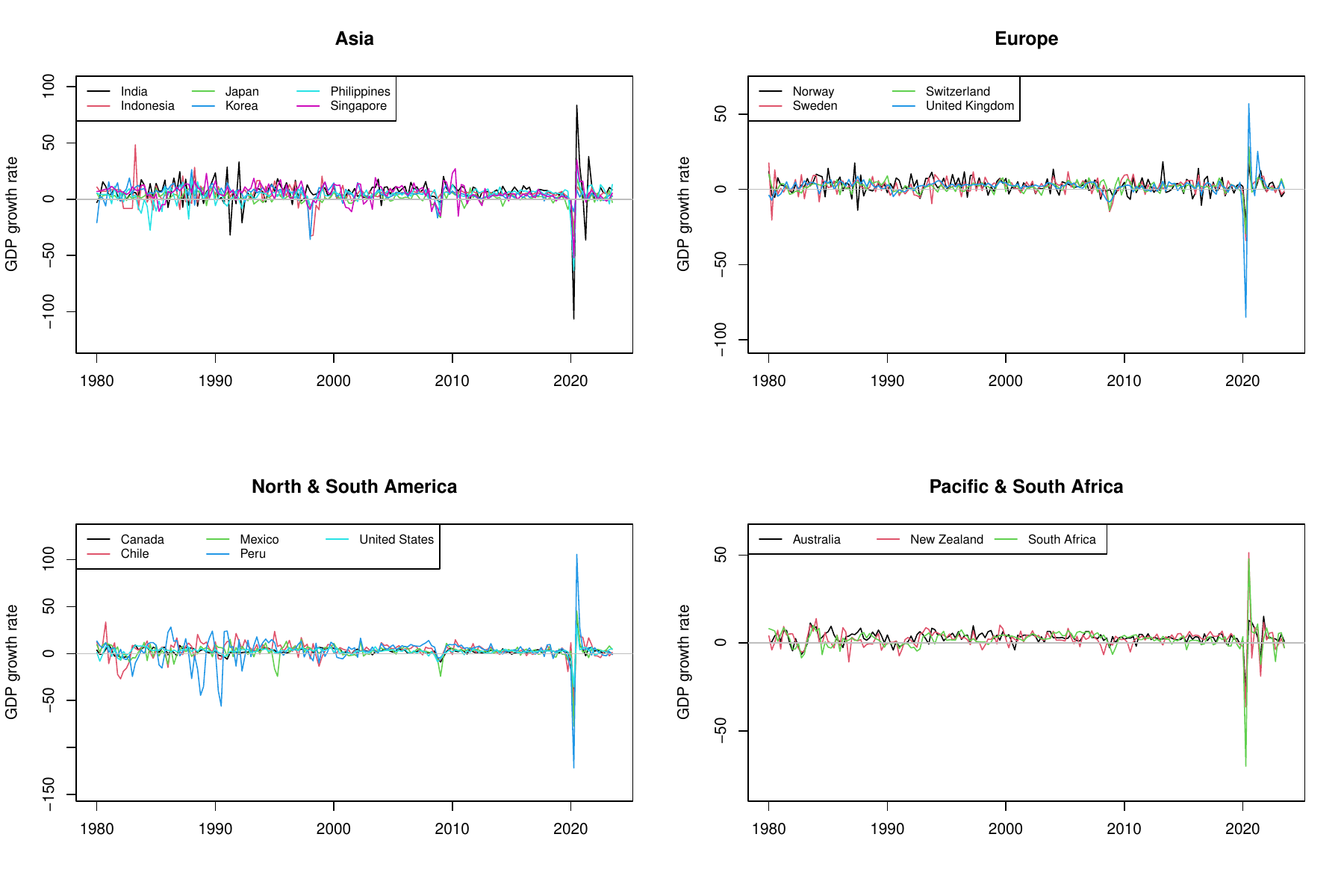}
    \caption{Annualised quarterly growth rates ($h=1$) of the eighteen countries divided by the regions.}
    \label{fig:gar_data_1}
\end{figure}

Now we demonstrate the proposed FDRQS for forecasting global GDP growth rate quantiles. 
We use the dataset compiled by \cite{Maha}. 
Out of 25 countries included in the dataset, we have excluded those where quarterly real GDP is not directly available in the early data periods and use the data for the remaining $N=18$ countries: Australia, Canada, Chile, India, Indonesia, Japan, Korea, Mexico, New Zealand, Norway, Peru, Philippines, Singapore, South Africa, Sweden, Switzerland, 
the United Kingdom and the United States. 
As in Section~\ref{sec:iar}, we let $y_t=400\log(Y_t/Y_{t-h})/h$ where $Y_t$ is the quarterly real GDP with $h=1$ and $4$. 
Figure~\ref{fig:gar_data_1} presents the time series plots of the annualised quarterly growth rates ($h=1$) of the eighteen countries divided into the four regions, Asia, Europe, North and South America, and Pacific and South Africa, between 1980Q4 and 2023Q3.  
It is observed that the countries have been growing over time, but they have also experienced occasional shocks that can be global or limited to subsets of the countries. 
See also Figure~\ref{fig:supp_gar_data_4} of the Supplementary Materials for the average annual growth rate ($h=4$).

We consider $J=4$ agent models in this analysis: three DQLMs fitted separately to each series and a factor quantile BART model (FQBART) proposed by \cite{Clark24}. 
FQBART is a Bayesian quantile regression model designed for GDP growth rate for multiple countries, and includes a nonparametric function of predictors based on BART and factor structure with stochastic volatility to account for multiple countries and quantiles simultaneously (see Section~\ref{sec:fqbart} for details). 

For DQLMs, we consider three models with different combinations of predictors, inflation rate, $\Dp_{it}$, log of nominal equity price index deflated by CPI, $\eq_{it}$, log of the exchange rate of $i$th country in US dollars deflated by CPI of $i$th country, nominal short-term interest rate per quarter, $\rr_{it}$, nominal long-term interest rate per quarter, $\lr_{it}$, and their lags. 
The predictor specifications are as follows: 
$\x=(1,y_{i,t-1}, \allowbreak \Dp_{it}, \allowbreak \ep_{it},  \allowbreak \eq_{it}, \allowbreak \rr_{it}, \allowbreak \lr_{it})'$ for DQLM1 and FQBART,
$\x=(1,y_{i,t-1}, y_{i,t-2}, \allowbreak \Dp_{it}, \allowbreak \ep_{it},  \allowbreak \eq_{it}, \allowbreak \rr_{it}, \allowbreak \lr_{it})'$ for DQLM2 and
$\x=(1, \allowbreak y_{i,t-1}, \allowbreak \Dp_{it}, \Dp_{i,t-1}, \allowbreak \ep_{it}, \allowbreak \eq_{it}, \allowbreak \rr_{it},\rr_{i,t-1}, \allowbreak \lr_{it})'$ for DQLM3. 
The discount factor for DQLMs is set to $0.9$. 

We first fitted DQLMs and FQBART using the data between 1980Q3 and 1994Q4 to obtain the one-step ahead quantile forecasts for 1995Q1 for $\tau=0.05,0.1,\dots,0.9,0.95$ and then expanded the forecast window recursively up to 2023Q3. 
FDRQS and DRQS are initially estimated using the data from 1995Q1 to 2009Q4, with the first out-of -sample forecast generated for 2010Q1. 
Then we expanded the forecast window recursively up to 2023Q3.  

The hyper parameters for DRQS are the same as those in Section~\ref{sec:iar} other than that the discount factors are set to $\delta=\beta=0.85$ such that the model can quickly adapt to the change in the agent models' performance. 
The hyper parameters used for FDRQS are similar to those for DRQS  as follows. 
For the hyper parameters for the initial state $\u_0$, we set $\m_0=c(\zero_{L}, 1/4 \mathbf{1}_{LJ})'$, where $\zero_{L}$ and $\mathbf{1}_{LJ}$ denote the $L$ vector of zeros and $LJ$ vector of ones, respectively, and $\C_0=\diag(1000,\dots,1000,1,\dots,1)$, where the first $L$ diagonal elements are $1000$ and the remaining $LJ$ elements are $1$. 
For the time varying scale parameters, we set $n_{i0}=s_{i0} =0.001$ for $i=1,\dots,N$. 
The discount factors for FDRQS are also set to $\delta=0.85$ and $\beta_i=0.85$ for $i=1,\dots,N$;  
the synthesis weights derived from $\u_t$ can vary sufficiently over time avoiding over smoothing. 
For the parameters in MGP, $\nu_j=3$, $a_{1j}=2.5$ and $a_{2j}=3.5$ for $j=0,1,\dots,J$. 
These values for $a_{1j}$ and $a_{2j}$ ensure that the variance of factor loading shrinks towards zero as $\ell$ increases \citep{DURANTE2017198}. 
We set the number of factors to be $L=5$.

For each model and forecast window, the MCMC algorithms were run to obtain 3000 posterior draws after 1000 burn-in period. 
For FDRQS and DRQS, we also used GNU Parallel \citep{tange} for the parallel computing over $\tau$'s. 

Similar to Section~\ref{sec:iar}, we compare the forecasting performance for quantiles based on the cumulative CSRPS for $i$th country and summed over countries (RTCS; Total RCS) relative to FQBART, respectively, defined by
\[
\RCS_{it^*}^{(m)}=\frac{\sum_{t=t_s}^{t^*}\CRPS_{it}^{(m)}}{\sum_{t=t_s}^{t^*}\CRPS_{it}^{(\FQBART)}}, \quad
\RTCS_{t^*}^{(m)}=\frac{\sum_{t=t_s}^{t^*}\sum_{i=1}^N\CRPS_{it}^{(m)}}{\sum_{t=t_s}^{t^*}\sum_{i=1}^N\CRPS_{it}^{(\FQBART)}}, 
\]
where $t_s=\text{2010Q1}$ and $t^*$ ranges from 2010Q1 to 2023Q3 in this analysis. 
The RCS or RTCS for a model below one indicate that the performance of the quantile forecasting of the corresponding model is superior to that of FQBART.

\subsection{Result}
\subsubsection{Quantile forecasting performance}
Table~\ref{tab:gar_crps} presents RTCS for $h=1$ and $4$, and RCS for each country in 2023Q3 for $h=1$. 
The table shows that the FDRQS produced the smallest RTCS,  indicating its superior performance for quantile forecasting over the agent models and separately fitted DRQS. 
FDRQS also resulted in the smallest RCS for most countries. 
Table~\ref{tab:gar_crps4} exhibits similar patterns in the case of $h=4$, which shows the profound underperformance of FQBART.

\begin{figure}[H]
    \centering
    \includegraphics[width=\textwidth]{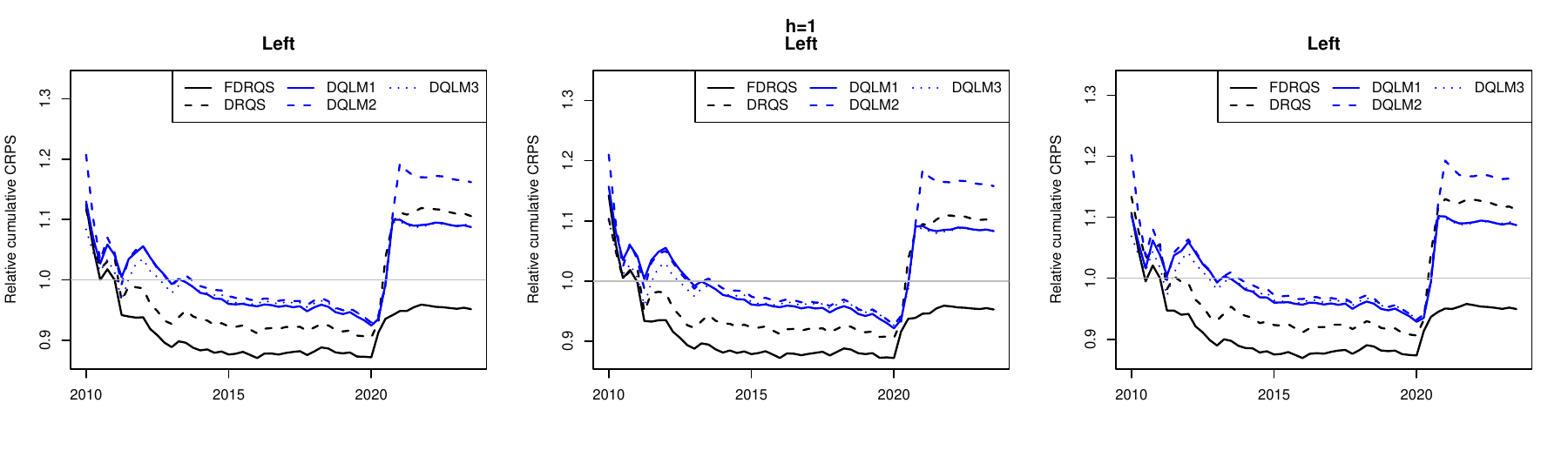}
    \includegraphics[width=\textwidth]{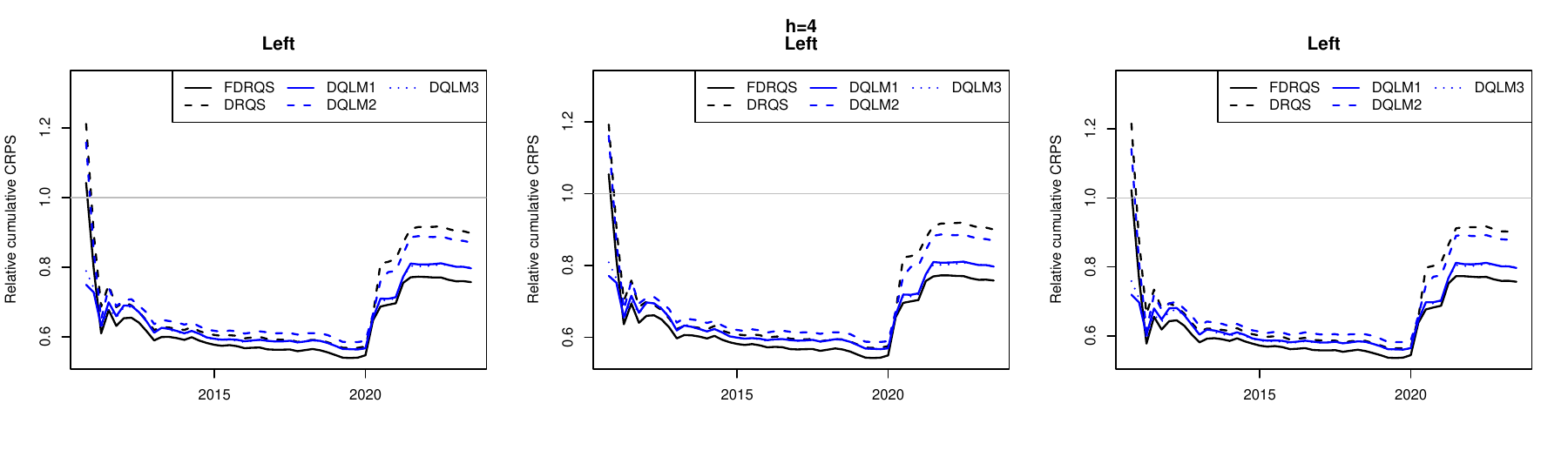}
    \caption{Time series plots of the cumulative total CRPS  relative to FQBART. }
    \label{fig:gar_crps_total}
\end{figure}

To see how RTCS and RCS changed over time in detail, Figure~\ref{fig:gar_crps_total} presents the time series plots of RTCS for $h=1$ and $4$. 
Overall, the figure shows that all models experienced the sudden deterioration in quantile forecasting in 2020 where the COVID pandemic caused global contraction resulting in the massive drops in the growth rates as plotted in Figures~\ref{fig:gar_data_1} and \ref{fig:supp_gar_data_4}. 
Nonetheless, FDRQS retains the smallest RTCS even after 2020 for $h=1$ and $4$. 
That for DRQS is the second smallest in the most of the period before 2020, after which it, however, suddenly incurred larger RCS  than some of the agent models. 
For $h=1$, while DQLMs performed better than FQBART for most of the 2010's, the relative performance sharply deteriorated after 2020, similar to DRQS. 

\begin{figure}[H]
    \centering
    \includegraphics[height=0.98\textheight]{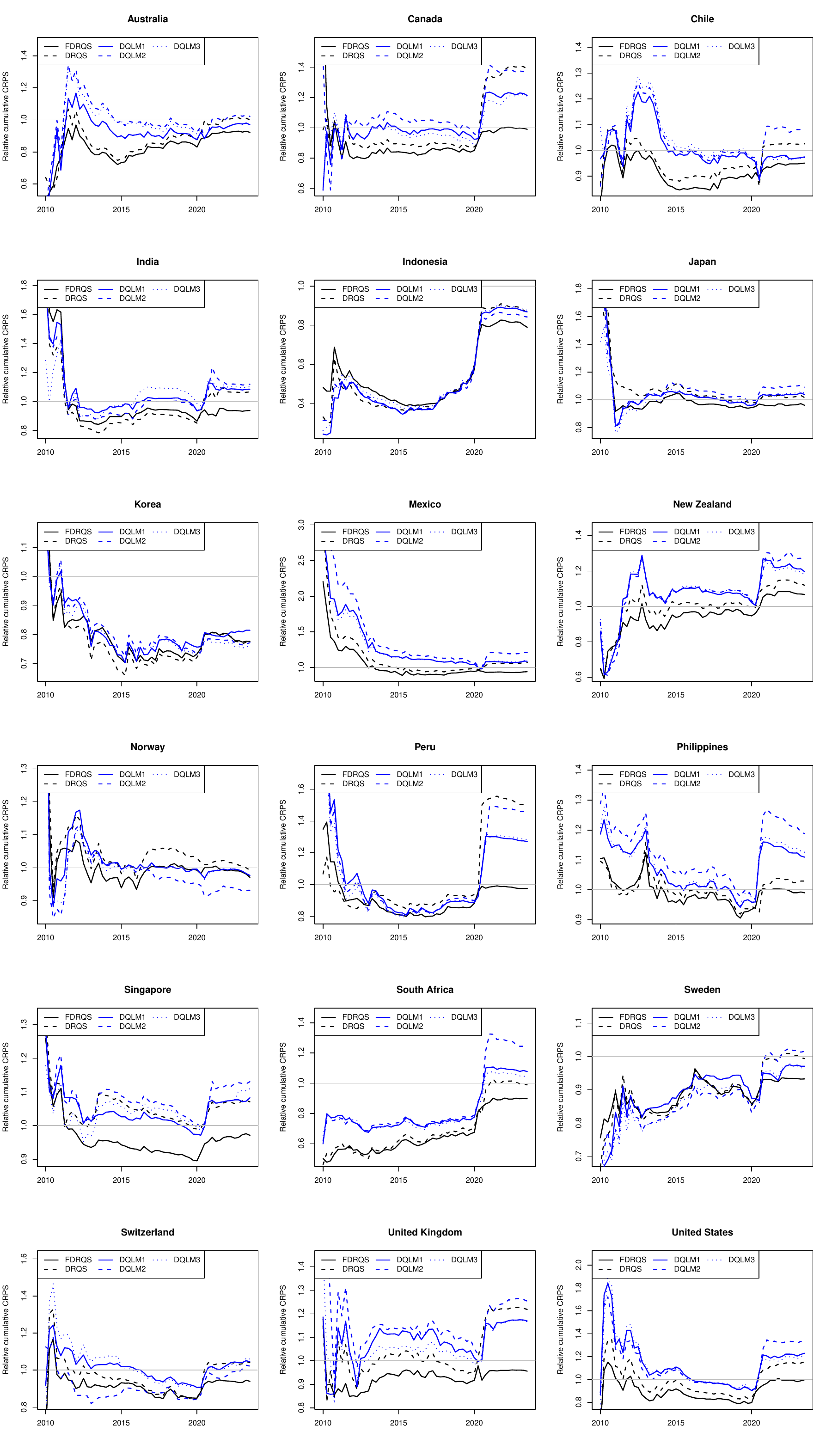}
    \caption{Time series plots of the cumulative CRPS with \texttt{none} relative to FQBART for each contry ($h=1$). }
    \label{fig:gar_crps}
\end{figure}

Figure~\ref{fig:gar_crps} presents the RCS with \texttt{none} for the eighteen countries for $h=1$. 
Consistent with our aggregate findings, FDRQS yielded the smallest cumulative CRPS for majority of countries throughout the data period. 
While the performance of the individual DQLMs and univariate DRQS was substantially destabilized by the COVID-19 shock, FDRQS exhibited notable resilience. 
Wimilar results were obtained for the tail-weighted cases \texttt{right} and \texttt{left}. 
Figure~\ref{fig:supp_gar_crps_4} presents the country-specific RCS with \texttt{none} for $h=4$. 
In this case, FDRQS, DRQS and DQLMs performed comparably prior to 2020; however, FDRQS again demonstrated superior robustness, maintaining stable performance in the post-pandemic period where other models faltered.

\begin{figure}[H]
    \centering
    \includegraphics[width=\textwidth]{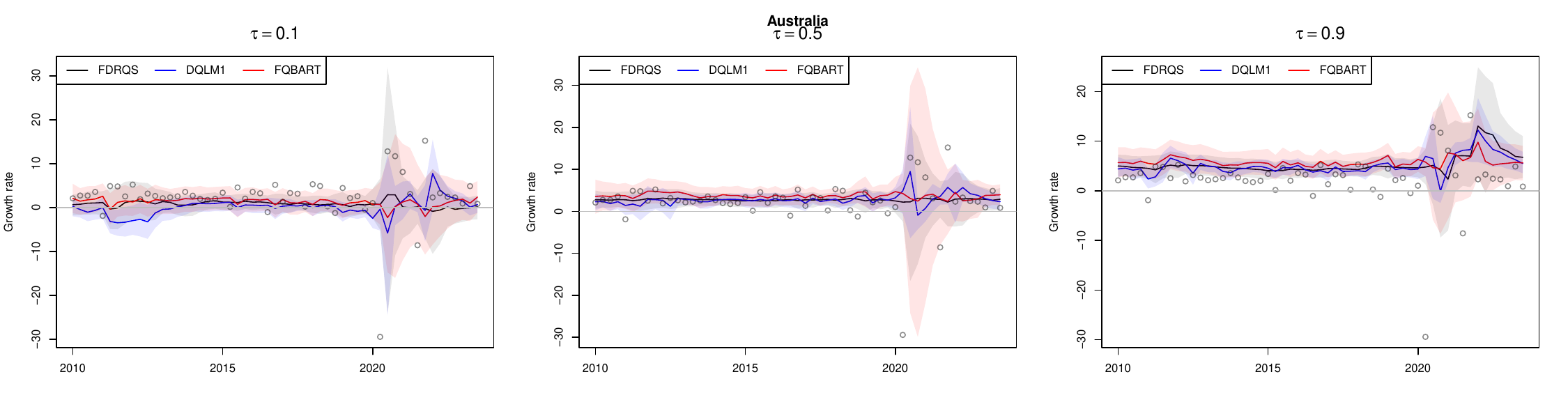}
    \includegraphics[width=\textwidth]{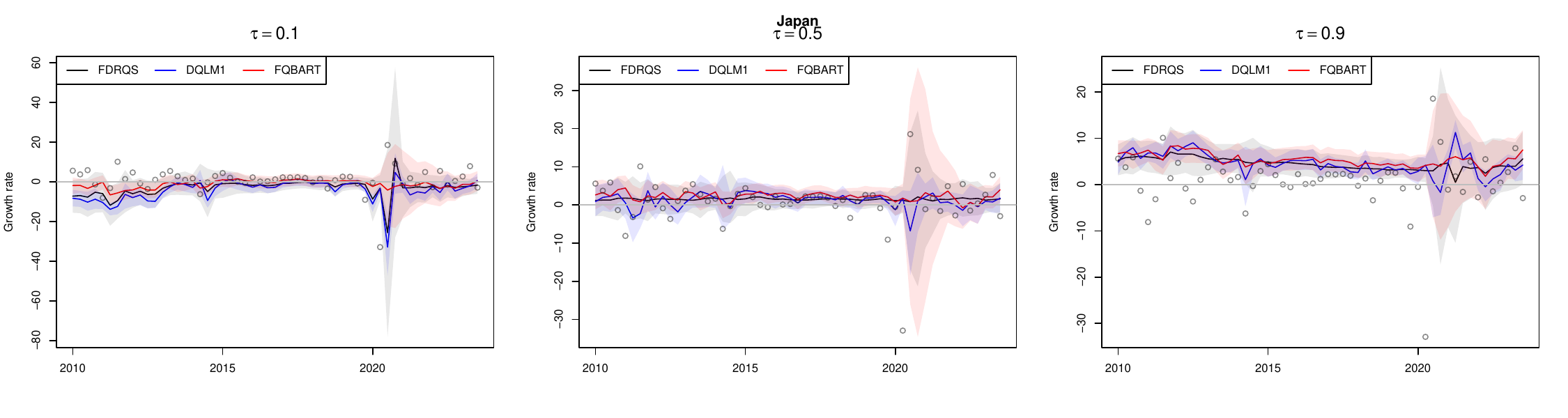}
    \includegraphics[width=\textwidth]{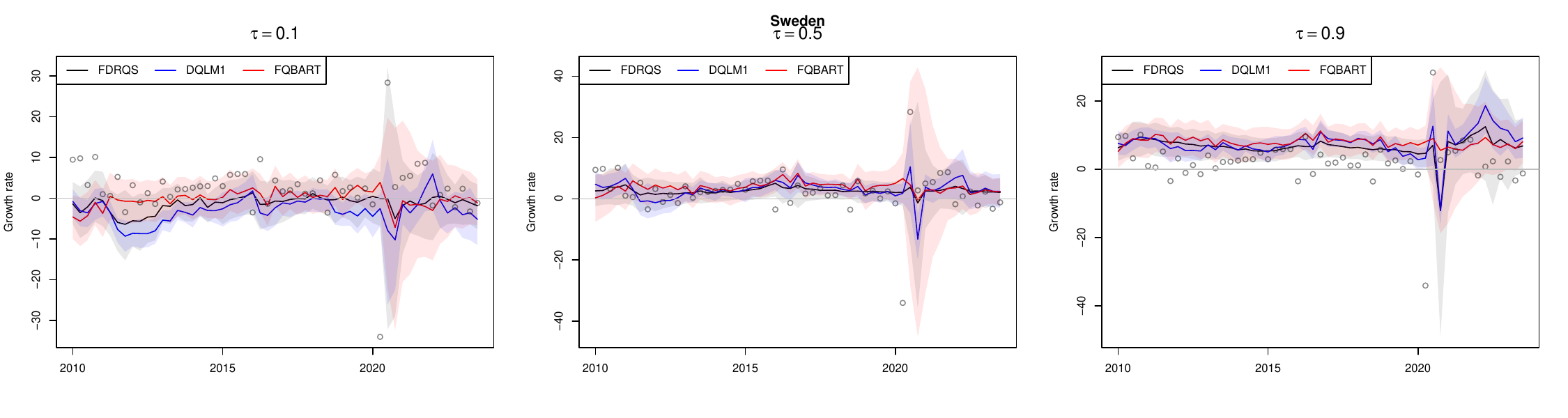}
    \includegraphics[width=\textwidth]{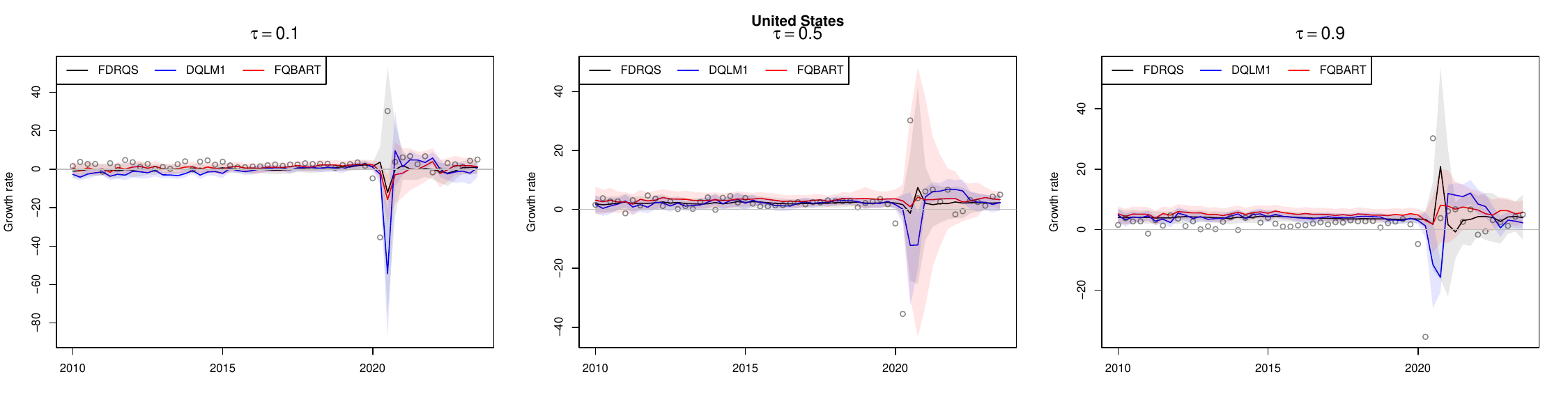}
    \caption{One-step ahead forecast of quantiles (solid lindes) with 95\% intervals (shaded areas) for $\tau=0.1$, $0.5$ and $0.9$ for FDRQS, DQLM1 and FQBART for Australia, Japan, Sweden and the United States for $h=1$. The points indicate the observed growth rates.}
    \label{fig:gar_pred_1}
\end{figure}

Figure~\ref{fig:gar_pred_1} presents the one-step ahead forecast of quantiles and 95\% intervals for $\tau=0.1$, $0.5$ and $0.9$ for FDRQS, DQLM1 and FQBART for some arbitrarily selected countries, Australia, Japan, Sweden and the United States for $h=1$. 
The results for the subset of the models are shown for visibility.  
DQLM1 was selected for the best performance among the DQLMs and its parsimonious model specification, and FQBART was selected as it contains a model component of a different class. 

The quantile forecasts under FDRQS generally lie between those of DQLM1 and FQBART. 
While the 95\% intervals are also intermediate in most periods, they become the widest in the immediate aftermath of 2020 for $\tau=0.1$ and $0.9$. 
Conversely, the forecasts of FQBART, especially for $\tau=0.1$, appear notably higher than those of the other models are associated with the wider 95\% intervals that persist long after the initial COVID-19 shock. 
Figure~\ref{fig:supp_gar_pred_4} of the Supplementary Materials reveals similar trends for $h=4$, though the discrepancies for the are more profound. 

\begin{figure}[H]
    \centering
    {\tiny 2023Q3 ($\tau=0.5$, $h=1$)}\\
    \includegraphics[width=0.3\textwidth]{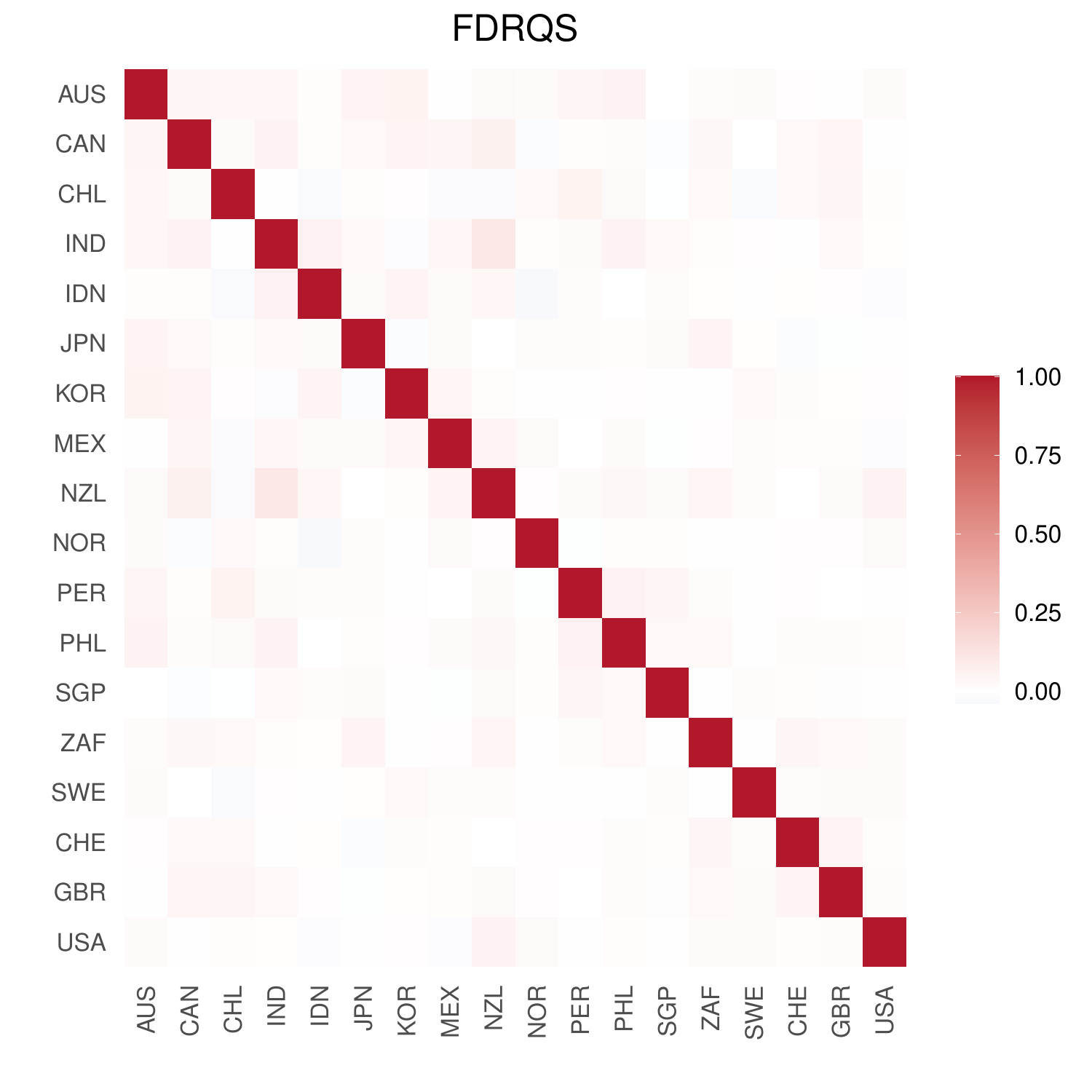}
    \includegraphics[width=0.3\textwidth]{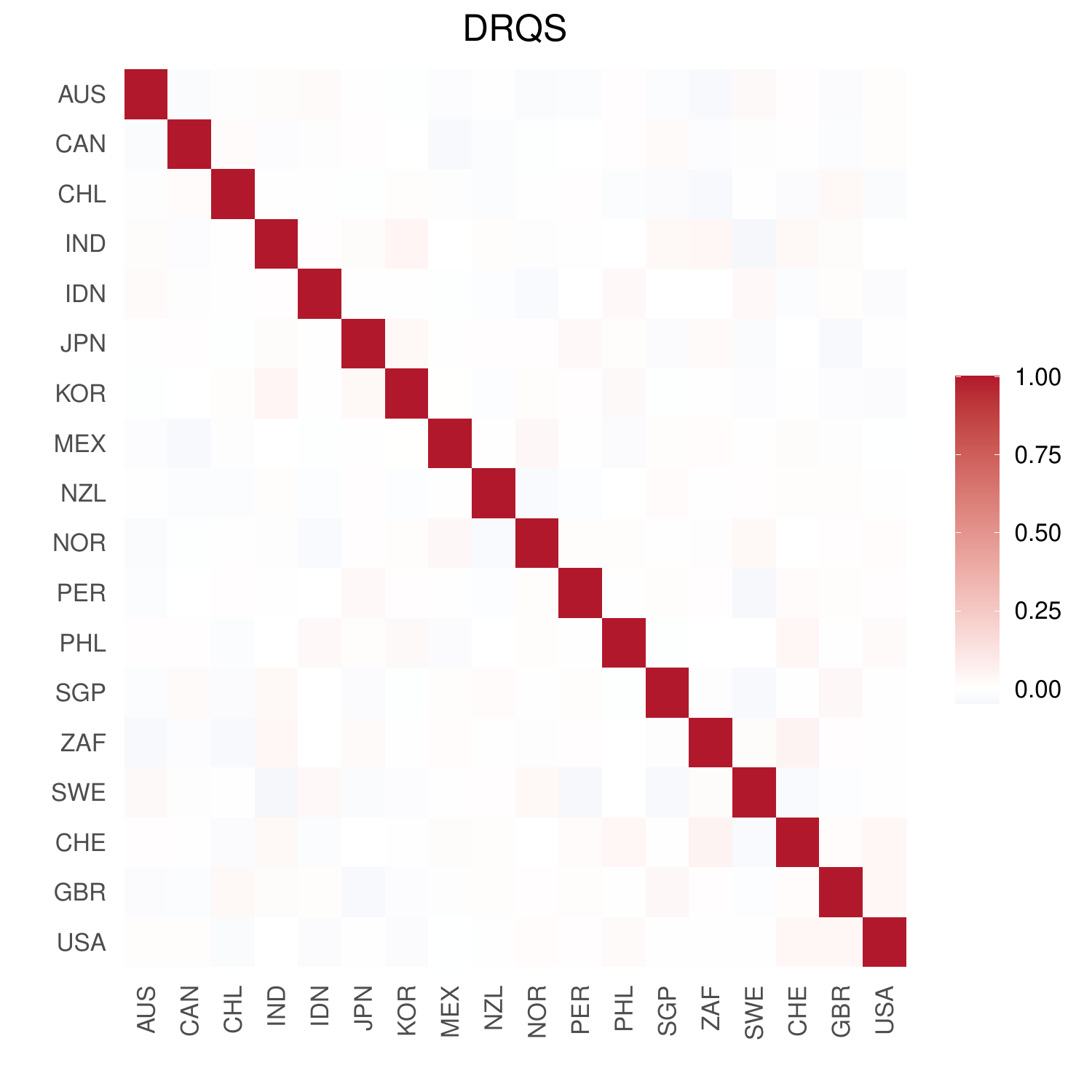}
    \includegraphics[width=0.3\textwidth]{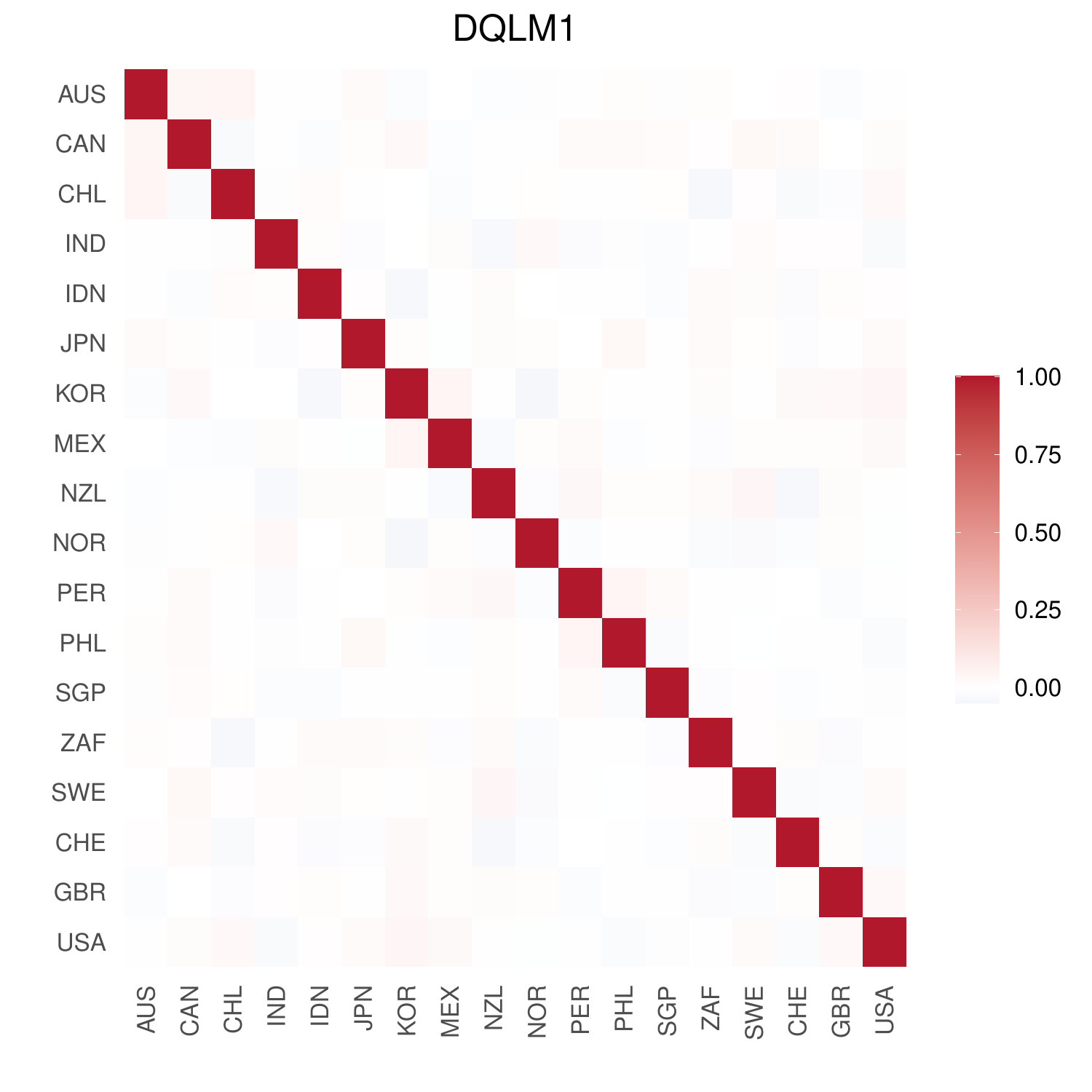}
    
    {\tiny 2015Q1 ($\tau=0.9$, $h=4$)}\\
    \includegraphics[width=0.3\textwidth]{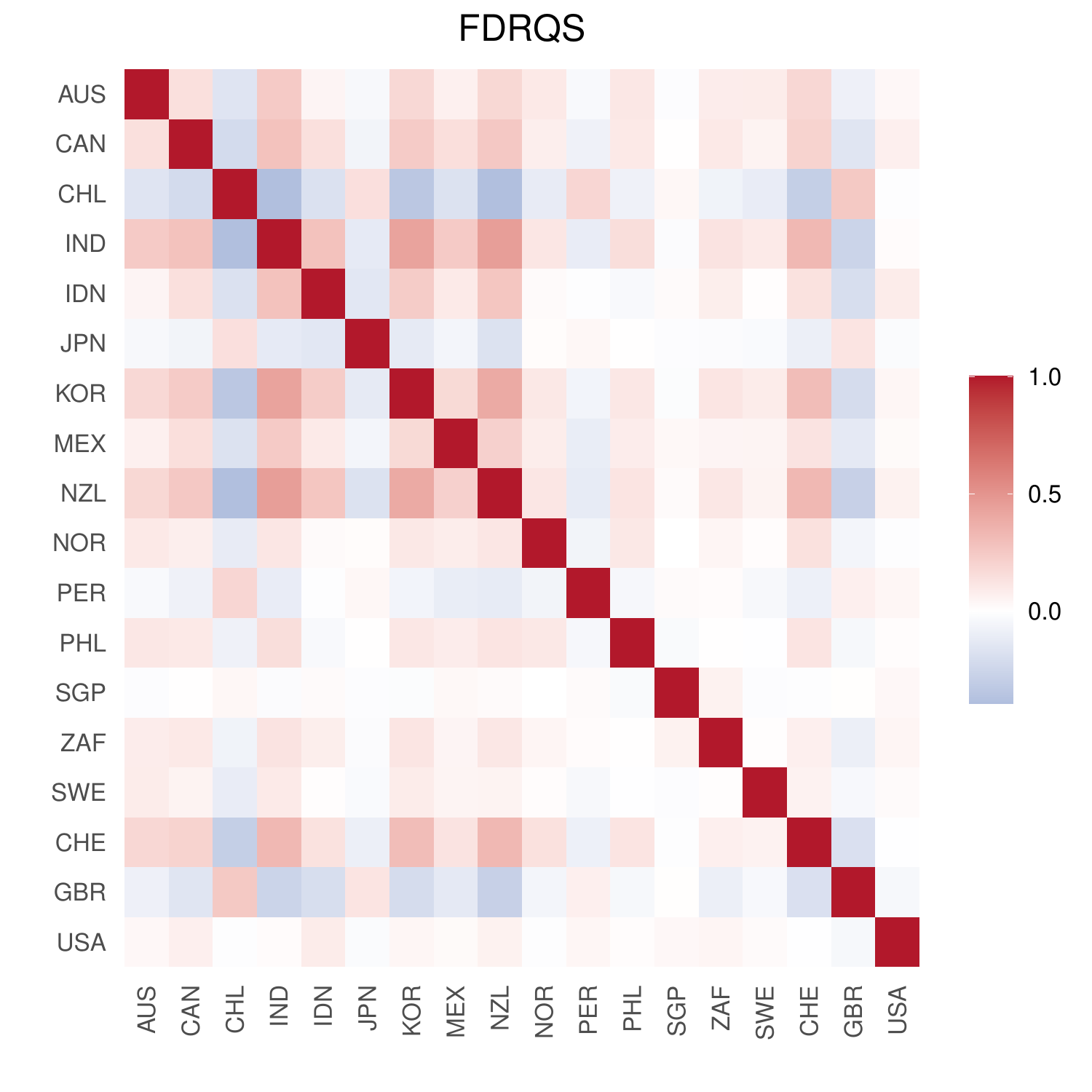}
    \includegraphics[width=0.3\textwidth]{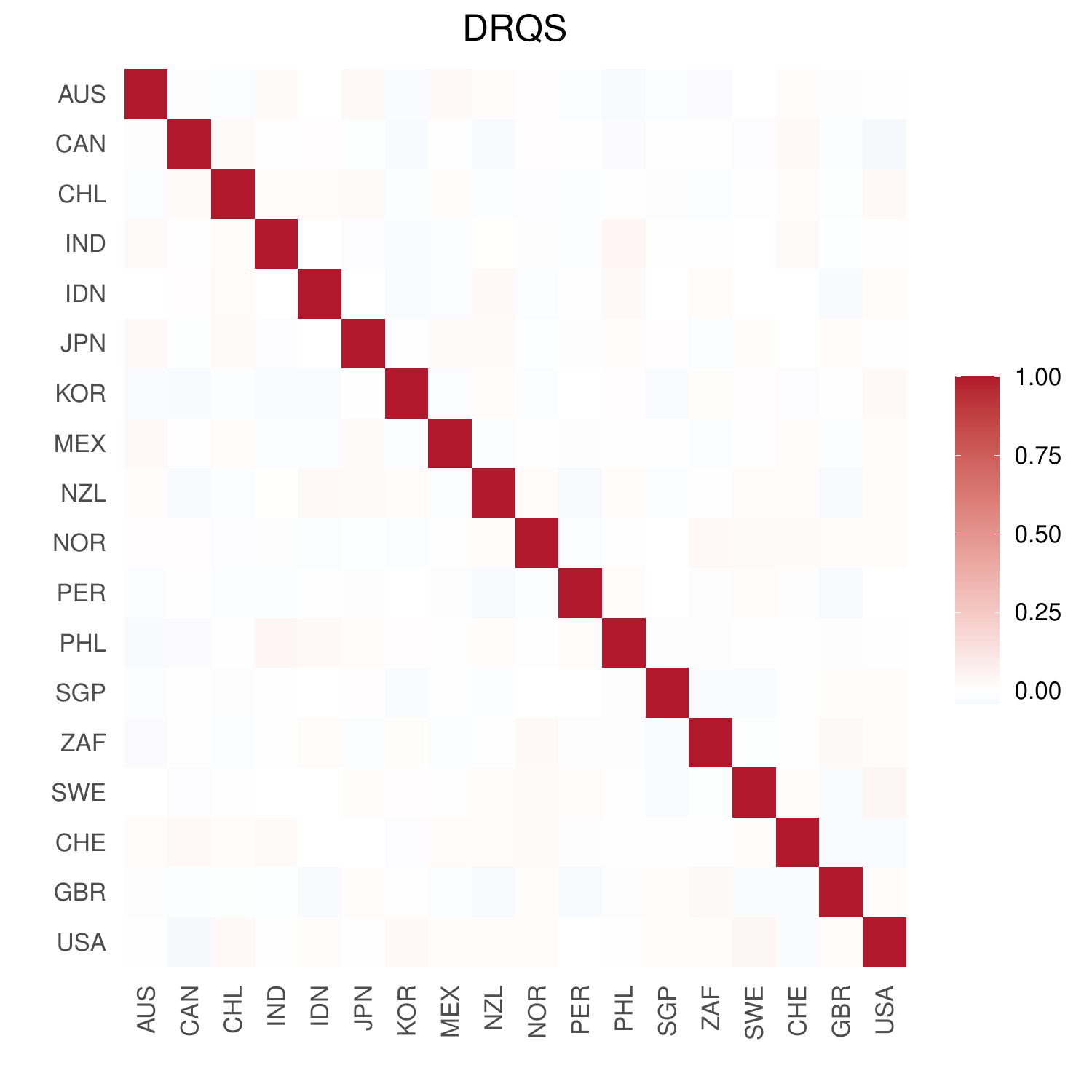}
    \includegraphics[width=0.3\textwidth]{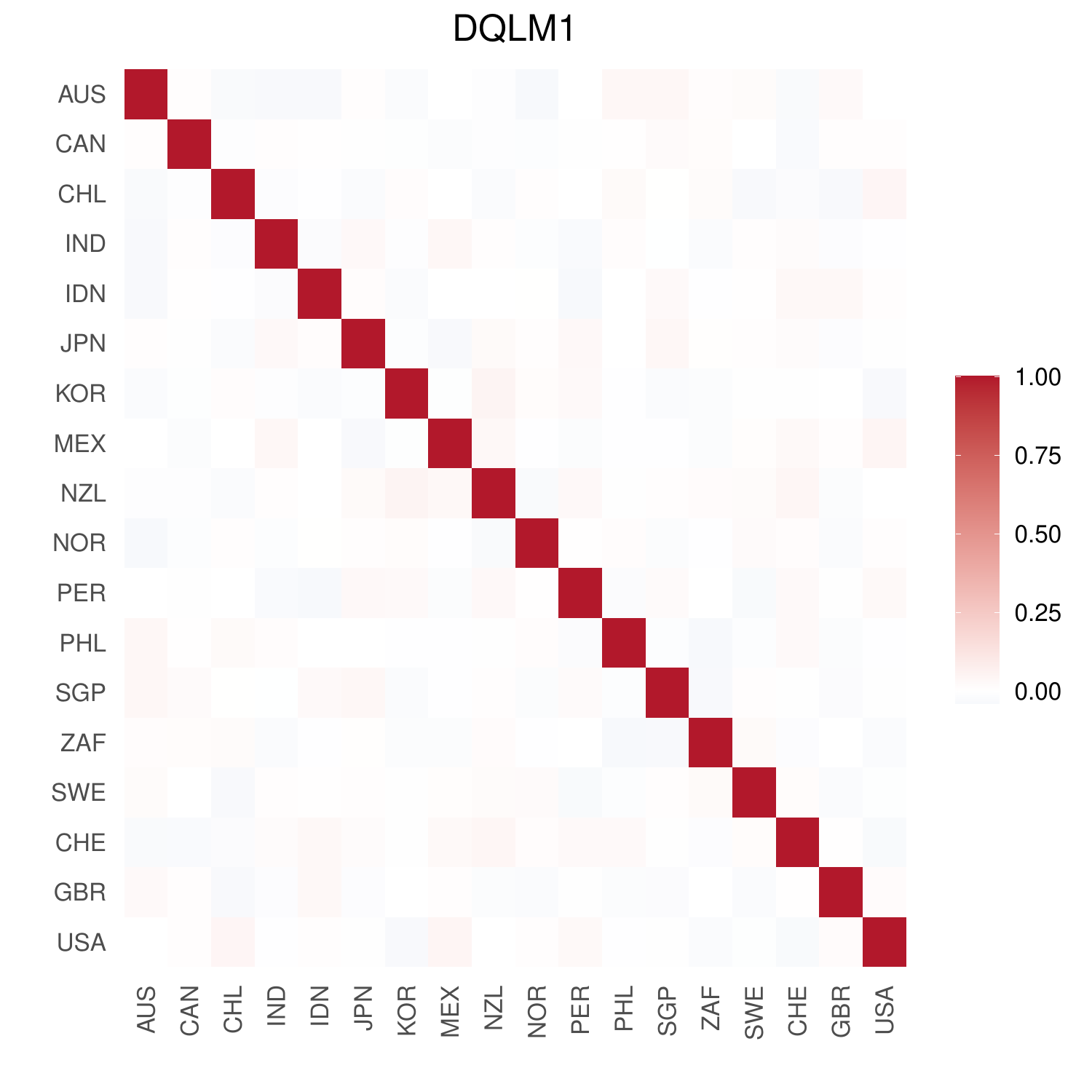}
    \caption{Correlation matrices the quantile forecasts between the countries under FDRQS, DRQS and DQLM1.}
    \label{fig:gar_cor}
\end{figure}

Finally, Figure~\ref{fig:gar_cor} presents the correlation matrices computed for the posterior predictive draws of the quantiles under FDRQS, DRQS and DQLM1. 
By design, the factor structure in FDRQS introduces cross-sectional dependencies that are absent in independent specifications. 
As shown in the left bottom panel, the predictive draws from the posterior predictive distributions for 2015Q1, $\tau=0.9$ and $h=4$ exhibit a clear, albeit mild,  correlation structure. 
In contrast, DRQS and DQLM1, fitted independently for each country, yielded entirely uncorrelated draws.  
While these patterns emerge sporadically, they demonstrate the model's unique capacity to leverage global information to characterize the joint distribution of growth rate across countries.

\begin{table}[H]
    \centering
    \caption{Cumulative total CRPS for $h=1$ and $4$, and cumulative CRPS for the countries ($h=1$) with \texttt{none} relative to FQBART at the end of the data period 2023Q3 for $h=1$.}
    \begin{tabular}{llrrrrrr}\toprule
                & FDRQS & DRQS & DQLM1 & DQLM2 & DQLM3\\\hline
Total ($h=1$)  & 0.925 & 1.030 & 1.034 & 1.077 & 1.034 \\
Total ($h=4$)  & 0.757 & 0.897 & 0.796 & 0.870 & 0.798 \\\hline
Australia      & 0.922 & 1.000 & 0.970 & 1.022 & 0.972 \\
Canada         & 0.990 & 1.387 & 1.213 & 1.354 & 1.207 \\
Chile          & 0.951 & 1.026 & 0.975 & 1.084 & 0.972 \\
India          & 0.938 & 1.066 & 1.085 & 1.120 & 1.102 \\
Indonesia      & 0.789 & 0.863 & 0.868 & 0.842 & 0.876 \\
Japan          & 0.960 & 1.016 & 1.038 & 1.092 & 1.049 \\
Korea          & 0.777 & 0.773 & 0.815 & 0.772 & 0.763 \\
Mexico         & 0.940 & 1.069 & 1.088 & 1.209 & 1.097 \\
New Zealand    & 1.068 & 1.119 & 1.197 & 1.256 & 1.177 \\
Norway         & 0.971 & 0.992 & 0.977 & 0.932 & 0.980 \\
Peru           & 0.975 & 1.498 & 1.271 & 1.455 & 1.284 \\
Philippines    & 0.990 & 1.029 & 1.109 & 1.188 & 1.126 \\
Singapore      & 0.971 & 1.071 & 1.083 & 1.131 & 1.110 \\
South Africa   & 0.897 & 0.989 & 1.077 & 1.231 & 1.040 \\
Sweden         & 0.932 & 0.994 & 0.971 & 1.015 & 0.967 \\
Switzerland    & 0.940 & 1.040 & 1.045 & 1.021 & 1.060 \\
United Kingdom & 0.956 & 1.218 & 1.169 & 1.254 & 1.166 \\
United States  & 0.999 & 1.156 & 1.230 & 1.333 & 1.205 \\\bottomrule
\end{tabular}
    \label{tab:gar_crps}
\end{table}

\subsubsection{Parameter estimates}
In this subsection, we investigate the posterior distributions of the parameters and latent variables of the proposed FDRQS. 

\begin{figure}[H]
    \centering
    \includegraphics[width=\textwidth]{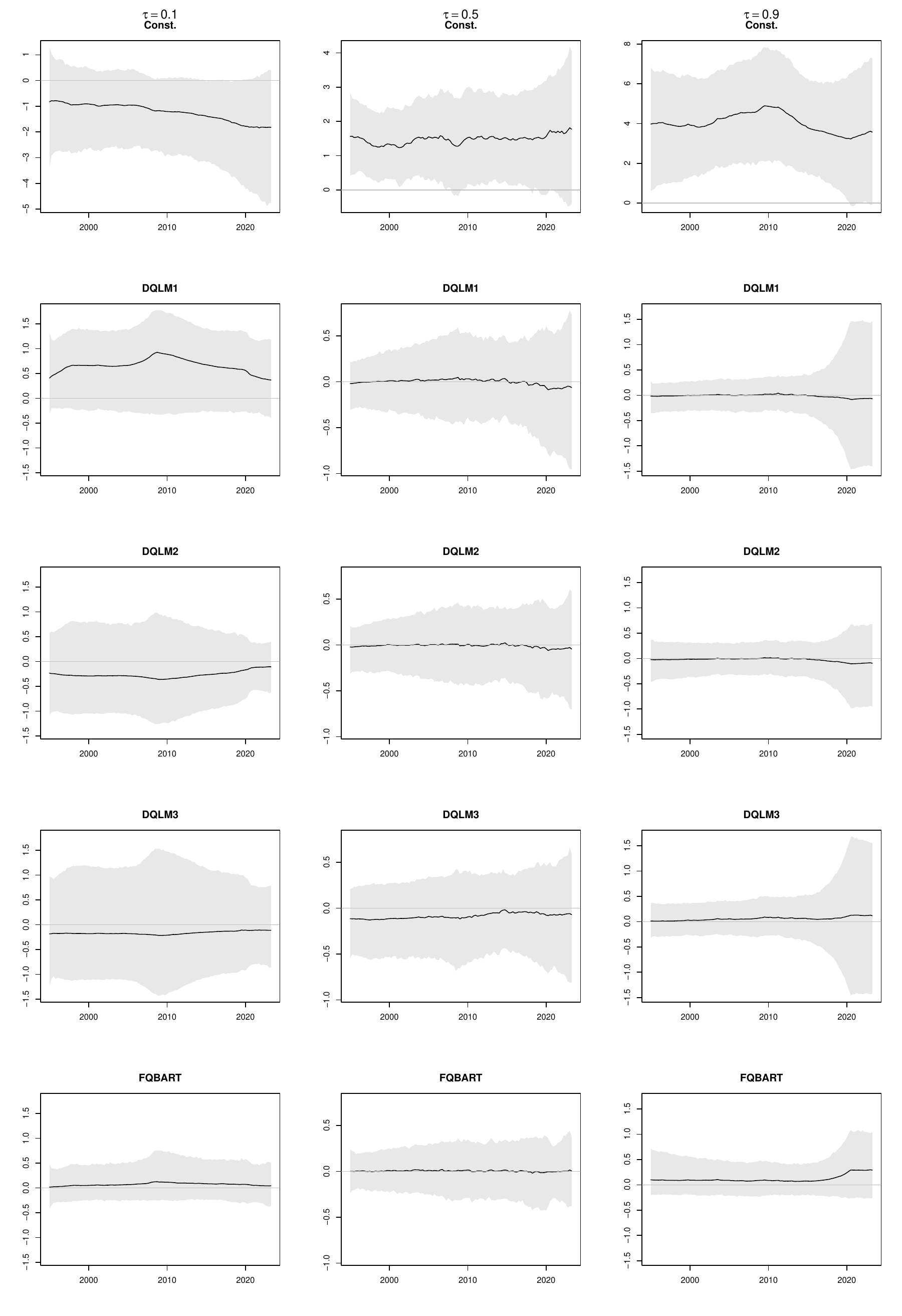}
    \caption{Posterior means (solid lines) and 95\% credible intervals (shaded areas) of the synthesis weights $\theta_{\tau,itj}$ for Japan ($h=1$) for $\tau=0.1$, $0.5$ and $0.9$ using the data up to 2023Q3.}
    \label{fig:gar_theta_jpn}
\end{figure}

\begin{figure}[H]
    \centering
    \includegraphics[width=\textwidth]{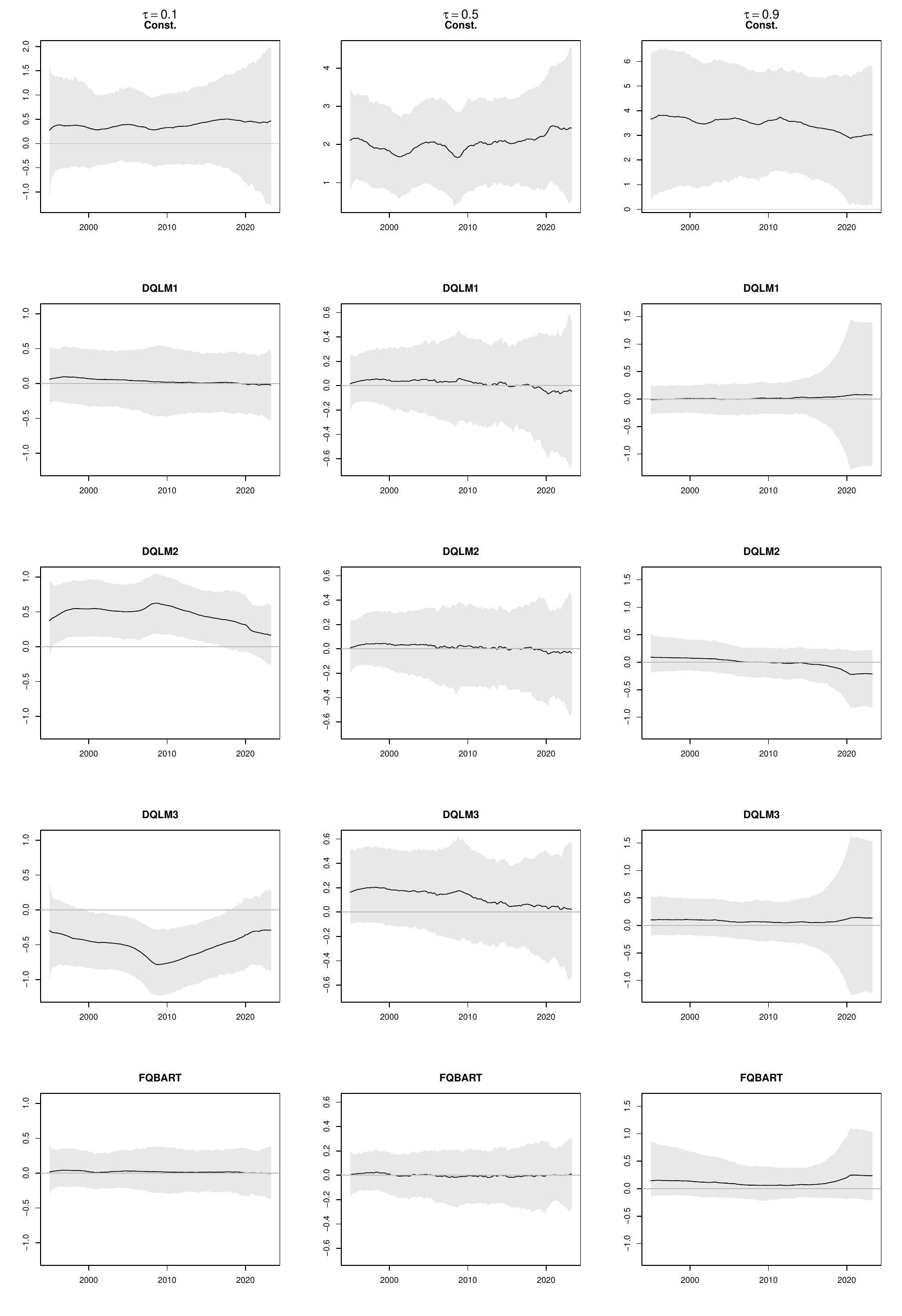}
    \caption{Posterior means (solid lines) and 95\% credible intervals (shaded areas) of the synthesis weights $\theta_{\tau,itj}$ for the United States ($h=1$) for $\tau=0.1$, $0.5$ and $0.9$ using the data up to 2023Q3.}
    \label{fig:gar_theta_us}
\end{figure}

Figures~\ref{fig:gar_theta_jpn} and \ref{fig:gar_theta_us} present the posterior means and 95\% credible intervals of the synthesis weights for $\tau=0.1$, $0.5$ and $0.9$ estimated using the data up to 2023Q3 for, respectively, Japan and the United States for $h=1$. 

In the case of Japan for $\tau=0.1$, Figure~\ref{fig:gar_theta_jpn} reveals that DQLM1 maintained a consistently positive posterior mean weight whereas the weights for DQLM2 and DQLM3 remained near zero, while their credible intervals covered zero throughout period.  
For $\tau=0.9$, we observed a marked widening of the 95\% credible intervals for the agent models following the 2020 shock, reflecting the heightened uncertainty in agent reliability. 
Notably, the posterior mean of the weight for FQBART adaptively pivoted toward positivity to navigate the post-pandemic recovery, a transition facilitated through the underlying factor structure. 

In the case of the United States shown in Figure~\ref{fig:gar_theta_us}, the synthesis wieghts exhibit a markedly different profile.. 
For $\tau=0.1$, the weight for DQLM2 remains positive for throughout the sample, with the credible intervals that exclude zero for the duration of the 2000s and most of 2010s.  
It is also seen that DQLM3 received a credibly negative weight in the same period, indicating its role as a corrective component within the synthesis. 
For $\tau=0.9$, we observe a shift in agent uncertainty and increased allocation toward FQBART similar to the pattern seen in Japan. 

Figures~\ref{fig:supp_gar_theta_jpn_4} and \ref{fig:supp_gar_theta_us_4} present the results for Japan and the United States in the case of $h=4$. 
We observe quite different patterns from those in Figures~\ref{fig:gar_theta_jpn} and \ref{fig:gar_theta_us}. 
For both countries, DQLM2 emerges as the primary contributing agent for all three quantiles, maintaining positive posterior mean weights and credible intervals that exlude zero for some proportion of the data period. 
While, DQLM1 and DQLM3 also resulted in the positive posterior means in certain periods, their credible intervals include zero throughout the data period. 
In Japan, for instance, DQLM2 dominates the down side risk ($\tau=0.1$), whereas the synthesis benefits from a more diversified allocation across DQLM1 and DQLM3 for $\tau=0.5$. 

\begin{figure}[H]
    \centering
    \includegraphics[width=\textwidth]{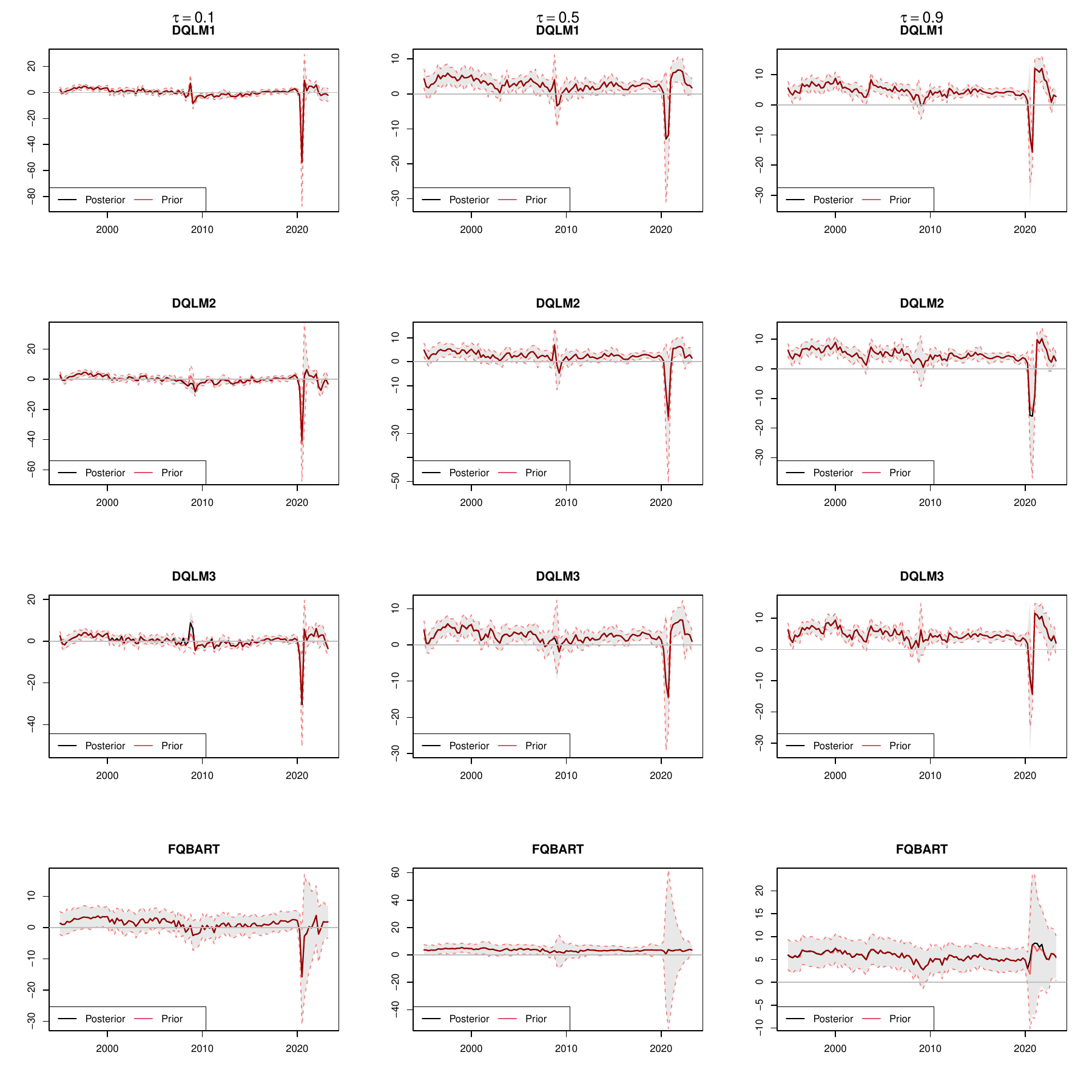}
    \caption{Posterior means, 95\% credible intervals, prior means and 95\% prior intervals of the latent factors $f_{\tau,itj}$ for $\tau=0.1$, $0.5$ and $0.9$ for the United States ($h=1$). }
    \label{fig:gar_f_us}
\end{figure}

Figure~\ref{fig:gar_f_us} presents the posterior means, 95\% credible intervals, prior means and 95\% prior intervals of the latent predictors $f_{itj}$ for the United States for $\tau=0.1$, $0.5$ and $0.9$ for $h=1$. 
It is seen that the posterior and prior distributions coincide for most periods. 
However, on rare occasions such as the 2009 finaicial crisis or the 2020 pandemic, the posterior distribution is seen to  deviate from the prior distribution. 

\begin{figure}[H]
    \centering
    \includegraphics[width=\linewidth]{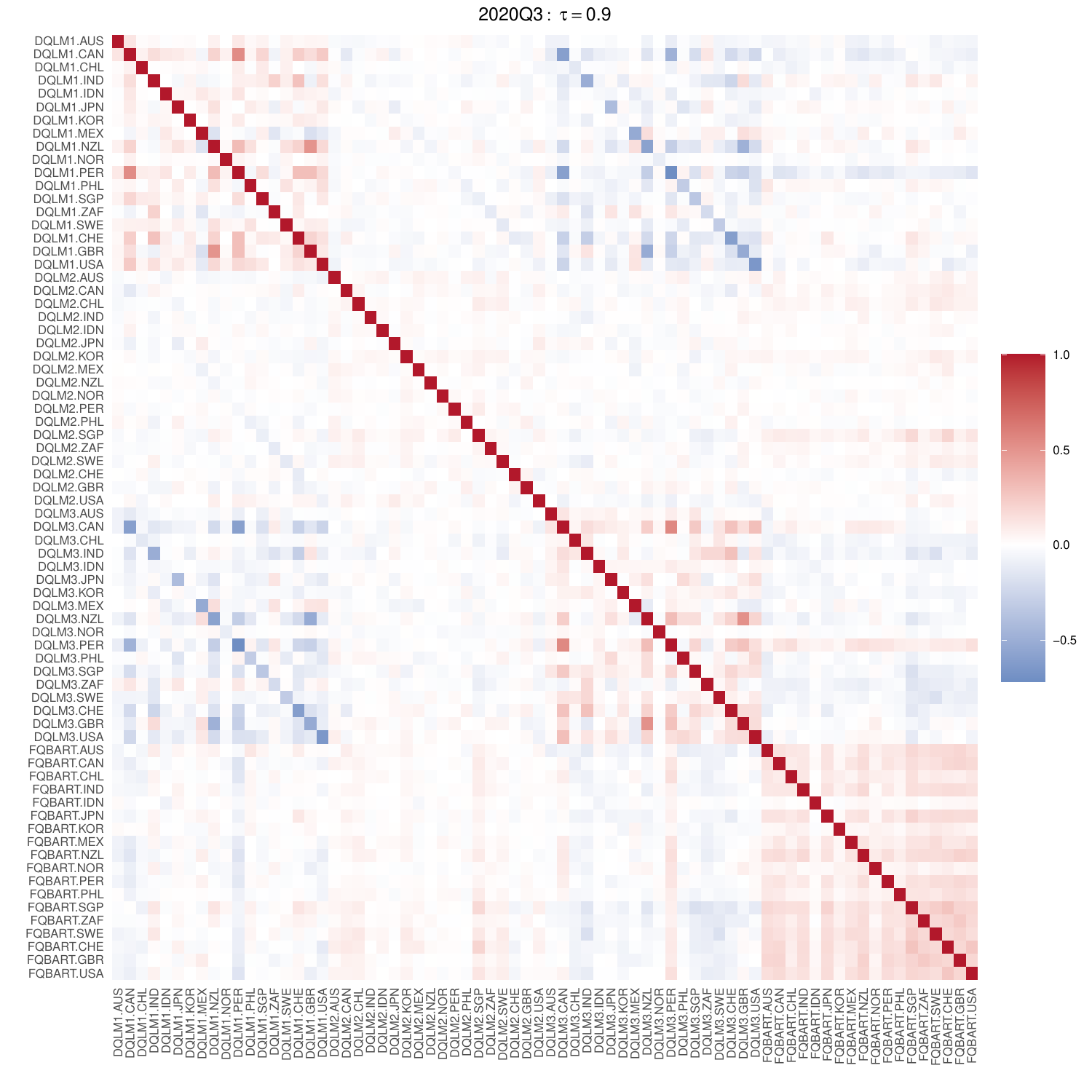}
    \caption{Correlation matrix of the posterior draws of the latent factors $f_{\tau,itj}$ for $\tau=0.9$ in 2020Q3 ($h=1$).}
    \label{fig:gar_fcor}
\end{figure}

It is important to note that because the agent models are fitted independently, the latent predictors are independent a priori. 
We investigate the posterior correlation matrix to see if the FDRQS can detect any emergent structure among them. 
For the majority of the sample, we find little evidence of clear correlation patterns, suggesting the model remains disciplined and largely adheres to its prior specifications. 
However, as shown in Figure~\ref{fig:gar_fcor}, the model does reveal latent dependencies during rare periods of extreme economic stress. 
In 2020Q3 ($\tau=0.9$, $h=1$), for example, we observe positive correlations within the same agent models—capturing a synchronized global response—alongside negative correlations between DQLM1 and DQLM3. 
This indicates that while the model generally avoids imposing artificial linkages, it is flexible enough to capture necessary cross-agent balancing when the data warrants it.

\begin{figure}[H]
    \centering
    \includegraphics[height=0.95\textheight]{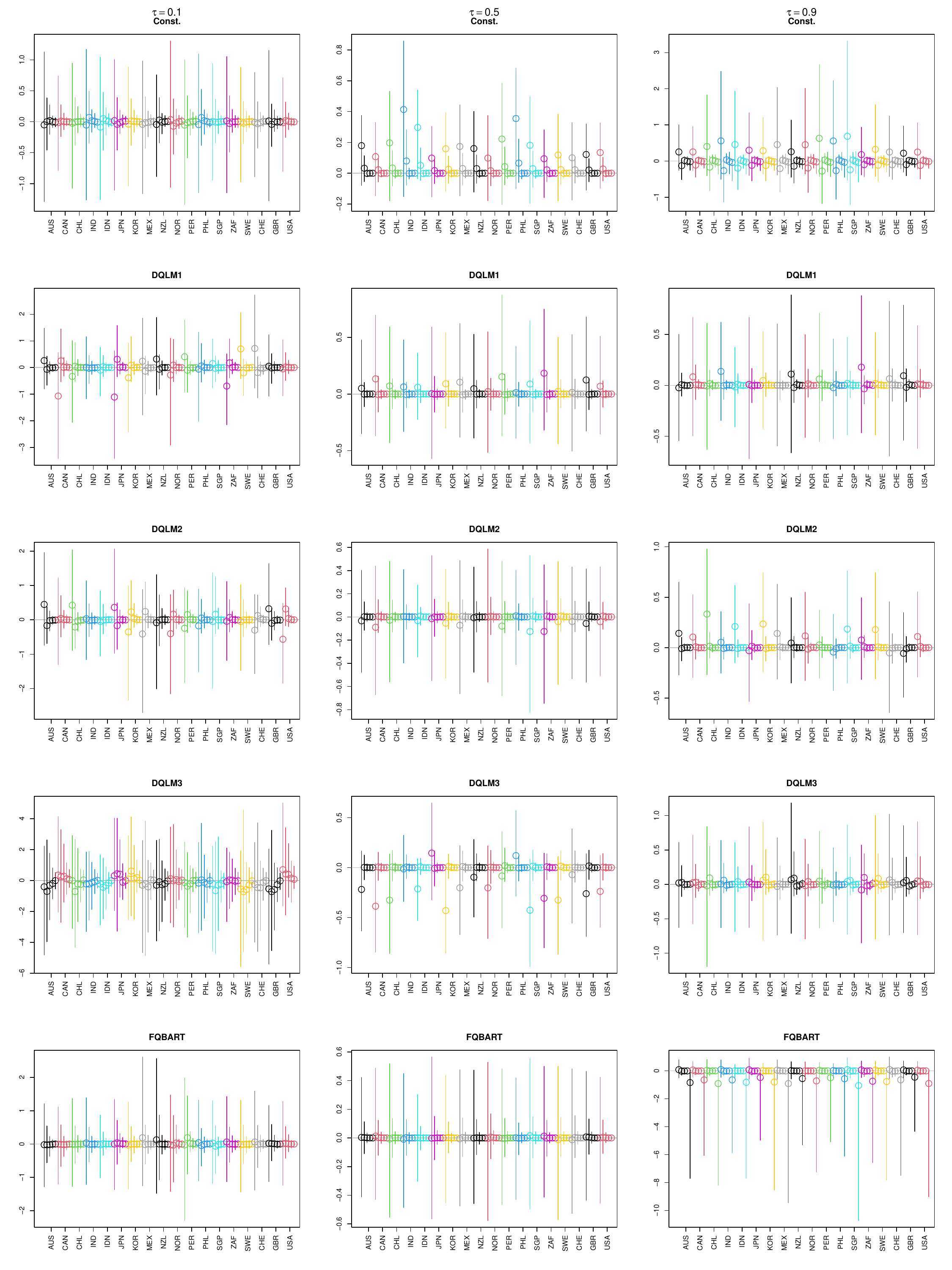}
    \caption{Posterior means (points) and 95\% credible intervals (line segments) of the factor loading $\lambda_{\tau,i\ell j}$ for $\tau=0.1$, $0.5$ and $0.9$ for $h=1$. For each country plotted in the same colour, the plots indicate the results for the factor loadings for $\ell=1,\dots,L$ from the left to right. }
    \label{fig:gar_lam}
\end{figure}

Figure~\ref{fig:gar_lam} presents the posterior means and 95\% credible intervals of the factor loadings, estimated using the data up to 2023Q3 for $\tau=0.1$, $0.5$ and $0.9$ for $h=1$. 
While we do not seek to assign a structural identity to these latent factors, as evidenced by the credible intervals generally encompassing zero—the empirical patterns are noteworthy. 
Consistent with the MGP prior, the first factor typically exhibits the largest posterior mean magnitudes and the widest intervals, with subsequent factors progressively shrinking toward zero. 
We observe interesting regional variations; for DQLM1 ($\tau=0.1$), countries such as Canada and Japan show negative loadings on the first factor, while Sweden and Switzerland show positive ones. 
A singular exception occurs for FQBART at $\tau=0.9$, where the final factor loadings are consistently negative across all countries with asymmetric intervals leaning toward negativity. 
This suggests a global adjustment mechanism within the synthesis that selectively moderates the upper-tail behavior of the non-parametric agent.

Figure~\ref{fig:supp_gar_lam_4} of the Supplementary Materials presents the results for $h=4$. 
In this case, the first factor factor appear to be more influential, while the remaining factors exhibit even stronger shrinkage toward zero. 
For example, in the case of DQLM2 for $\tau=0.5$, the posterior means of the first factor loadings are consistently positive across all countries, 
The 95\% credible intervals exclude zero with the exceptions of Australia, Chile, New Zealand and Norway.

\subsubsection{Predictive distributions of growth rates}
With the quantile forecasts $\hat{Q}_{it}(\tau)$ obtained on the grid, $\tau=0.05,\dots,0.95$, it is possible to construct a one-step ahead predictive distribution of the growth rate $y_{i,{t^*}}$. 
Following the semiparametric approach of \cite{Mitchell24}, which builds upon \cite{Adrian}, 
we generated $R=10000$ draws from the one-step ahead predictive distributions of the growth rates. 
See Section~\ref{sec:mitchell} of the Supplementary Materials for the algorithm of \cite{Mitchell24}. 

\begin{figure}[H]
    \centering
    \includegraphics[width=0.32\textwidth]{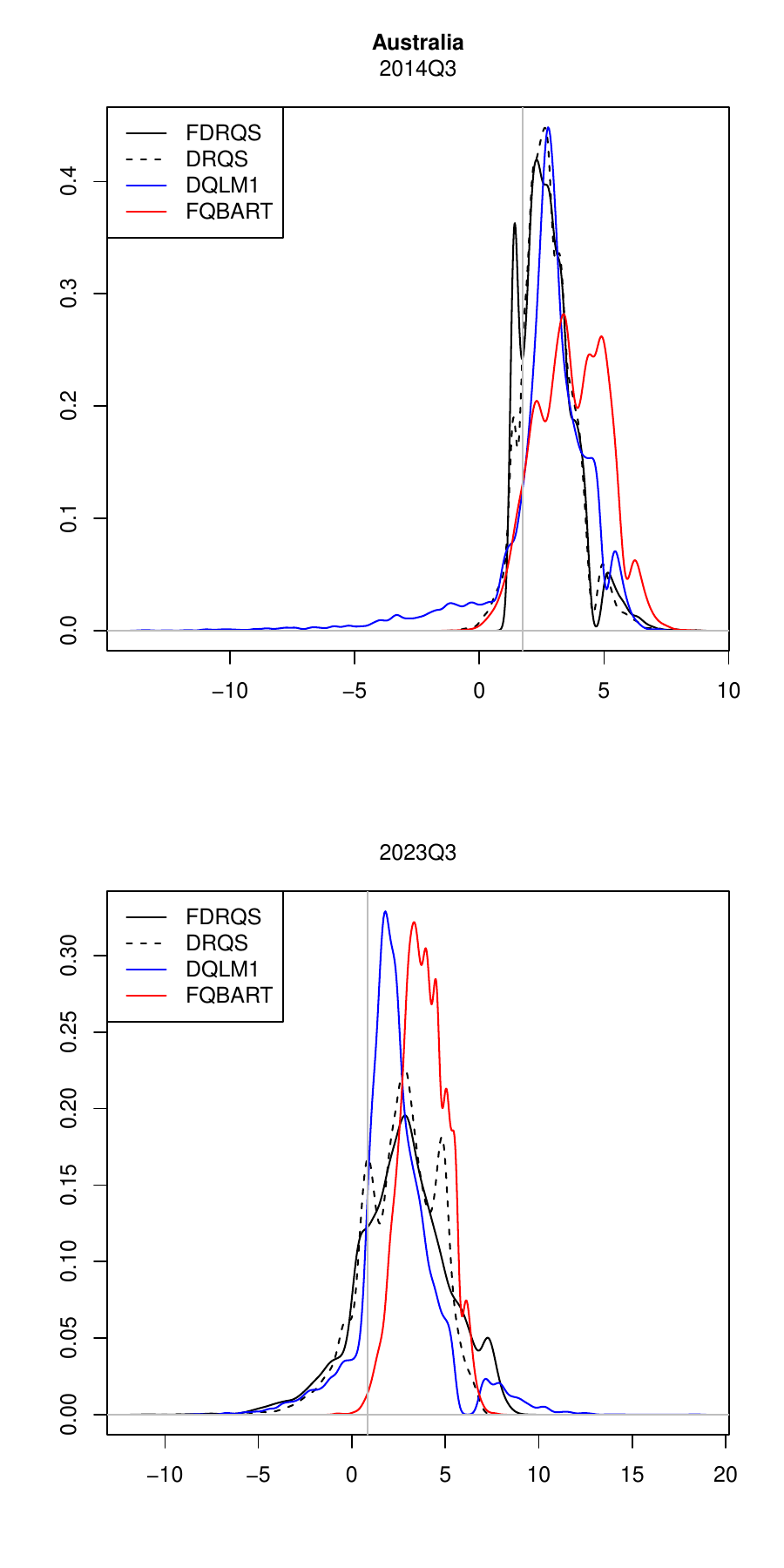}
    \includegraphics[width=0.32\textwidth]{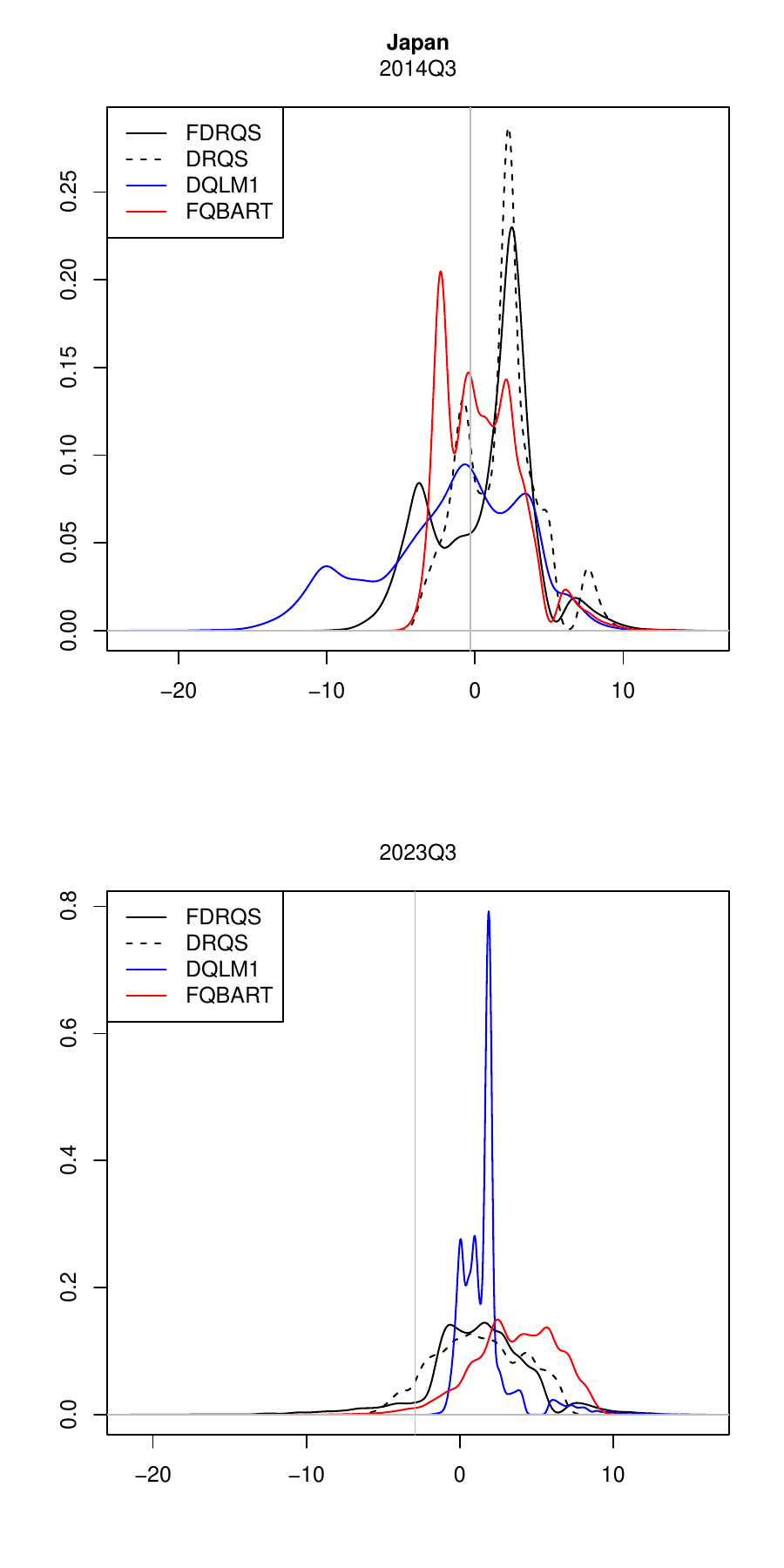}
    \includegraphics[width=0.32\textwidth]{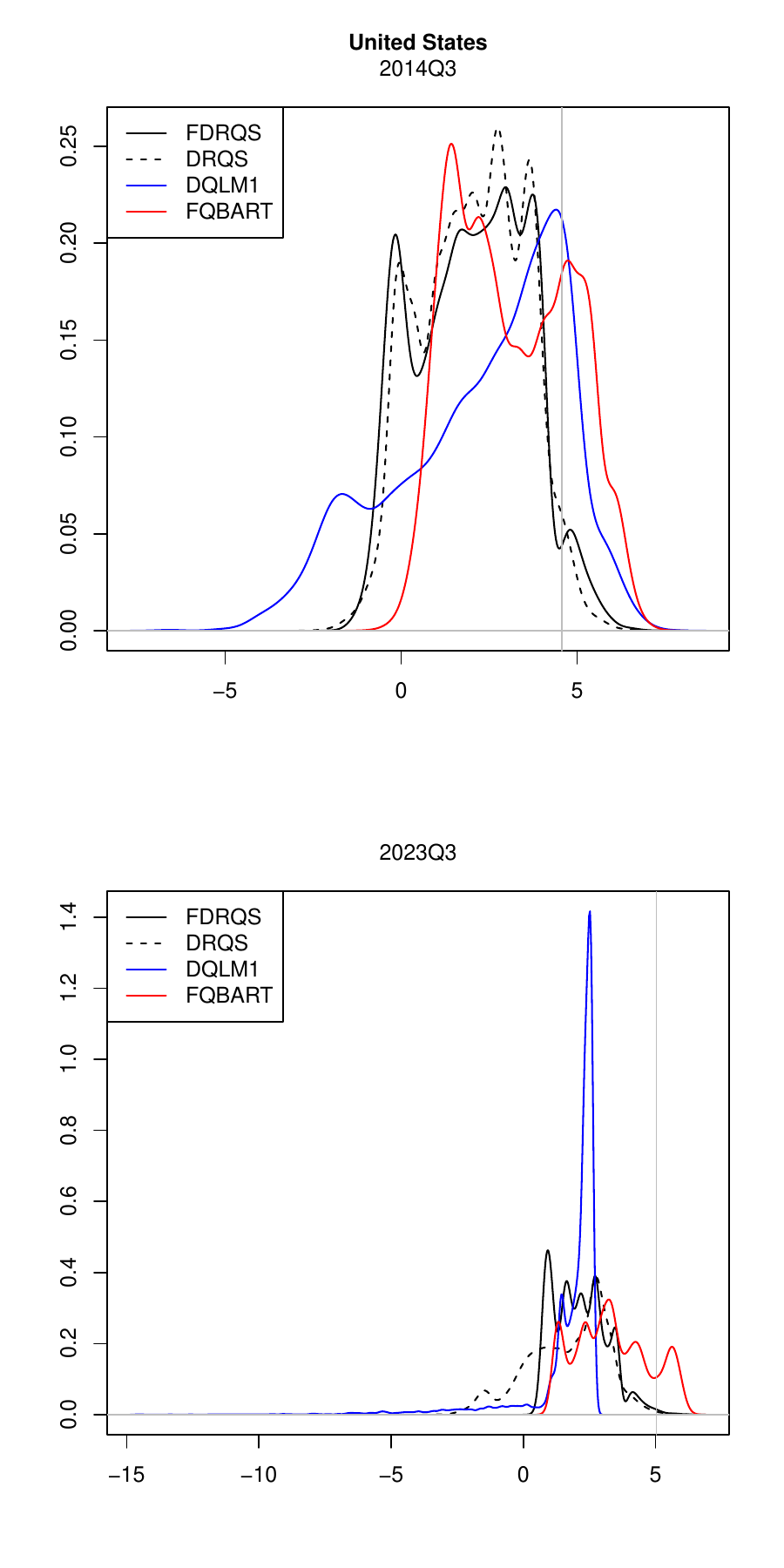}
    \caption{Predictive distribution of growth rate under FDRQS, DRQS, DQLM1 and FQBART in 2014Q3 and 2023Q3 for Australia, Japan, and the United States ($h=1$). The vertical lines indicate the observed growth rates. }
    \label{fig:gar_pred_dist}
\end{figure}

Figure~\ref{fig:gar_pred_dist} illustrates the one-step ahead predictive densities of the growth rates under FDRQS, DRQS, DQLM1 and FQBART for Australia, Japan and the United States in the arbitrarily selected period 2014Q3 and 2023Q3 for $h=1$. 
Those for $h=4$ are also presented in Figure~\ref{fig:supp_gar_pred_dist_4} of the Supplementary Materials. 
In each panel, the vertical lines indicate the observed growth rates. 
Note that the multimodality is due to the semiparametric nature of the local-linear method. 
The distributions are generally centered around the observed growth rates, though their dispersion varies significantly by model and period. 
For instance, in 2023Q3, the DQLM1 distributions for Japan and the United States appear much more peaked than those of the other models. 
Conversely, the $h=4$ results in the Supplementary Materials reveal that FQBART occasionally suffers from poor calibration, with observed growth rates falling into the extreme tails of its distribution (e.g., Australia in 2015Q1 and Japan in 2023Q3). 

To visually assess the overall goodness of fit of the predictive densities, we utilize the probability integral transforms (PIT), defined by \[
\PIT_{it}^{(m)}\allowbreak =\frac{1}{R}\sum_{r=1}^R I\left(y_{it}\leq y_{i{t}}^{(m,r)}\right), 
\]
where $y_{i{t}}^{(r,m)}$ is $r$th draw from the predictive distribution of $y_{it}$ obtained from $m$th model. 

\begin{figure}[H]
    \centering
    \includegraphics[width=\textwidth]{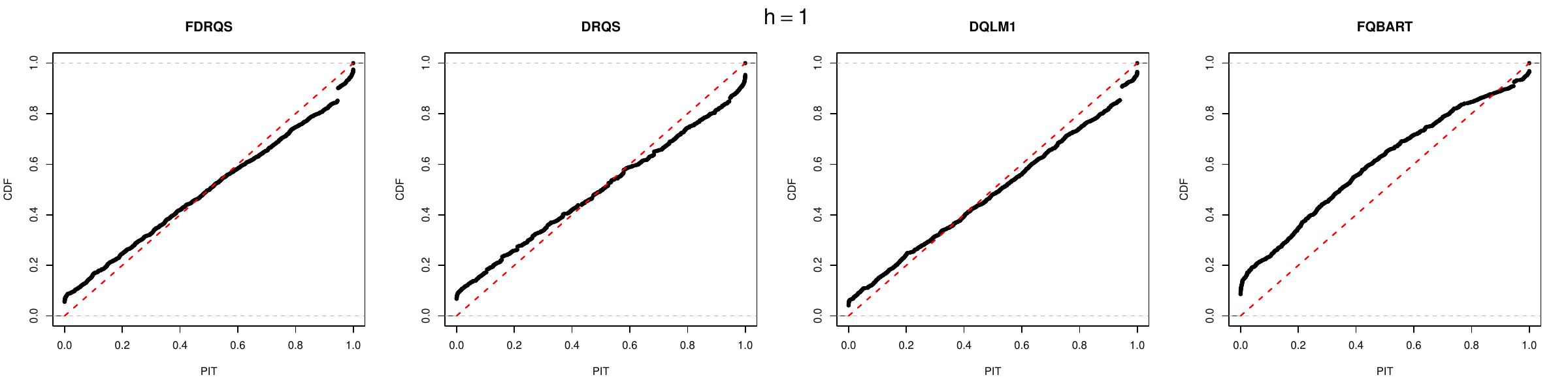}
    \includegraphics[width=\textwidth]{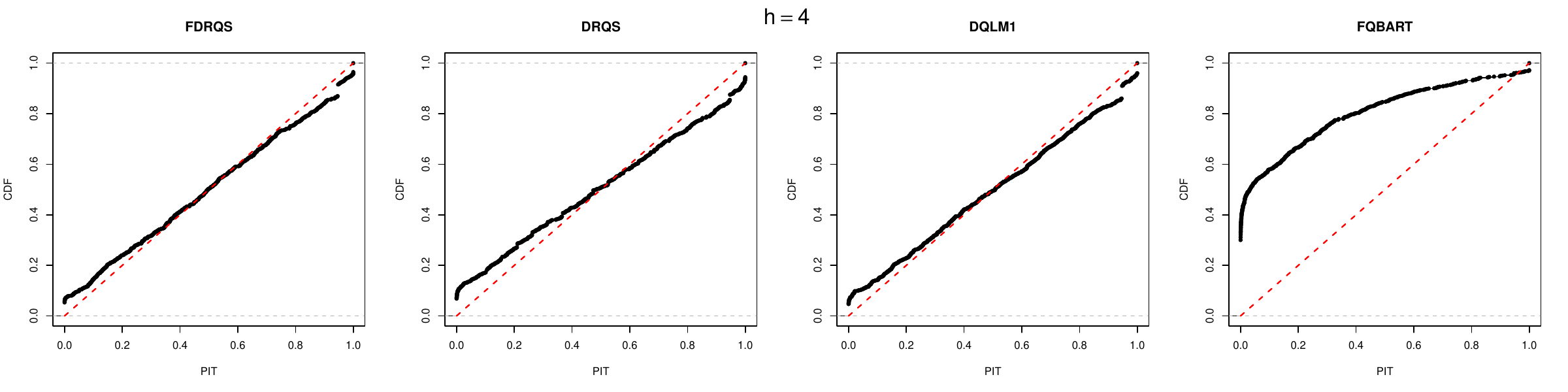}
    \caption{Empirical cumuldative distribution functions (CDF) of PIT for the predictive distribution for FDRQS, DRQS, DQLM1 and FQBART for $h=1$ and $h=4$. The red dashed lines indicate the 45 degree lines. }
    \label{fig:gar_pit}
\end{figure}

Figure~\ref{fig:gar_pit} presents the empirical cumulative distribution function (CDF) of PIT over all countries and periods for each model for $h=1$ and $h=4$. 
An empirical CDF of PIT closely tracks the 45 degree line, plotted in the red dashed line, indicates good fit of the predictive distribution to the data. 
The figure shows that the empirical CDF of the PITs for FDRQS, DRQS, and DQLM1 closely tracks the 45-degree line, indicating a reliable fit across countries and horizons. In contrast, FQBART exhibits a pronounced deviation, particularly at $h=4$, confirming that its predictive distributions are systematically miscalibrated at longer horizons. 

\section{Concluding remarks}\label{sec:conc}
We have proposed a novel method for synthesizing regression quantiles within the framework of Bayesian Predictive Synthesis (BPS). 
By utilizing the asymmetric Laplace distribution for the synthesis function, we cast the model as a dynamic quantile linear model incorporating latent predictors, which  represent the draws from the predictive distributions of the synthesized agent models. 
We further extended the DRQS framework to a multivariate setting by introducing a factor structure for the synthesis weights, allowing for the simultaneous analysis of multiple time series.
The model is efficiently estimated using a Gibbs sampler with data augmentation and Forward-Filtering Backward-Sampling (FFBS). 

Through empirical analyses of US inflation and global GDP growth, we have demonstrated that this approach significantly improves quantile forecasting performance. Notably, the FDRQS exhibits greater robustness than both univariate DRQS and individual agent models following significant economic crises. 
This resilience is achieved by adaptively reweighting agent contributions and leveraging shared information across the series through the latent factor structure

\paragraph{Acknowledgements}
This work is partially supported by Japan Society for Promotion of Science (JSPS) KAKENHI (\#22K13376,  \#24K00244 and \#25H00546) and JSPS Postdoctoral Fellowship (\#P25314).

\newpage
\appendix
\renewcommand{\thesection}{S.\arabic{section}}
\setcounter{section}{0}

\renewcommand{\theequation}{S.\arabic{equation}}
\setcounter{equation}{0}

\renewcommand{\thefigure}{S.\arabic{figure}}
\setcounter{figure}{0}

\renewcommand{\thetable}{S.\arabic{table}}
\setcounter{table}{0}

\begin{center}
{\Large\textbf{Supplementary Materials for `Dynamic Bayesian regression quantile synthesis for forecasting outlook-at-risk' }
}

\vspace{0.5cm}
Genya Kobayashi, Shonosuke Sugasawa, Yuta Yamauchi and Dongu Han
\end{center}

\section{Posterior computation}
\subsection{Gibbs sampler for DRQS}\label{sec:gibbs_s}Using the location scale mixture representation of the asymmetric Laplace distribution \citep{kozumi11}, the model is given by
\[
\begin{split}
y_{t} &= \F_{t}'\bthe_{t} + \kappa_1v_{t} + e_{t},\quad \epsilon_{t}\sim N(0,\sigma_{t}\kappa_2v_{t}),\\
\bthe_{t} &= \bthe_{t-1} + \omega_{t},\quad \omega_{t}\sim N(\zero,\sigma_{t}\W_{t}),\\
\sigma^{-1}_{t}&=\frac{\beta}{\gamma_t}\sigma^{-1}_{t-1},\quad \gamma_{t}\sim Beta\left(\frac{\beta n_{t-1}}{2},\frac{(1-\beta) n_{t-1}}{2}\right).
\end{split}
\]
where  $\F_{t}=(1,\f_{t}')'$, $v_{t}\sim Exp(\sigma_{t})$, $\kappa_1=\frac{1-2\tau}{\tau(1-\tau)}$ and $\kappa_2=\frac{2}{\tau(1-\tau)}$. 
The full conditional distributions required for the proposed Gibbs sampler are described as follows. 

\begin{itemize}
    \item 
    The full conditional distribution of $v_{t}$ is the generalised inverse Gaussian distribution, denoted by $GIG(\lambda,\chi,\psi)$. 
    Since 
    \[
    \begin{split}
        p(v_{t}|-)&\propto v_{t}^{-1/2}\exp\left\{-\frac{(y_{t}-\F_{t}'\bthe_{t}-\kappa_1v_t)^2}{2\sigma_{t}\kappa_2v_t}\right\}\exp\left\{-\frac{v_{t}}{\sigma_{t}}\right\}\\
        &\propto v_{t}^{-1/2}\exp\left\{-\frac{1}{2}\left(\frac{(y_{t}-\F_{t}'\bthe_{t})^2}{\sigma_{t}\kappa_2} v_{t}^{-1}+\left(\frac{2}{\sigma_{t}}+\frac{\kappa_1^2}{\sigma_{t}\kappa_2}\right)v_{t}\right)\right\},
    \end{split}
    \]
    $v_t$ is sampled from $GIG\left(\frac{1}{2},\frac{(y_{t}-\F_{t}'\bthe_{t})^2}{\sigma_{t}\kappa_2}, \frac{2}{\sigma_{t}}+\frac{\kappa_1^2}{\sigma_{t}\kappa_2}\right)$ for $t=1,\dots,T$. 
    
    \item 
    Assuming $f_{tj}\sim N(a_{tj},A_{tj})$ for $j=1,\dots,J$, the full conditional distribution of $\f_t=(f_{t1},\dots,f_{tJ})'$ is 
    \[
    \begin{split}
        p(\f_{t}|-)&\propto \exp\left\{-\frac{(y_{t}-\theta_{t0}-\f_{t}'\bthe_{t+}-\kappa_1\v_{t})^2}{2\sigma_{t}\kappa_2v_{t}}\right\}
        \exp\left\{-\frac{1}{2}(\f_{t}-\a_{t})'\A_{t}^{-1}(\f_{t}-\a_{t})\right\}\\
        &\propto \exp\left\{-\frac{1}{2}(\f_{t}-\hat{\a}_{t})'\hat{\A}_{t}^{-1}(\f_{t}-\hat{\a}_{t})\right\},\\
        \hat{\A}_{t}&=\left[\frac{\bthe_{t+}\bthe_{t+}'}{\sigma_{t}\kappa_2v_{t}}+\A_t^{-1}\right]^{-1},
        \quad 
        \hat{\a}_{t}=\hat{\A}_{t}\left[\frac{\bthe_{t+}(y_{t}-\theta_{t0}-\kappa_1v_{t})}{\sigma_{t}\kappa_2 v_{t}} + \A_{t}^{-1}\a_{t}\right],
    \end{split}
    \]
    where $\bthe_{t+}=(\theta_{t1},\dots,\theta_{tJ})'$, $\a_t=(a_{t1},\dots,a_{tJ})'$ and $\A_t=\diag(A_{t1},\dots,A_{tJ})$. 
    Therefore, $\f_t$ is sampled from $N(\hat{\a}_t,\hat{\A}_t)$ for $t=1,\dots,T$. 
    
    \item 
    Conditionally on $v_{t}$ and $\f_{t}$, we run FFBS for sampling $\bthe_{1:T}$ and $\varphi_{1:T}^{-1}=\sigma_{1:T}$ in one block. 
    The FFBS algorithm is obtained by modifying the one described in \cite{MW19}. See also \cite{WH97} and \cite{Prado}. 

    Let us denote $D_{t}=(y_{1:t},v_{1:t})$. 
    \begin{itemize}
        \item Forward filtering: for $t=1,\dots,T$
        \begin{enumerate}
            \item Time $t-1$ posterior:
            \[
            \begin{split}
                &\bthe_{t-1}|\varphi_{t-1},D_{t-1}\sim N(\m_{t-1},\C_{t-1}/(\varphi_{t-1}s_{t-1}))\\
                &\varphi_{t-1}|D_{t-1}\sim G(n_{t-1}/2,n_{t-1}s_{t-1}/2)
            \end{split}
            \]
            \item Update to time $t$ prior:
            \[
            \begin{split}                
                &\bthe_{t}|\varphi_{t},D_{t-1}\sim N(\m_{t-1},\R_t /(\varphi_{t}s_{t-1})),\quad \R_{t}=\C_{t-1}/\delta\\
                &\varphi_{t}|D_{t-1}\sim G(\beta n_{t-1}/2,\beta n_{t-1}s_{t-1}/2)
            \end{split}
            \]
            \item 
            One step ahead predictive distributions:
            \[
            \begin{split}
            &y_t|\varphi_t,v_t,D_{t-1}\sim N(f_t+\kappa_1v_t,Q_t/(\varphi_{t}s_{t-1})),\\
            &v_t|\varphi_t,D_{t-1}\sim Exp(1/\varphi_t),\\
            &f_t={\F_t}'\m_{t-1},\quad Q_t={\F}_t'\R_t{\F}_t+s_{t-1}\kappa_2v_t
            \end{split}
            \]
            \item 
            Filtering update to time $t$ posterior:
            %
            \[
            \begin{split}
                &\bthe_t|\varphi_t,D_t\sim N(\m_{t},\C_t/(\varphi_t s_t))\\
                &\varphi_t|D_t\sim G(n_t/2,n_ts_t/2),\\
    &n_t=\beta n_{t-1}+3,\quad s_t=r_ts_{t-1}, \\
    &r_t=(\beta n_{t-1}+e_t^2/Q_t+2v_t/s_{t-1})/n_t,\\
    &e_t=y_t-f_t-\kappa_1v_t, \quad \A_t=\R_t\F_t/Q_t\\
    &\m_t=\m_{t-1}+\A_t e_t, \quad \C_t=r_t(\R_t-Q_t\A_t\A_t')
            \end{split}
            \]

        \end{enumerate}

        \item 
        Backward sampling:
        \begin{enumerate}
            \item At time $T$: Sample $\varphi_T$ from $G(n_T/2,n_Ts_T/2)$ and then sample $\bthe_T$ from $N(\m_T,\C_T/(\varphi_Ts_T))$. 
            \item 
            For $t=T-1,\dots,1$: 
            \begin{enumerate}
                \item 
                Sample $\eta_t$ from $G((1-\beta)n_t/2,n_ts_t/2)$ and then let $\varphi_{t}=\beta\varphi_{t+1}+\eta_t$. 
                \item 
                Sample $\theta_t$ from $N(\m_t+\delta(\bthe_{t+1}-\m_{t}), (1-\delta)\C_t/(\varphi_t s_t))$.
            \end{enumerate}            
        \end{enumerate}
        
    \end{itemize}
\end{itemize}

\subsection{Gibbs sampler for FDRQS}\label{sec:gibbs_f}
The observation equation is rewritten using the mixture representation:
\[
\begin{split}
\y_{t} &= \F_{t}\bthe_t+\kappa_1\v_t+\e_t,\quad \e_{t}\sim N(\zero, \bSig_t),\\
\end{split}
\]
where $\v_t=(v_{1t},\dots,v_{Nt})'$ with $v_{it}\sim Exp(\sigma_{it})$ for $i=1,\dots,N$, $\bSig_t=\diag(\kappa_2\sigma_{1t}v_{1t},\dots,\kappa_2\sigma_{Nt}v_{Nt})$, 
\[
\begin{split}
\F_t&=\begin{bmatrix}
 \I& \F_{t1}\dots\F_{tJ}
\end{bmatrix}
,\quad 
\F_{tj}=\diag(f_{1tj},\dots,f_{Ntj}),\quad j=1,\dots,J\\
\bthe_t&=(
    \bthe_{t0}', \bthe_{t1}',\dots,\bthe_{tJ}
)',\quad \bthe_{tj} = (\theta_{1tj},\dots,\theta_{Ntj})'.
\end{split}
\]
Since the synthesis weight has the factor structure given by $\theta_{itj}=\blam_{ij}'\u_{tj}$, where $\blam_{ij}=(\lambda_{i1j},\dots,\lambda_{iLj})'$, $\u_{tj}=(u_{t1j},\dots,u_{tLj})'$, we can write 
\[
\bthe_{tj}=\bLam_j\u_{tj}, \quad 
\bLam_j'=
\begin{bmatrix}
    \blam_{1j} & \blam_{2j} &\cdots&\blam_{Nj}
\end{bmatrix},
\]
where $\bLam_j$ is the $N\times L$ matrix of the factor loadings. 
Then the Gibbs sampler is based on the further rewritten model given by
\[
\begin{split}
\y_t&=\F_t\bLam\u_t + \kappa_1 \v_t +\e_t,\quad \e_t\sim N(\zero, \bSig_t),\\
\u_t&=\u_{t-1}+\w_t,\quad \w_t\sim N(\zero, \W_t),\\
\sigma^{-1}_{it}&=\frac{\beta_i}{\gamma_{it}}\sigma^{-1}_{i,t-1},\quad \gamma_{it}\sim Beta\left(\frac{\beta n_{i,t-1}}{2},\frac{(1-\beta) n_{i,t-1}}{2}\right), \\
\lambda_{i\ell j}&\sim N(0,\phi_{i\ell j}^{-1}\omega_{\ell j}^{-1}),\quad \phi_{i\ell j}\sim Ga(\nu_j/2,\nu_j/2),\quad \omega_{\ell j}=\prod_{h=1}^\ell \delta_{hj},\\
\delta_{1j}&\sim Ga(a_{1j},1),\quad \delta_{\ell j}\sim Ga(a_{2j},1),\quad \ell\geq 2,
\end{split}
\]
where $\bLam$ is the $N(J+1)\times L(J+1)$ matrix with $\bLam_j$, $j=0,1,\dots,J$ on the diagonal blocks and $\u_t=(\u_{t0}',\u_{t1}',\dots,\u_{tJ}')'$. 
Then, the full conditional distributions required for the proposed Gibbs sampler are described as follows.

\begin{itemize}
    \item 
    Sampling of $v_{it}$ and $\f_{it}=(f_{it1},\dots,f_{itJ})'$ proceed in the same manner as in the case of DRQS for $i=1,\dots,N$ and $t=1,\dots,T$.  

    \item 
    The observation equation of $i$th series is rewritten as
    \[
    y_{it}=\blam_i'\tilde{\u}_{it}+\kappa_1v_{it}+e_{it},\quad e_{it}\sim N(0, \sigma_{it}\kappa_2v_{it}),
    \]
    where $\blam_i=(\blam_{i0}',\blam_{i1}',\dots,\blam_{iJ}')'$, $\tilde{\u}_{it}=(\u_{t0}',f_{it1}\u_{t1}',\dots,f_{itJ}\u_{tJ})'$. 
    The full conditional distribution of $\blam_i$ is given by
    \[
    \begin{split}
    p(\blam_i|-)&\propto\exp\left\{-\frac{1}{2}\sum_{t=1}^T\frac{(y_{it}-\blam_i'\tilde{\u}_t-\kappa_1 v_{it})^2}{\kappa_2\sigma_{it}v_{it}}\right\}\exp\left\{-\frac{1}{2}\blam_i'\H_i^{-1}\blam_i\right\}\\
    &\propto\exp\left\{-\frac{1}{2}(\blam_i-\hat{\h}_i)'\hat{\H}_i(\blam_i-\hat{\h}_i)\right\},
    \end{split}
    \]
    where 
    \[
    \begin{split}
    &\hat{\H}_i=\left[\sum_{t=1}^T\frac{\tilde{\u}_{it}\tilde{\u}_{it}'}{\kappa_2\sigma_{it}v_{it}}+\H_{i}^{-1} \right]^{-1},\quad 
    \hat{\h}_i=\hat{\H}_i\left[\sum_{t=1}^T\frac{\tilde{\u}_{it}(y_{it}-\kappa_1v_{it})}{\kappa_2\sigma_{it}v_{it}}\right],\\
    &\H_i^{-1}=\diag(\phi_{i10}\omega_{10},\dots,\phi_{iL0}\omega_{L0},
    \phi_{i11}\omega_{11},\dots,\phi_{iL1}\omega_{L1},\dots,
    \phi_{i1J}\omega_{1J},\dots,\phi_{iLJ}\omega_{LJ}).
    \end{split}
    \]
    Therefore, $\blam_i$ is sampled from $N(\hat{\h}_i,\hat{\H}_i)$ for $i=1,\dots,N$. 

    \item 
    For $i=1,\dots,N$, $\ell=1,\dots,L$, $j=0,1,\dots,J$, the full conditional distribution of $\phi_{i\ell j}$ is given by 
    \[
    \begin{split}        
    p(\phi_{i\ell j}|-)&\propto \phi_{i\ell j}^{1/2}\exp\left\{-\frac{\phi_{i\ell j}\omega_{\ell j}\lambda_{i\ell j}^2}{2}\right\}\phi_{i \ell j}^{\nu_j/2-1}\exp\left\{-\frac{\nu_{j}}{2}\phi_{i\ell j}\right\}\\
    &\propto\phi_{i\ell j}^{(\nu_j+1)/2-1}\exp\left\{-\frac{\omega_{\ell j}\lambda_{i\ell j}^2+\nu_{j}}{2}\phi_{i\ell j}\right\}. 
    \end{split}
    \]
    Then $\phi_{i\ell j}$ is sampled from $Ga\left(\frac{\nu_j+1}{2}, \frac{\omega_{\ell j}\lambda_{i\ell j}^2+\nu_{j}}{2}\right)$.  
    \item 
    For $j=0,1,\dots,J$, the full conditional distribution of $\delta_{1j}$ is given by
    \[
    \begin{split}
        p(\delta_{1j}|-)&\propto \prod_{i=1}^N\prod_{\ell=1}^L\left[\delta_{1j}^{1/2}\exp\left\{-\frac{\phi_{i\ell j}\lambda_{i\ell j}^2\omega^{(1)}_{\ell, j}\delta_{1j}}{2}\right\}\right]
        \delta_{1 j}^{a_{1 j}-1}\exp\left\{-\delta_{1 j}\right\},\quad \omega^{(h)}_{\ell,j}=\prod_{s=1,s\neq h}^\ell\delta_{sj}\\
        &=\delta_{1j}^{NL/2+a_{1j}-1}\exp\left\{-\left(1 + \frac{1}{2}\sum_{\ell=1}^L\omega_{\ell,j}^{(1)}\sum_{i=1}^N\phi_{i\ell j}\lambda^2_{i\ell j} \right)\delta_{1j}\right\}. 
    \end{split}
    \]
    Then $\delta_{1j}$ is sampled from $Ga\left(\frac{NL}{2}+a_{1j}, 1 + \frac{1}{2}\sum_{\ell=1}^L\omega_{\ell,j}^{(1)}\sum_{i=1}^N\phi_{i\ell j}\lambda^2_{i\ell j} \right)$. 
    \item 
    Similarly, for $h=2,\dots,L$ and $j=0,1,\dots,J$, the full conditional distribution of $\delta_{h j}$ is given by
    \[
    \begin{split}
        p(\delta_{hj}|-)&\propto \prod_{i=1}^N\prod_{\ell=h}^L\left[\delta_{h j}^{1/2}\exp\left\{-\frac{\phi_{i\ell j}\lambda_{i\ell j}^2\omega^{(h)}_{\ell, j}\delta_{hj}}{2}\right\}\right]
        \delta_{h j}^{a_{2 j}-1}\exp\left\{-\delta_{h j}\right\}\\
        &=\delta_{hj}^{N(L-h+1)/2+a_{2j}-1}\exp\left\{-\left(1+\frac{1}{2}\sum_{\ell=h}^L\omega_{\ell,j}^{(h)}\sum_{i=1}^N\phi_{i\ell j}\lambda_{i\ell j}^2\right)\delta_{hj}\right\}
    \end{split}
    \]
    Then $\delta_{hj}$ is sampled from $Ga\left(\frac{N(L-h+1)}{2}+a_{2j}, 1+\frac{1}{2}\sum_{\ell=h}^L\omega_{\ell,j}^{(h)}\sum_{i=1}^N\phi_{i\ell j}\lambda_{i\ell j}^2 \right)$.

    \item 
    Conditionally on all the other variables, we implement FFBS for sampling $\u_{1:T}$ in one block, similar to \cite{McAlinn20}. 

    \begin{itemize}
        \item Forward filtering for $t=1,\dots,T$
        \begin{enumerate}
            \item 
            Time $t-1$ posterior:
            \[
            \u_{t-1}|\y_{1:t-1} \sim N(\m_{t-1}, \C_{t-1})
            \]
            \item 
            Time $t$ prior: 
            \[
            \u_t|\y_{1:t-1}\sim N(\m_{t-1}, \R_t),\quad \R_t=\C_{t-1}/\delta
            \]
            \item 
            One-step ahead predictive distribution
            \[
            \begin{split}
            &\y_t|\y_{1:t-1} \sim N(\f_t,\Q_t),\\
            &\f_t=\kappa_1\v_t+\tilde{\F}_t\m_{t-1},\quad
            \tilde{\F}_t=\F_t\bLam,\quad 
            \Q_t= \tilde{\F}_t\R_t\tilde{\F}_t'+\bSig_t.
            \end{split}
            \]
            \item 
            Time $t$ posterior:
            \[
            \begin{split}
            &\u_t|\y_{1:t}\sim N(\m_t, \C_t),\\
            &\m_t=\m_{t-1}+\A_t\e_t,\quad \e_t=\y_t-\f_t,\quad \C_t=\R_t-\A_t\Q_t\A_t'.
            \end{split}
            \]
        \end{enumerate}
        \item 
        Backward sampling
        \begin{enumerate}
            \item Sample $\u_T$ from $N(\m_T, \C_T)$. 
            \item For $t=T-1,\dots,1$, sample $\u_t$ from $N(\m_t+\delta(\u_{t+1}-\m_{t}),(1-\delta)\C_t)$. 
        \end{enumerate}
    \end{itemize}

    \item
    For $i=1,\dots,N$, we implement FFBS to sample $\varphi_{i,1:T}=\sigma^{-1}_{i,1:T}$ in one block conditionally on all the other variables.  
    \begin{enumerate}
        \item Forward filtering:
        \[
        \begin{split}
        &\varphi_{it}|\y_{1:t}\sim Ga(n_{it}/2,d_{it}/2)\\
        &n_{it}=\beta n_{i,t-1}+3,\quad 
        d_{it}=\beta d_{i,t-1} + \frac{(y_{it}-\tilde{f}_{it}-\kappa_1 v_{it})^2}{\kappa_2 v_{it}} + 2 v_{it},
        \end{split}
        \]
        where $\tilde{f}_{it}$ is the $i$th element of $\tilde{\F_t}\u_t$. 

        \item 
        Backward sampling: 
        \begin{itemize}
            \item 
            Sample $\varphi_{iT}$ from $Ga(n_{iT}/2,d_{iT}/2)$.
            \item 
            For $t=T-1,\dots,1$, sample $\eta_{it}$ from $Ga((1-\beta)n_{it}/2,d_{it}/2)$ and let $\varphi_{it}=\beta\varphi_{i,t+1}+\eta_{it}$. 
            
        \end{itemize}
    
    \end{enumerate}
\end{itemize}

\section{Agent models}
\subsection{DQLM}\label{sec:dqlm}
DQLM of \cite{DQLM} can be regarded as a special case of the DRQS with the fixed predictors $\x_t$ and $\sigma$. 
The model is given by
%
\[
\begin{split}
y_t &= \x_{t}'\bbe_{t} + \epsilon_{t},\quad \epsilon_t\sim AL(\tau, \sigma),\\
\bbe_{ t} &= \bbe_{t-1} + \w_{t},\quad \w_{t}\sim N(\zero, \W_{t}), 
\end{split}
\]
where $\x_{t}$ is the fixed $P$ vector of predictors, $\bbe_{t}$ is the time varying coefficient vector, $\W_t$ is the covariance matrix of the state equation. 
The model is estimated using the Gibbs sampler with FFBS. 
A discount factor is used for $\W_t$. 
Throughout this paper, we use the following prior distributions: $\bbe_0\sim N(\zero, 1000\I)$ and $\sigma\sim IG(0.01,0.01)$.

\subsection{FQBART}\label{sec:fqbart}
For $i$th series at time $t$, we consider the version of FQBART given by
\[
y_{it}=g_{\tau,i}(\x_{it}) + \lambda_{i\tau}f_{\tau, t} + \epsilon_{\tau, it},\quad \epsilon_{\tau, it} \sim AL(\tau,\sigma_{\tau,i}),
\]
for $i=1,\dots,N$ and $t=1,\dots,T$, where $\x_{it}$ is the $P$ vector of predictors, $g_{\tau,i}:\mathbb{R}^P\rightarrow\mathbb{R}$ is the series specific unknown function, $\lambda_{\tau,i}$ is the series specific factor loading, $f_{\tau,t}$ is the corresponding factor. 
The unknown function $g_{\tau,i}$ is approximated using BART. 
The factor includes the stochastic volatility structure: $f_{\tau,t}|h_{\tau,t}\sim N(0,\exp(h_{\tau,t}))$ and the log volatility $h_{\tau,t}$ follows the AR(1) process: $h_{\tau,t}=\rho_\tau h_{\tau,t-1}+\varsigma_\tau u_{\tau,t}$, $u_{\tau,t}\sim N(0,1)$. 
For this rather complex model, we used the default prior distributions set by the authors. 
For the parameters in BART, the number of ensemble trees is set to $S=250$, we set $\alpha=0.95$ and $\zeta=2$ such that the probability that the node at depth $d$ is partitioned as non-terminal node with probability proportional to $\alpha(1+d)^{-\zeta}$, the node parameter follows the normal distribution with mean zero and variance equal to $(\max_ty_{it}-\min_{t}{y_{it}})/(2\gamma/\sqrt{S})$ with $\gamma=2$. 
For the remaining parameter, $\sigma_{\tau,i}\sim IG(0.5,0.5)$, $\lambda_{\tau,t}\sim N(0,1)$, $(\rho_\tau+1)/2\sim Be(5,1.5)$, $\varsigma_\tau\sim Ga(0.5,0.5)$. 

To fit the model, we used the R code provided by the authors: \url{https://github.com/mpfarrho/qf-bart}.

\section{Algorithm of \cite{Mitchell24}}\label{sec:mitchell}
Here we describe the algorithm of \cite{Mitchell24} to obtain samples from the predictive distribution. 
We suppress the subscript $i$, as it can be implemented to each series independently. 
Given the quantile forecasts $\hat{Q}_{t}(\tau)$ of $y_t$ on the grid of quantiles, $\tau=\tau_1,\dots,\tau_K$, we can generate the sample size of $R$ from the predictive distributions can be obtained through the following steps. 
\begin{enumerate}
    \item 
    Rearrange $\hat{Q}_{t}(\tau_k)$'s in an ascending order in $\tau$ in case they are not monotonoic. 
    \item 
    For $k=2,\dots,K$, generate  $(\tau_k-\tau_{k-1})R$ draws from $U\left(\hat{Q}_t(\tau_{k-1}), \hat{Q}_t(\tau_k)\right)$. 
    \item 
    \begin{enumerate}
        \item 
    Fit the Gaussian distributions to the left and right tails by solving the following for $(\mu_{1t},\sigma_{1t})$ and $(\mu_{2t},\sigma_{2t})$:
    \[\begin{split}        
    &\Phi\left(\hat{Q}_{t}(\tau_1);\mu_{1t},\sigma_{1t}\right)=\tau_1, \quad \Phi\left(\hat{Q}_{t}(\tau_2);\mu_{1t},\sigma_{1t}\right)=\tau_2,\\
    &\Phi\left(\hat{Q}_{t}(\tau_{K-1});\mu_{2t},\sigma_{2t}\right)=\tau_{K-1}, \quad \Phi\left(\hat{Q}_{t}(\tau_K);\mu_{2t},\sigma_{2t}\right)=\tau_K,
    \end{split}
    \]
    where $\Phi(\cdot;\mu,\sigma)$ denotes the distribution function of $N(\mu,\sigma^2)$. 
    \item 
    Draw $R\tau_1$ observations from $N(\mu_{1t},\sigma_{1t}^2)$ and $R(1-\tau_K)$ observations from $N(\mu_{2t},\sigma_{2t}^2)$. 
    \end{enumerate}
\end{enumerate}
In the analysis of Section~\ref{sec:gar} of the main text, we set $R=10000$. 

\section{Additional results}
\subsection{US inflation-at-risk}\label{sec:supp_iar}
\begin{figure}[H]
    \centering
    \includegraphics[width=0.45\linewidth]{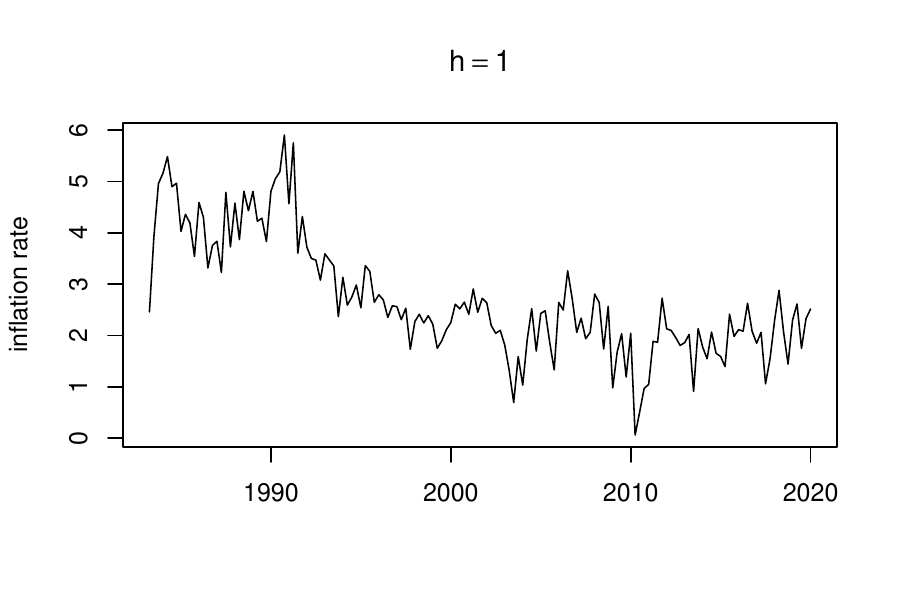}
    \includegraphics[width=0.45\linewidth]{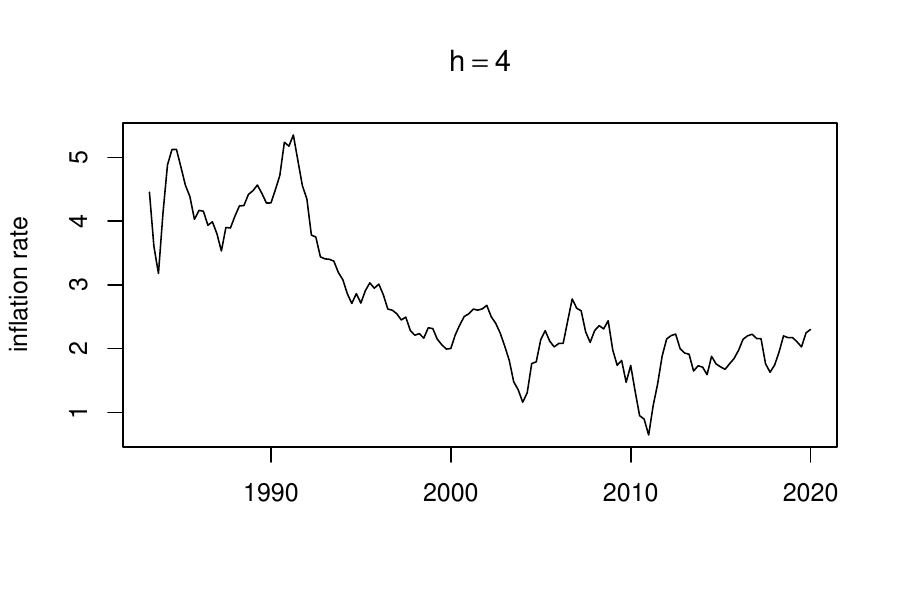}
    \caption{US annualised quarterly inflation rate ($h=1$) and average annual inflation rate ($h=4$) between 1983Q1 and 2019Q4}
    \label{fig:supp_iar_data}
\end{figure}

\begin{figure}[H]
    \centering
    \includegraphics[width=\textwidth]{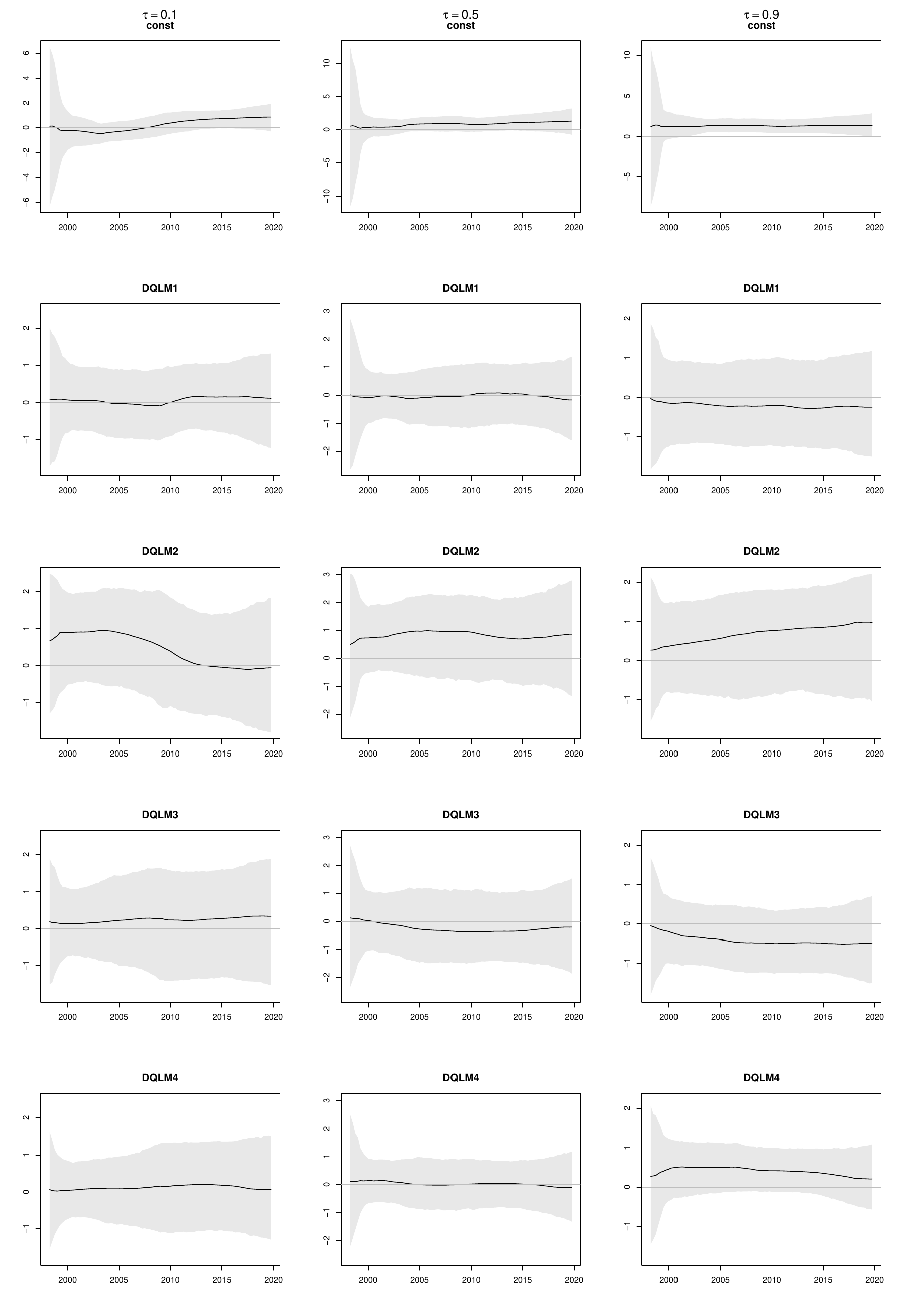}
    \caption{Posterior means (solid lines) and 95\% credible intervals (shaded areas) of synthesis weights $\theta_{tj}$ at 2019Q3 for $\tau=0.1$ , $0.5$ and $0.9$  for $h=1$. }
    \label{fig:supp_iar_theta_1}
\end{figure}

\begin{figure}[H]
    \centering
    \includegraphics[width=\textwidth]{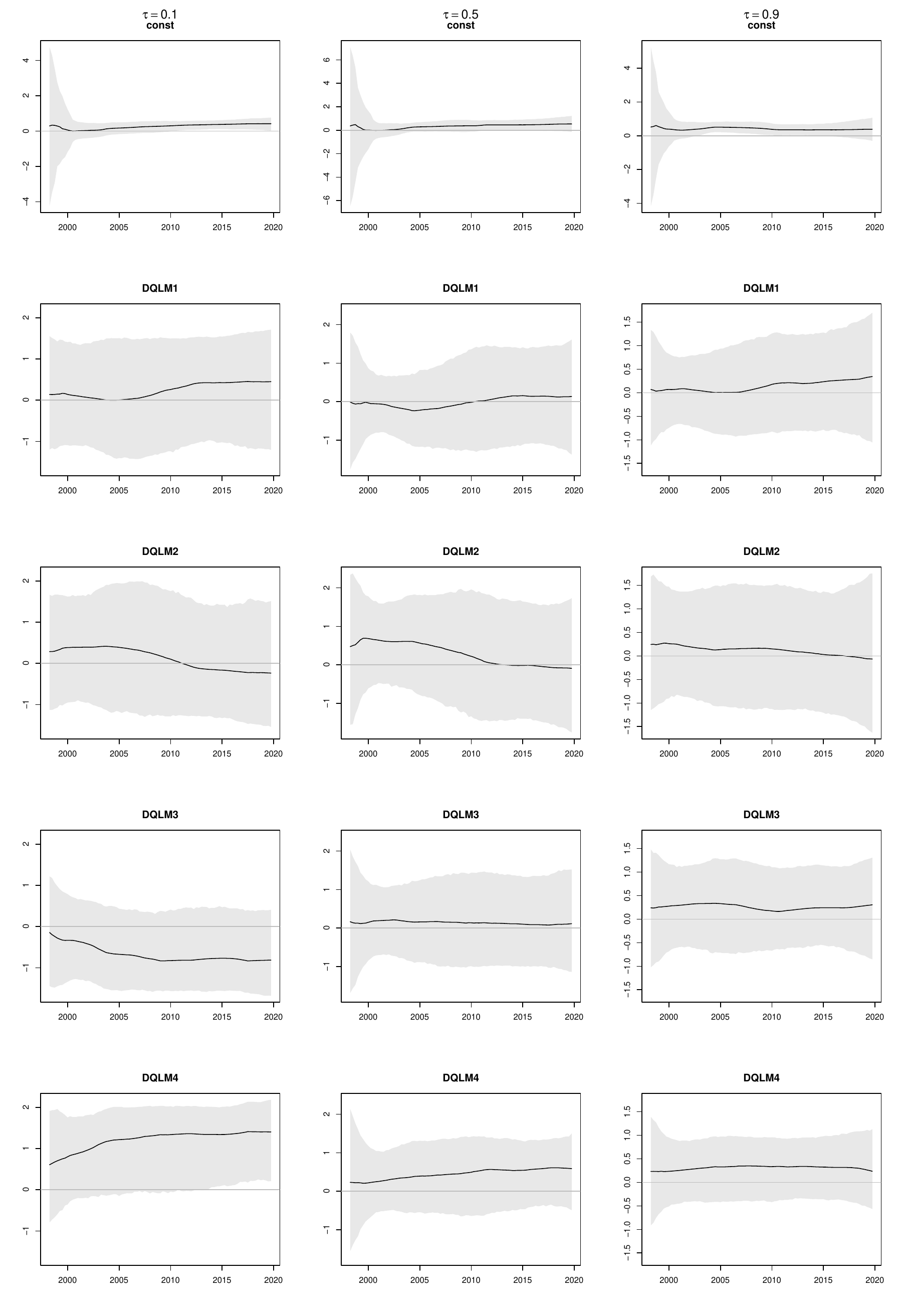}
    \caption{Posterior means (solid lines) and 95\% credible intervals (shaded areas) of synthesis weights $\theta_{tj}$ at 2019Q3 for $\tau=0.1$, $0.5$  and $0.9$  for $h=4$. }
    \label{fig:supp_iar_theta_4}
\end{figure}

\subsection{Global growth-at-risk}\label{sec:supp_gar}

\begin{figure}[H]
    \centering
    \includegraphics[width=\textwidth]{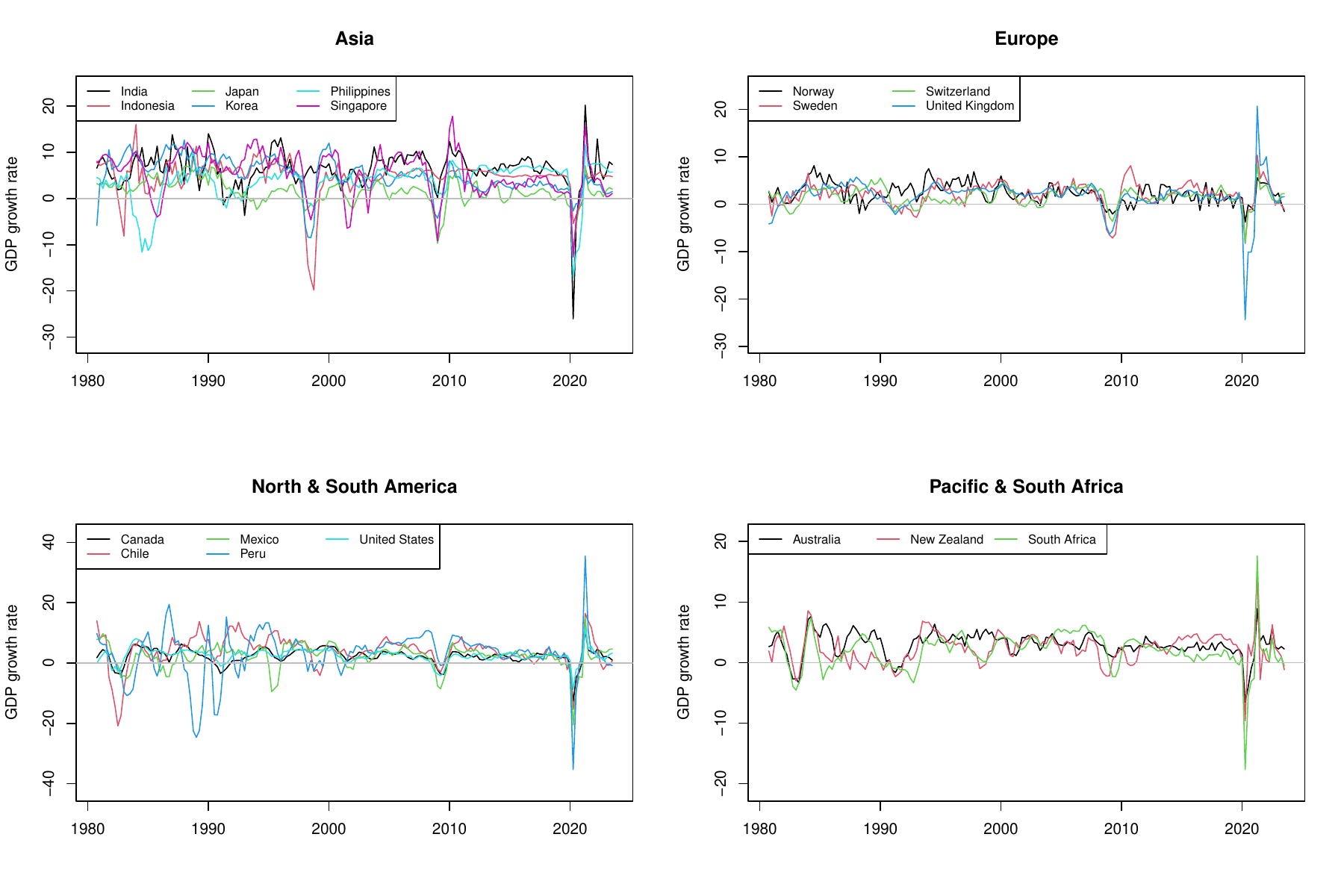}
    \caption{Average annual growth rates ($h=4$) of the eighteen countries divided by the regions.}
    \label{fig:supp_gar_data_4}
\end{figure}

\begin{table}[H]
    \centering
    \caption{Cumulative total CRPS for $h=1$ and $4$, and cumulative CRPS for the countries ($h=1$) relative to FQBART at the end of the data period 2023Q3 for $h=4$.}
    \begin{tabular}{llrrrrrr}\toprule
                & FDRQS & DRQS & DQLM1 & DQLM2 & DQLM3\\\hline
Total          & 0.757 & 0.897 & 0.796 & 0.870 & 0.798\\\hline
Australia      & 0.622 & 0.644 & 0.658 & 0.713 & 0.682\\
Canada         & 0.769 & 1.022 & 0.767 & 0.912 & 0.750\\
Chile          & 0.758 & 0.769 & 0.740 & 0.772 & 0.741\\
India          & 0.787 & 1.001 & 0.847 & 0.917 & 0.870\\
Indonesia      & 0.494 & 0.615 & 0.491 & 0.546 & 0.506\\
Japan          & 0.650 & 0.707 & 0.731 & 0.722 & 0.736\\
Korea          & 0.445 & 0.481 & 0.480 & 0.494 & 0.469\\
Mexico         & 0.885 & 1.063 & 0.892 & 1.030 & 0.893\\
New Zealand    & 0.906 & 0.992 & 0.977 & 1.010 & 0.960\\
Norway         & 0.872 & 0.918 & 0.899 & 0.902 & 0.908\\
Peru           & 0.910 & 1.255 & 1.000 & 1.134 & 1.004\\
Philippines    & 0.773 & 1.075 & 0.767 & 0.924 & 0.765\\
Singapore      & 0.739 & 0.820 & 0.775 & 0.842 & 0.759\\
South Africa   & 0.649 & 0.670 & 0.764 & 0.861 & 0.740\\
Sweden         & 0.669 & 0.706 & 0.709 & 0.695 & 0.725\\
Switzerland    & 0.683 & 0.796 & 0.695 & 0.814 & 0.723\\
United Kingdom & 0.943 & 1.206 & 0.946 & 1.066 & 0.943\\
United States  & 0.700 & 0.851 & 0.782 & 0.825 & 0.777\\\bottomrule
\end{tabular}
    \label{tab:gar_crps4}
\end{table}

\begin{figure}[H]
    \centering
    \includegraphics[height=0.98\textheight]{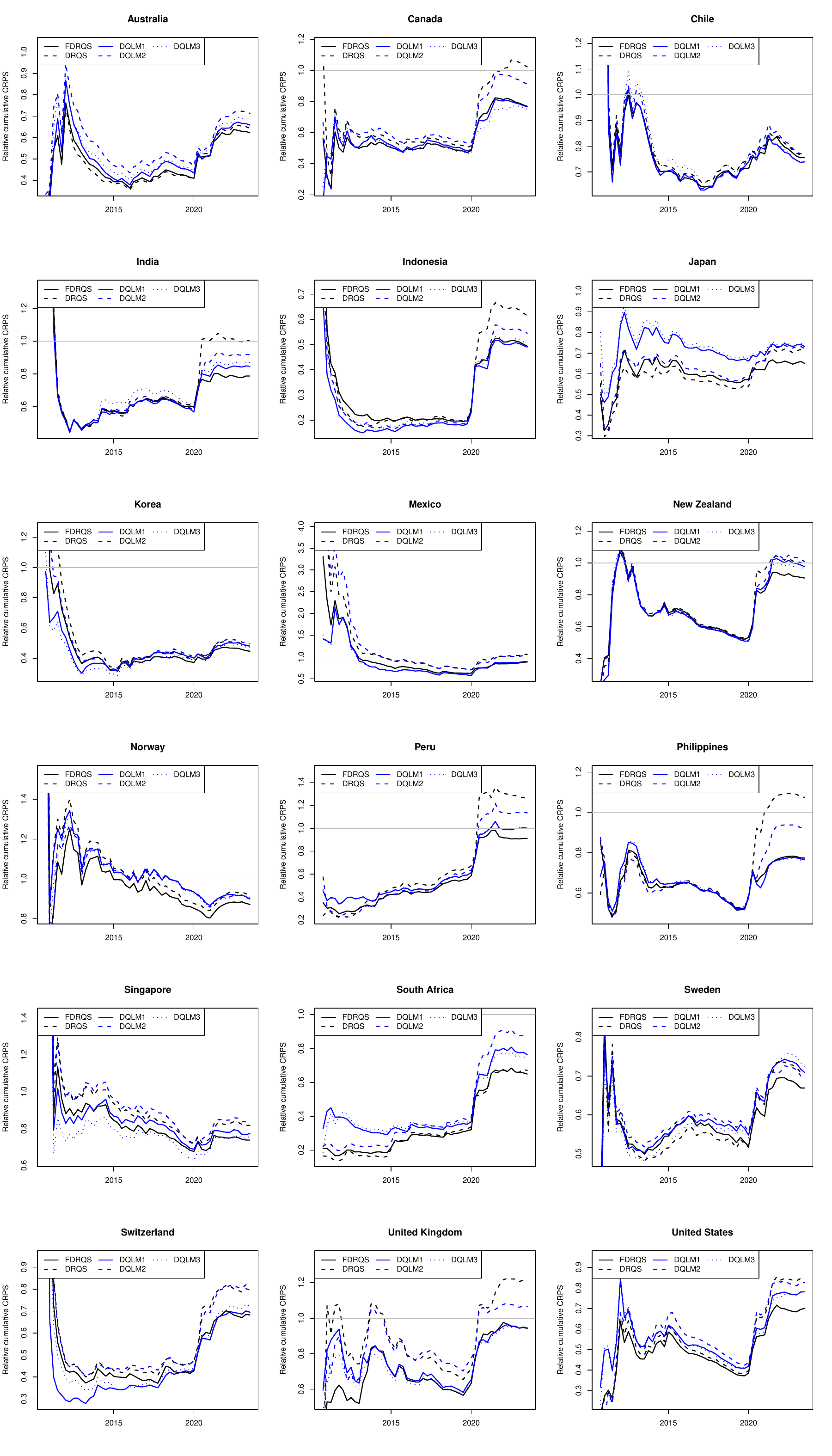}
    \caption{Time series plots of the cumulative CRPS with \texttt{none} relative to FQBART for each contry ($h=4$). }
    \label{fig:supp_gar_crps_4}
\end{figure}

\begin{figure}[H]
    \centering
    \includegraphics[width=\textwidth]{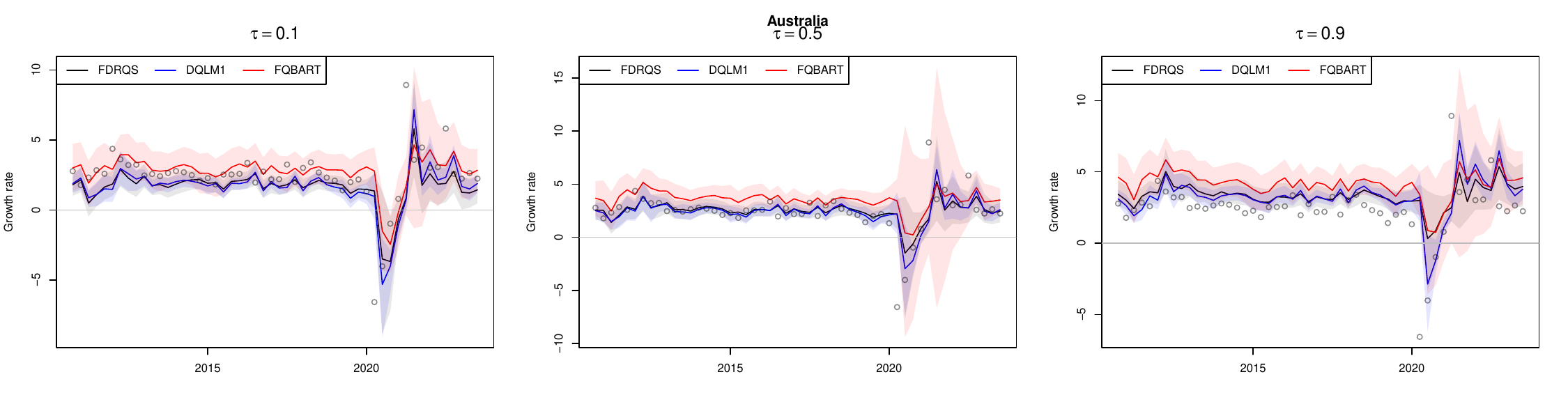}
    \includegraphics[width=\textwidth]{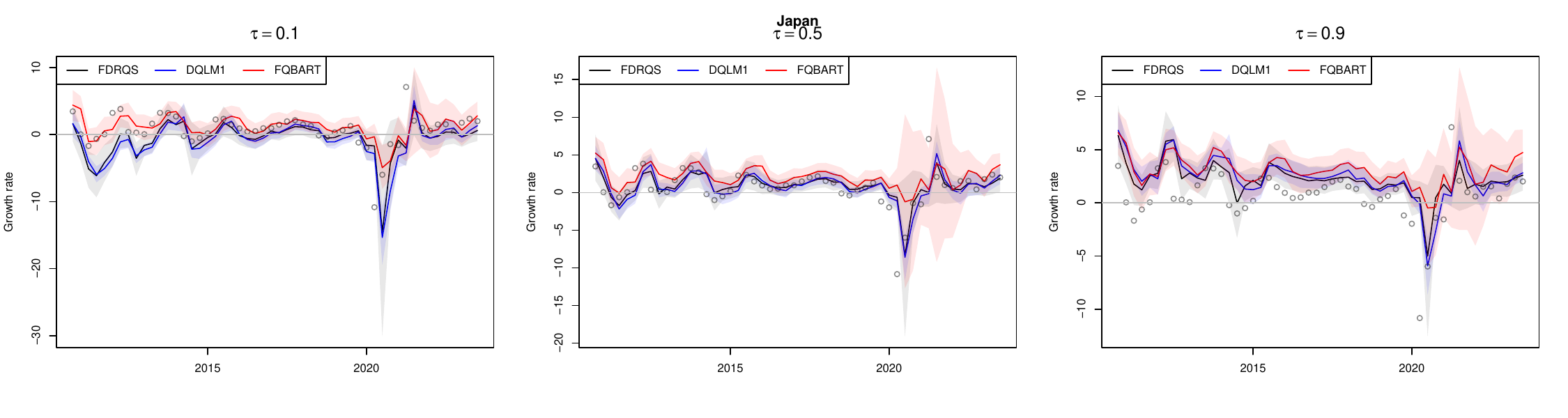}
    \includegraphics[width=\textwidth]{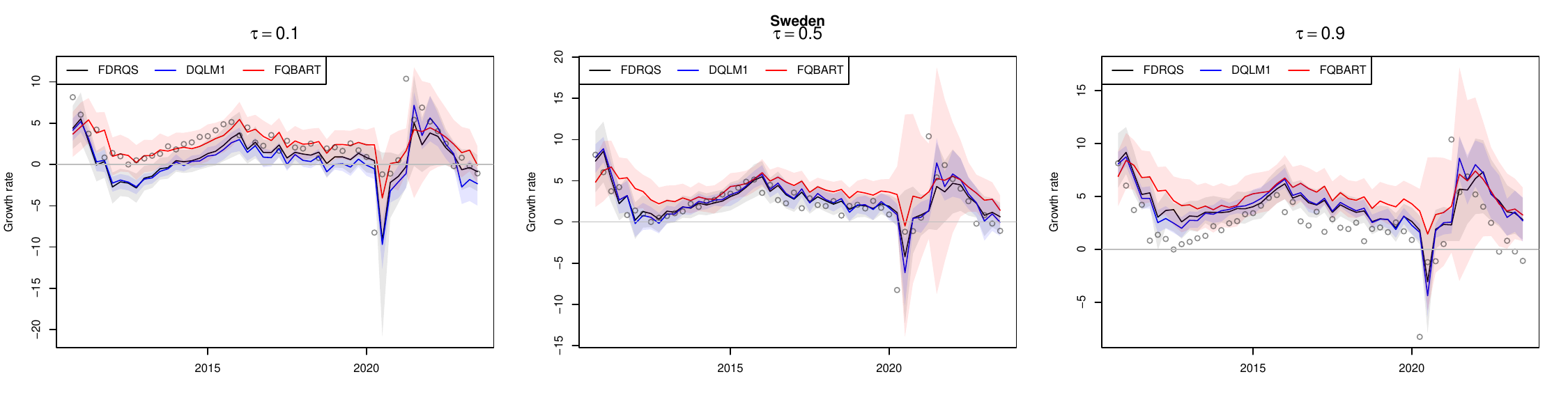}
    \includegraphics[width=\textwidth]{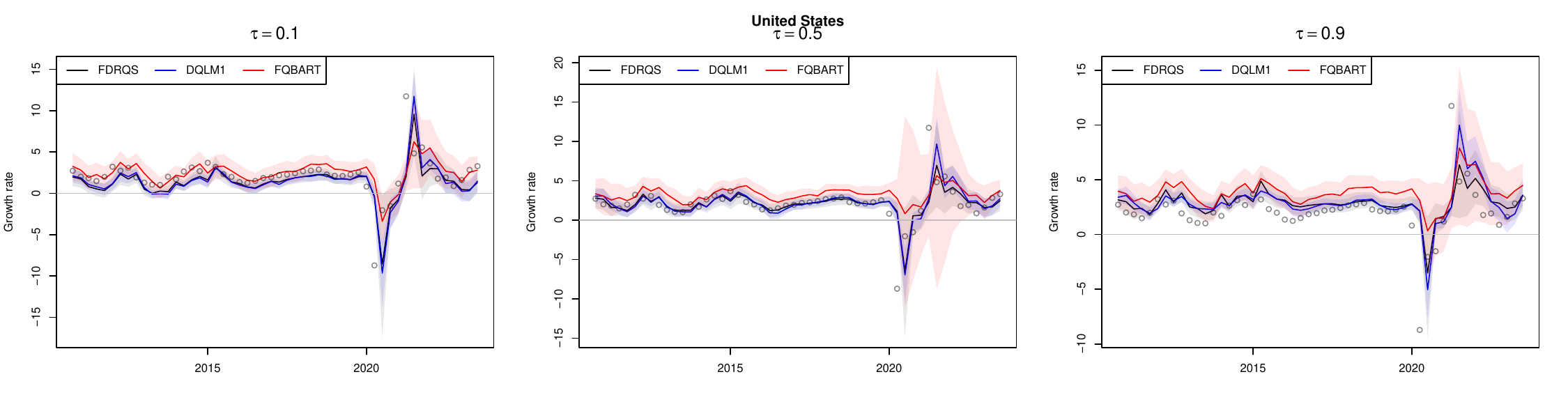}
    \caption{One-step ahead forecast of quantiles (solid lindes) with 95\% intervals (shaded areas) for $\tau=0.1$, $0.5$ and $0.9$ for FDRQS, DQLM1 and FQBART for Australia, Japan, Sweden and the United States for $h=1$. The points indicate the observed growth rates.}
    \label{fig:supp_gar_pred_4}
\end{figure}

\begin{figure}[H]
    \centering
    \includegraphics[width=\textwidth]{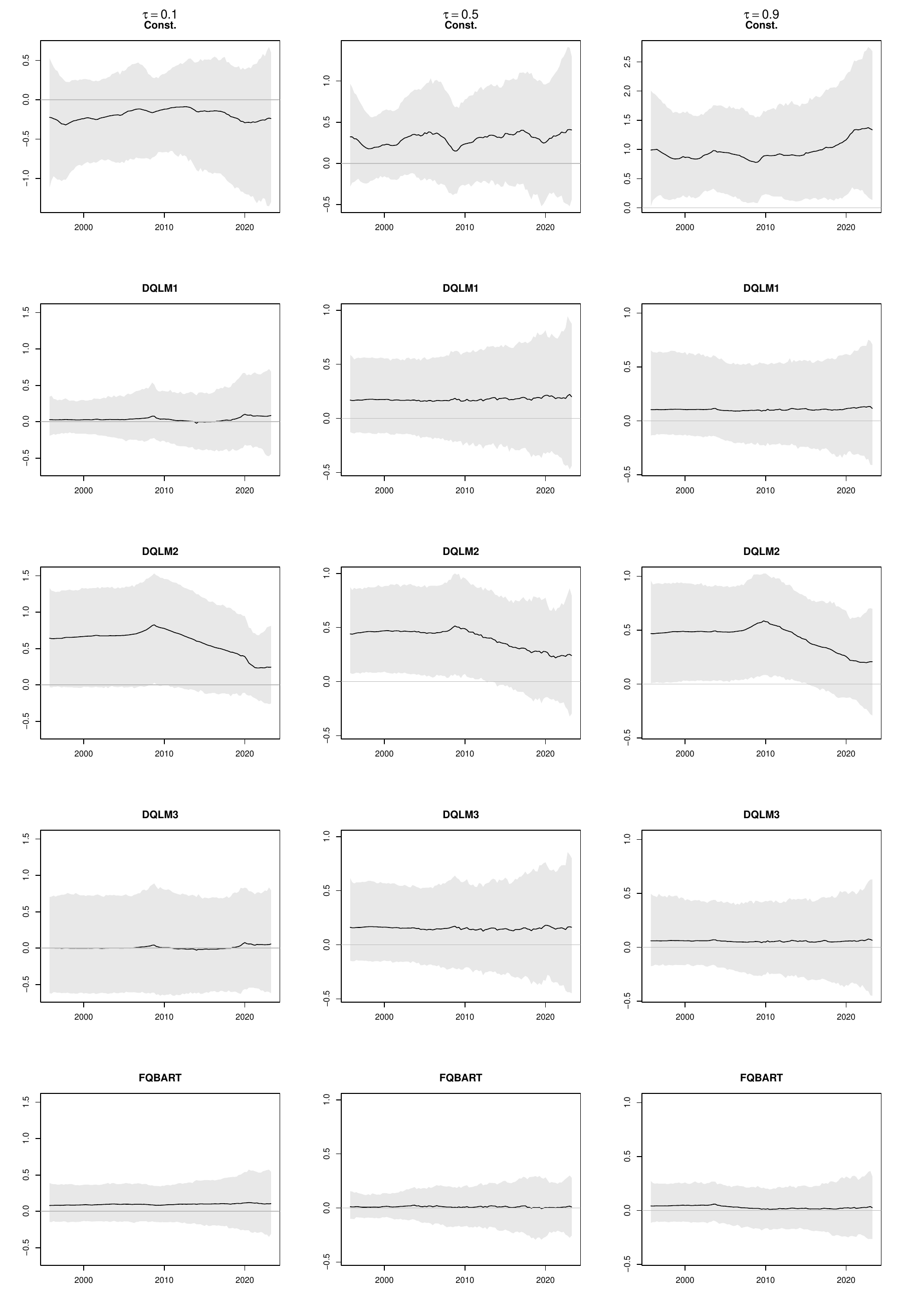}
    \caption{Posterior means (solid lines) and 95\% credible intervals (shaded areas) of the synthesis weights $\theta_{\tau,itj}$ for Japan ($h=1$) for $\tau=0.1$, $0.5$ and $0.9$ in 2023Q3.}
    \label{fig:supp_gar_theta_jpn_4}
\end{figure}

\begin{figure}[H]
    \centering
    \includegraphics[width=\textwidth]{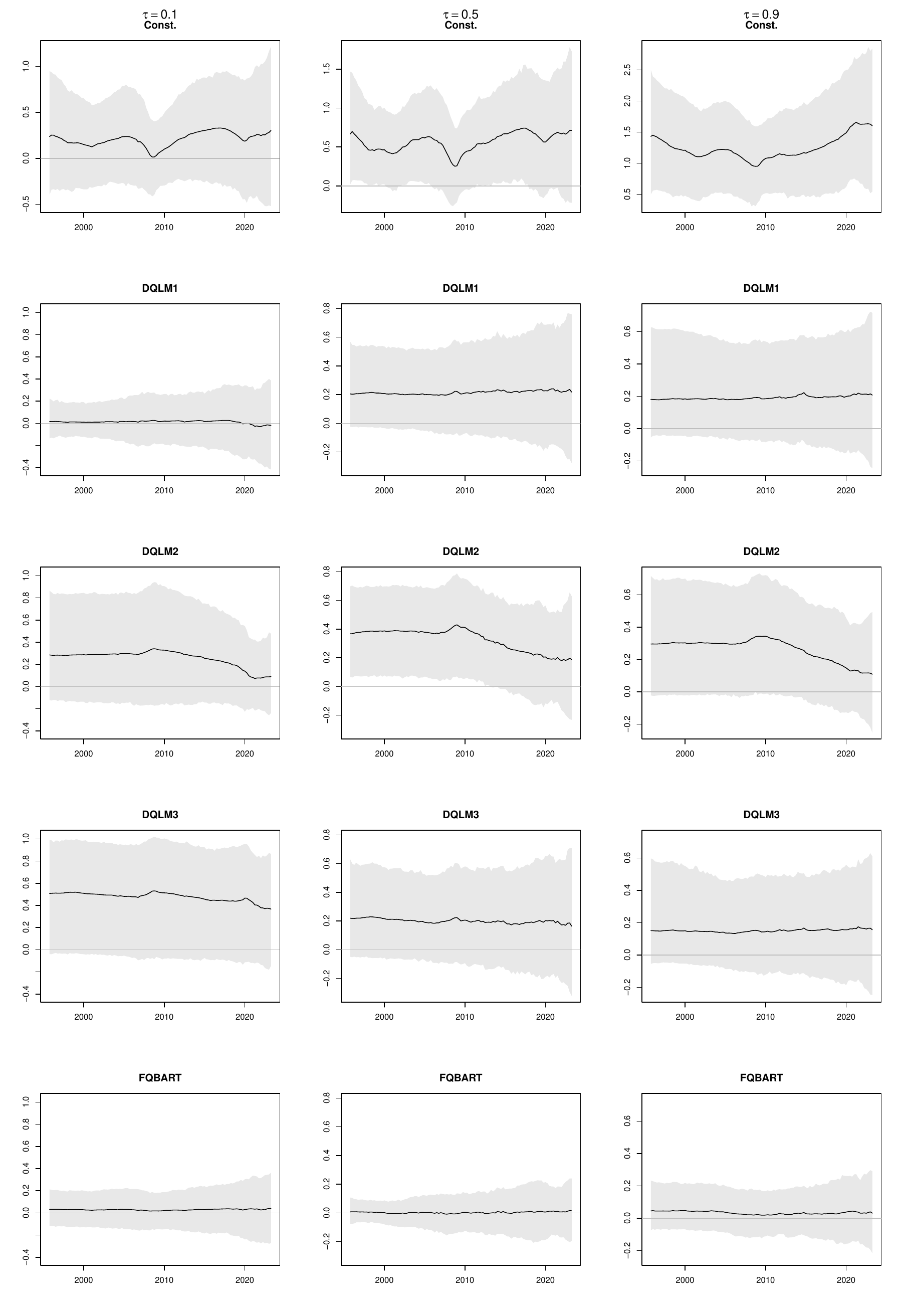}
    \caption{Posterior means (solid lines) and 95\% credible intervals (shaded areas) of the synthesis weights $\theta_{\tau,itj}$ for the United States ($h=4$) for $\tau=0.1$, $0.5$ and $0.9$ in 2023Q3.}
    \label{fig:supp_gar_theta_us_4}
\end{figure}

\begin{figure}[H]
    \centering
    \includegraphics[height=0.95\textheight]{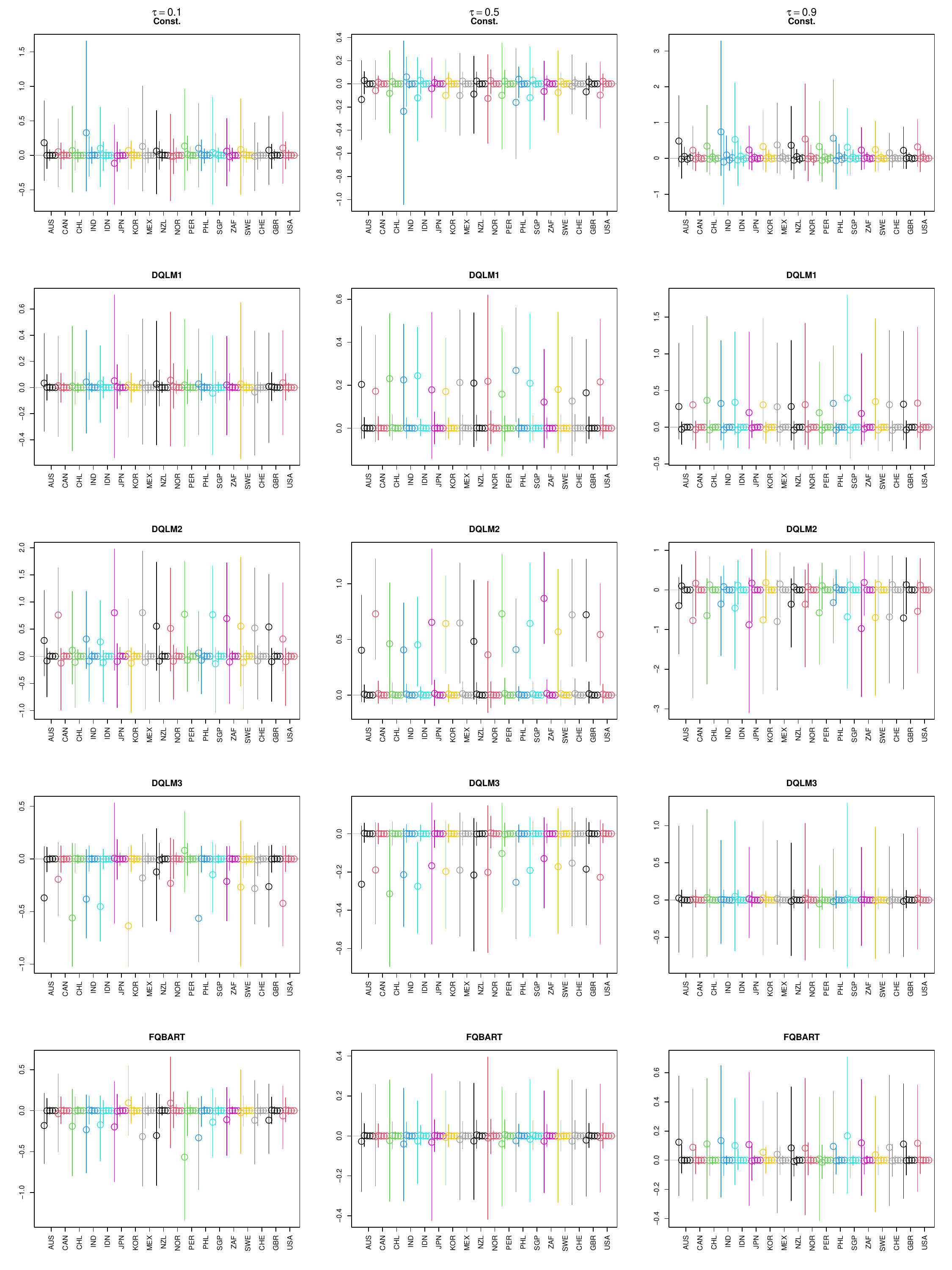}
    \caption{Posterior means (points) and 95\% credible intervals (line segments) of the factor loading $\lambda_{\tau,ij}$ for $\tau=0.1$, $0.5$ and $0.9$ for $h=4$. }
    \label{fig:supp_gar_lam_4}
\end{figure}

\begin{figure}
    \centering
    \includegraphics[width=0.32\textwidth]{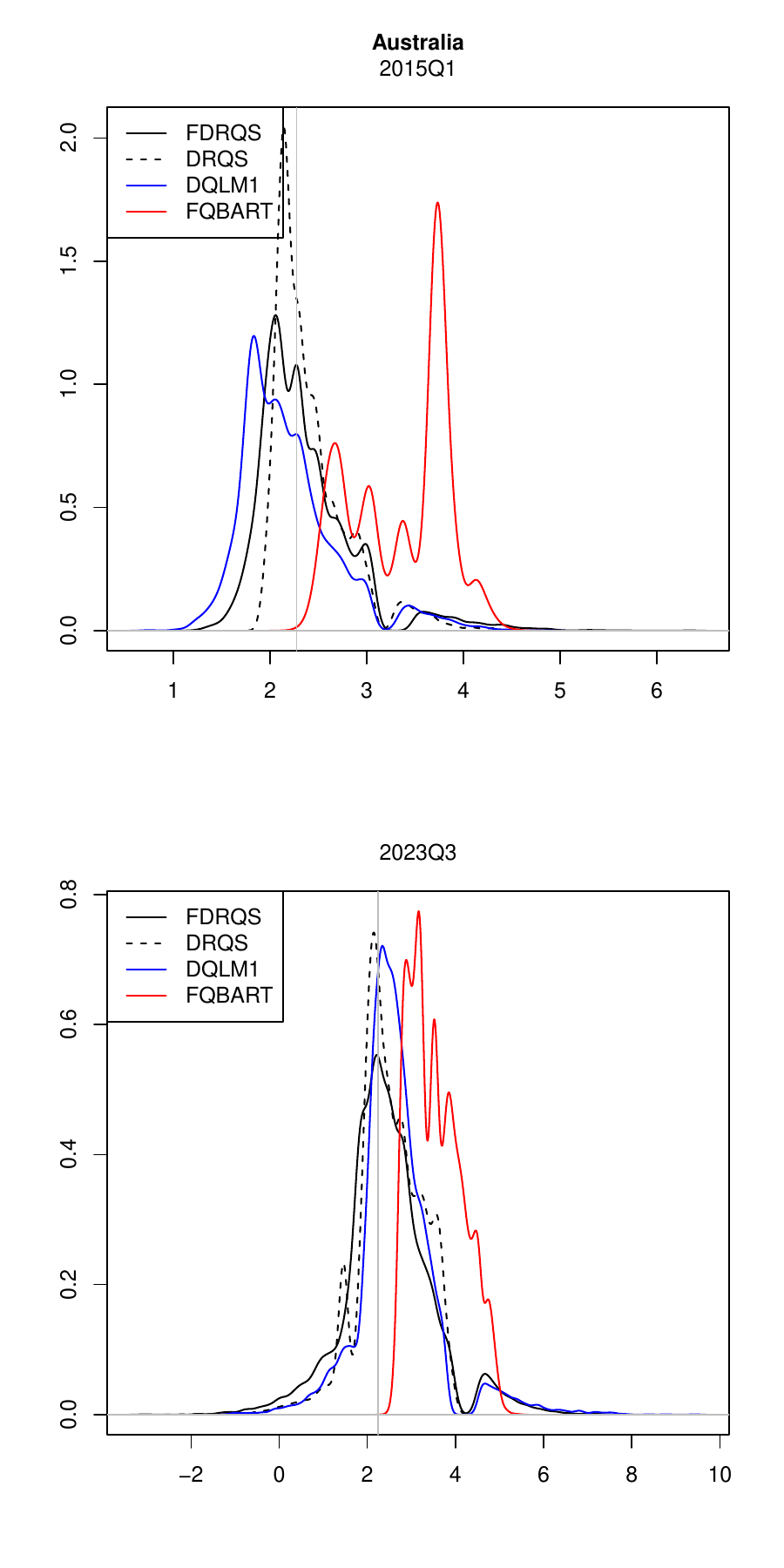}
    \includegraphics[width=0.32\textwidth]{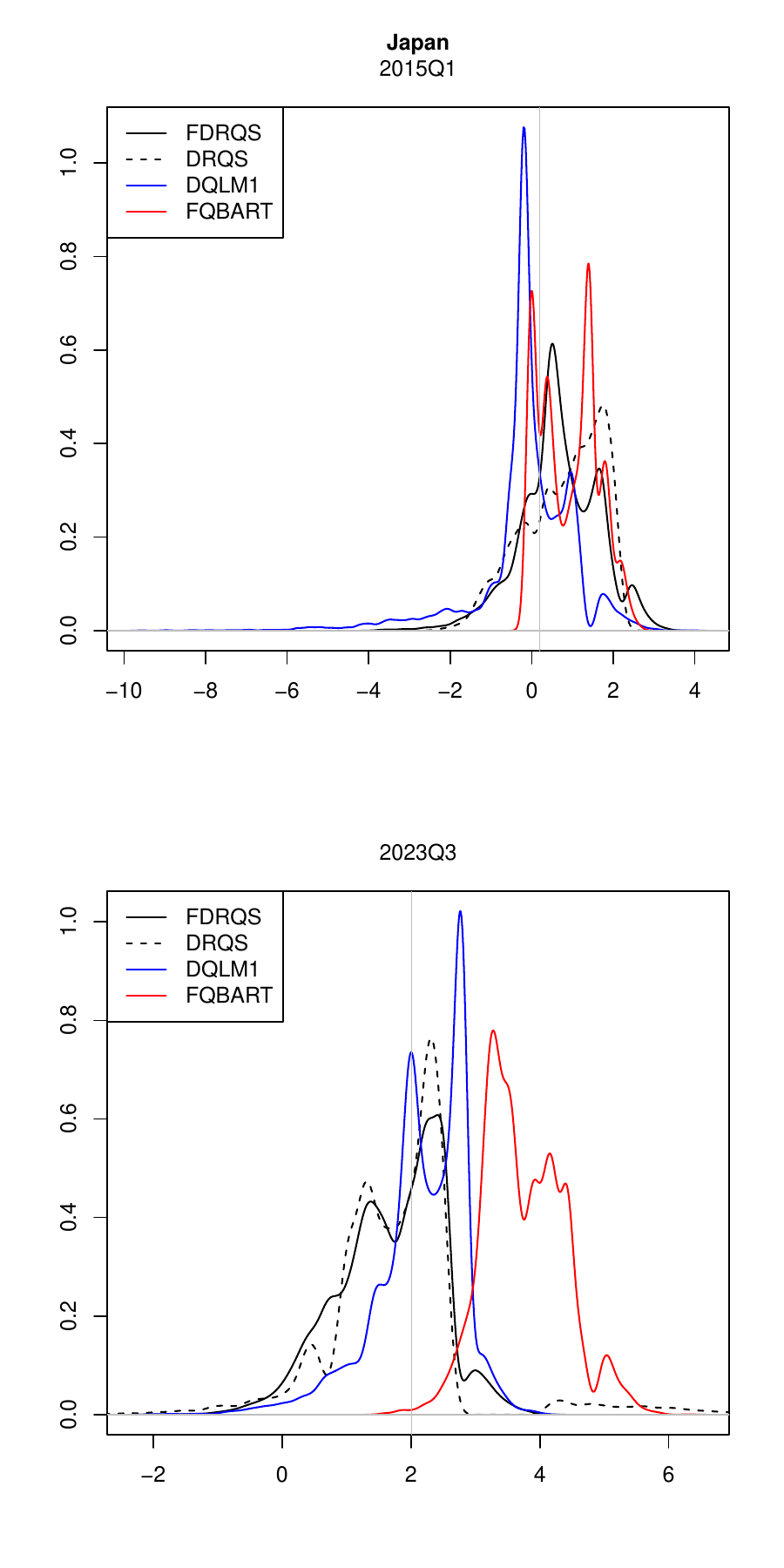}
    \includegraphics[width=0.32\textwidth]{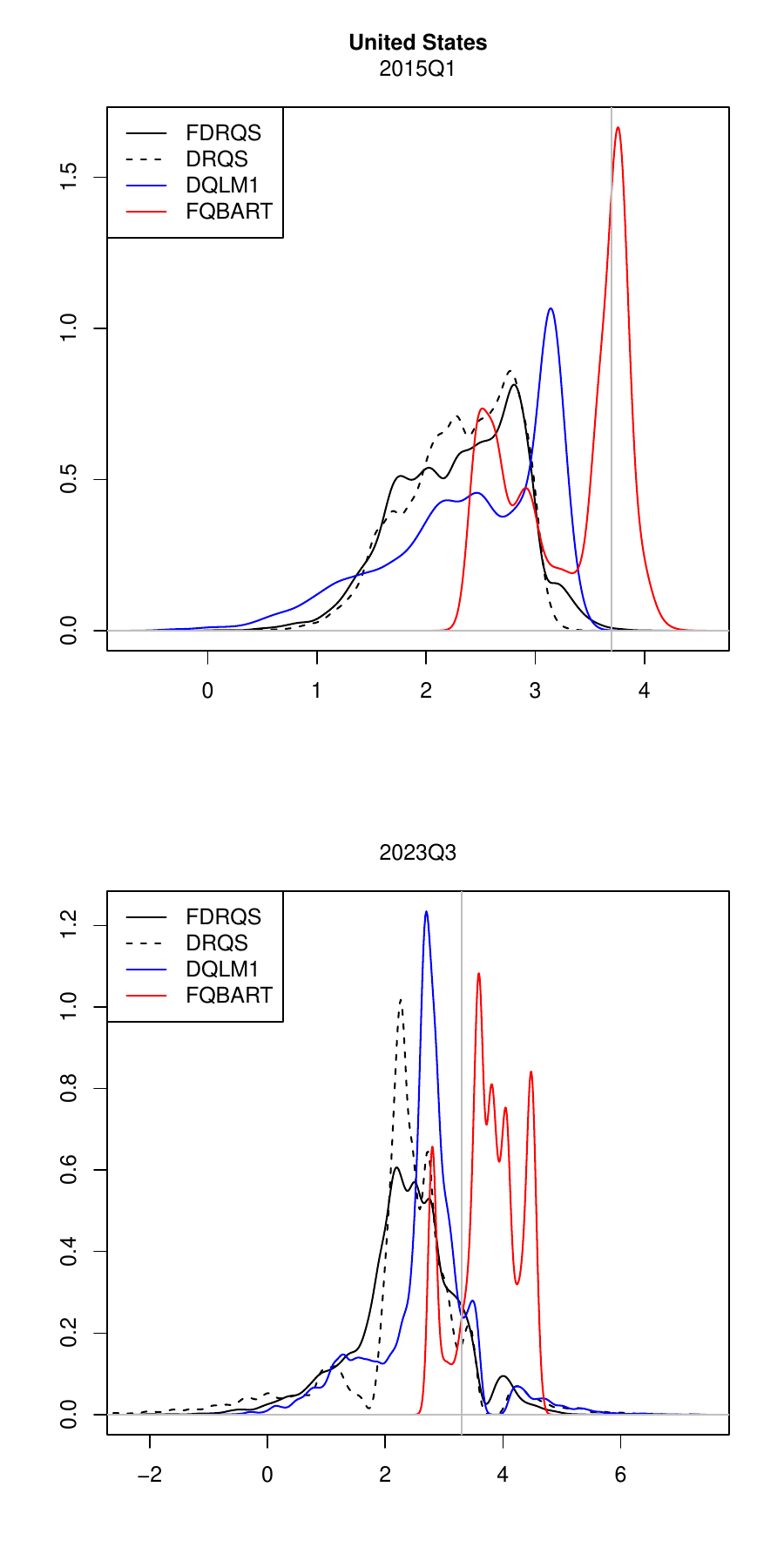}
    \caption{Predictive distribution of growth rate under FDRQS, DRQS, DQLM1 and FQBART in 2014Q3 and 2023Q3 for Australia, Japan, and the United States ($h=4$). The vertical lines indicate the observed growth rates. }
    \label{fig:supp_gar_pred_dist_4}
\end{figure}

\bibliographystyle{chicago}
\bibliography{Ref}

@article{Adrian,
 ISSN = {00028282, 19447981},
 URL = {https://www.jstor.org/stable/26637205},
 author = {Tobias Adrian and Nina Boyarchenko and Domenico Giannone},
 journal = {The American Economic Review},
 number = {4},
 pages = {1263--1289},
 publisher = {American Economic Association},
 title = {Vulnerable Growth},
 urldate = {2024-11-06},
 volume = {109},
 year = {2019}
}

@article{QBMA,
	author = {Bernardi, Mauro and Casarin, Roberto and Maillet, Bertrand B. and Petrella, Lea},
	date = {2024/11/21},
	doi = {10.1007/s10479-024-06378-7},
	id = {Bernardi2024},
	isbn = {1572-9338},
	journal = {Annals of Operations Research},
	title = {Bayesian dynamic quantile model averaging},
	url = {https://doi.org/10.1007/s10479-024-06378-7},
	year = {2024}}

@article{MGP,
    author ={Bhattacharya, A. and Dunson, D.B.} ,
    title = {Sparse Bayesian infinite factor models},
    journal = {Biometrika},
    year = {2011},
    volume={98}, 
    number={2},
    pages={291-306},
}

@article{Chen21,
author = {Chen, Liang and Dolado, Juan J. and Gonzalo, Jesús},
title = {Quantile Factor Models},
journal = {Econometrica},
volume = {89},
number = {2},
pages = {875-910},
doi = {https://doi.org/10.3982/ECTA15746},
url = {https://onlinelibrary.wiley.com/doi/abs/10.3982/ECTA15746},
eprint = {https://onlinelibrary.wiley.com/doi/pdf/10.3982/ECTA15746},
year = {2021}
}

@article{Chernis23,
  title={Predictive Density Combination Using a Tree-Based Synthesis Function},
  author={Chernis, Tony and Hauzenberger, Niko and Huber, Florian and Koop, Gary and Mitchell, James},
  journal={arXiv preprint arXiv:2311.12671},
  year={2023}
}

@article{Chernis24,
url = {https://doi.org/10.1515/snde-2022-0108},
title = {Combining Large Numbers of Density Predictions with Bayesian Predictive Synthesis},
title = {},
author = {Tony Chernis},
pages = {293--317},
volume = {28},
number = {2},
journal = {Studies in Nonlinear Dynamics \& Econometrics},
doi = {doi:10.1515/snde-2022-0108},
year = {2024},
lastchecked = {2024-11-06}
}

@article{Clark23,
  title={Tail forecasting with multivariate Bayesian additive regression trees},
  author={Clark, Todd E and Huber, Florian and Koop, Gary and Marcellino, Massimiliano and Pfarrhofer, Michael},
  journal={International Economic Review},
  volume={64},
  number={3},
  pages={979--1022},
  year={2023},
  publisher={Wiley Online Library}
}

@article{Clark24,
author = {Todd E. Clark and Florian Huber and Gary Koop and Massimiliano Marcellino and Michael Pfarrhofer},
title = {Investigating Growth-at-Risk Using a Multicountry Nonparametric Quantile Factor Model},
journal = {Journal of Business \& Economic Statistics},
volume = {42},
number = {4},
pages = {1302--1317},
year = {2024},
publisher = {ASA Website},
doi = {10.1080/07350015.2024.2310020},
}

@article{Ferrara,
title = {High-frequency monitoring of growth at risk},
journal = {International Journal of Forecasting},
volume = {38},
number = {2},
pages = {582-595},
year = {2022},
issn = {0169-2070},
doi = {https://doi.org/10.1016/j.ijforecast.2021.06.010},
author = {Laurent Ferrara and Matteo Mogliani and Jean-Guillaume Sahuc},
}

@article{GIGLIO16,
title = {Systemic risk and the macroeconomy: An empirical evaluation},
journal = {Journal of Financial Economics},
volume = {119},
number = {3},
pages = {457-471},
year = {2016},
issn = {0304-405X},
doi = {https://doi.org/10.1016/j.jfineco.2016.01.010},
author = {Stefano Giglio and Bryan Kelly and Seth Pruitt}
}

@article{DQLM,
author = {Kelly C. M. Gon{\c{c}}alves and H{\'e}lio S. Migon and Leonardo S. Bastos},
title = {{Dynamic Quantile Linear Models: A Bayesian Approach}},
volume = {15},
journal = {Bayesian Analysis},
number = {2},
publisher = {International Society for Bayesian Analysis},
pages = {335 -- 362},
year = {2020},
doi = {10.1214/19-BA1156},
}

@article{Iacopini,
  title={Bayesian Multivariate Quantile Regression with alternative Time-varying Volatility Specifications},
  author={Matteo Iacopini and Francesco Ravazzolo and Luca Rossini},
  journal={arXiv preprint arXiv:2211.16121 },
  year={2024}
}

@book{Koenker2005,
  asin = {0521608279},
  author = {Koenker, Roger},
  isbn = {0521608279},
  publisher = {Cambridge University Press},
  series = {Econometric Society Monographs},
  title = {Quantile Regression},
  year = {2005},
}

@article{kobayashi2024predicting,
  title={Predicting COVID-19 hospitalisation using a mixture of Bayesian predictive syntheses},
  author={Kobayashi, Genya and Sugasawa, Shonosuke and Kawakubo, Yuki and Han, Dongu and Choi, Taeryon},
  journal={The Annals of Applied Statistics},
  volume={18},
  number={4},
  pages={3383--3404},
  year={2024},
  publisher={Institute of Mathematical Statistics}
}

@article{Korobilis24,
author = {Dimitris Korobilis and Maximilian Schröder},
title = {Probabilistic Quantile Factor Analysis},
journal = {Journal of Business \& Economic Statistics},
volume = {0},
number = {0},
pages = {1--14},
year = {2024},
publisher = {ASA Website},
doi = {10.1080/07350015.2024.2396956}
}

@article{kozumi11,
author = {Hideo Kozumi and Genya Kobayashi},
title = {Gibbs sampling methods for Bayesian quantile regression},
journal = {Journal of Statistical Computation and Simulation},
volume = {81},
number = {11},
pages = {1565--1578},
year = {2011},
publisher = {Taylor \& Francis},
doi = {10.1080/00949655.2010.496117},
}

@article{masuda2024proofs,
  title={On the proofs of the predictive synthesis formula},
  author={Masuda, Riku and Irie, Kaoru},
  journal={arXiv preprint arXiv:2409.09660},
  year={2024}
}

@article{MW19,
  title={Dynamic Bayesian predictive synthesis in time series forecasting},
  author={Kenichiro McAlinn and Mike West},
  journal={Journal of Econometrics},
  volume={210},
  pages={155--169},
  year={2019},
  publisher={Elsevier},
}

@article{McAlinn20,
  title={Multivariate Bayesian predictive synthesis in macroeconomic forecasting},
  author={Kenichiro McAlinn and Knut Are Aastveit and Jouchi Nakajima and Mike West},
  journal={Journal of the American Statistical Association},
  volume={115},
  pages={1092--1110},
  year={2020},
  publisher={Taylor \& Francis},
}

@inbook{Mitchell22,
	author = {Mitchell, James and Poon, Aubrey and Mazzi, Gian Luigi},
	booktitle = {Essays in Honor of M. Hashem Pesaran: Prediction and Macro Modeling},
	doi = {10.1108/S0731-90532021000043A004},
	editor = {Chudik, Alexander and Hsiao, Cheng and Timmermann, Allan},
	isbn = {978-1-80262-062-7, 978-1-80262-061-0},
	pages = {51--72},
	publisher = {Emerald Publishing Limited},
	series = {Advances in Econometrics},
	title = {Nowcasting Euro Area GDP Growth Using Bayesian Quantile Regression},
	url = {https://doi.org/10.1108/S0731-90532021000043A004},
	volume = {43A},
	year = {2022},
	year1 = {2022/01/01}}

@article{Mitchell24,
author = {Mitchell, James and Poon, Aubrey and Zhu, Dan},
title = {Constructing density forecasts from quantile regressions: Multimodality in macrofinancial dynamics},
journal = {Journal of Applied Econometrics},
volume = {39},
number = {5},
pages = {790-812},
doi = {https://doi.org/10.1002/jae.3049},
url = {https://onlinelibrary.wiley.com/doi/abs/10.1002/jae.3049},
eprint = {https://onlinelibrary.wiley.com/doi/pdf/10.1002/jae.3049},
year = {2024}
}

@article{Nicolo17,
author = {De Nicolò, Gianni and Lucchetta, Marcella},
title = {Forecasting Tail Risks},
journal = {Journal of Applied Econometrics},
volume = {32},
number = {1},
pages = {159-170},
doi = {https://doi.org/10.1002/jae.2509},
url = {https://onlinelibrary.wiley.com/doi/abs/10.1002/jae.2509},
eprint = {https://onlinelibrary.wiley.com/doi/pdf/10.1002/jae.2509},
year = {2017}
}

@article{PFARRHOFER22,
title = {Modeling tail risks of inflation using unobserved component quantile regressions},
journal = {Journal of Economic Dynamics and Control},
volume = {143},
pages = {104493},
year = {2022},
issn = {0165-1889},
doi = {https://doi.org/10.1016/j.jedc.2022.104493},
author = {Michael Pfarrhofer}
}

@article{wang2024inflation,
  title={Inflation Target at Risk: A Time-varying Parameter Distributional Regression},
  author={Wang, Yunyun and Oka, Tatsushi and Zhu, Dan},
  journal={arXiv preprint arXiv:2403.12456},
  year={2024}
}

@article{YU01,
title = {Bayesian quantile regression},
journal = {Statistics \& Probability Letters},
volume = {54},
number = {4},
pages = {437-447},
year = {2001},
issn = {0167-7152},
doi = {https://doi.org/10.1016/S0167-7152(01)00124-9},
url = {https://www.sciencedirect.com/science/article/pii/S0167715201001249},
author = {Keming Yu and Rana A. Moyeed},
}

@book{WH97,
    author = {West, M. and Harrison, P. J.},
    title ={Bayesian Forecasting and Dynamic Models},
    edition={2nd},
    publisher ={Springer-Verlag},
    year={1997},
}

@book{Prado,
    author = {Raquel Prado and Marco A. R. Ferreira and Mike West},
    title ={ Time Series
Modeling, Computation, and Inference, Second Edition},
    publisher ={Chapman and Hall/CRC},
    year={2021},
}

@article{Gneiting01072011,
author = {Tilmann Gneiting and Roopesh Ranjan and},
title = {Comparing Density Forecasts Using Threshold- and Quantile-Weighted Scoring Rules},
journal = {Journal of Business \& Economic Statistics},
volume = {29},
number = {3},
pages = {411--422},
year = {2011},
publisher = {ASA Website},
doi = {10.1198/jbes.2010.08110},
URL = { 
        https://doi.org/10.1198/jbes.2010.08110
},
eprint = { 
        https://doi.org/10.1198/jbes.2010.08110
}
}

@article{tallman2024bayesian,
  title={Bayesian predictive decision synthesis},
  author={Tallman, Emily and West, Mike},
  journal={Journal of the Royal Statistical Society Series B: Statistical Methodology},
  volume={86},
  number={2},
  pages={340--363},
  year={2024},
  publisher={Oxford University Press US}
}

@article{Maha,
    author={Mohaddes, K. and M. Raissi},
    year=2024,
    title={Compilation, Revision and Updating of the Global VAR (GVAR) Database, 1979Q2-2023Q3},
    publisher={University of Cambridge: Judge Business School (mimeo)},
}

@software{tange,
      author       = {Tange, Ole},
      title        = {GNU Parallel 20251222 ('Bondi')},
      month        = Dec,
      year         = 2025,
      note         = {{GNU Parallel is a general parallelizer to run
                       multiple serial command line programs in parallel
                       without changing them.}},
      publisher    = {Zenodo},
      doi          = {10.5281/zenodo.18039569},
      url          = {https://doi.org/10.5281/zenodo.18039569}
}

@article{fruhwirth1994data,
  title={Data augmentation and dynamic linear models},
  author={Fr{\"u}hwirth-Schnatter, Sylvia},
  journal={Journal of time series analysis},
  volume={15},
  number={2},
  pages={183--202},
  year={1994},
  publisher={Wiley Online Library}
}

@article{DURANTE2017198,
title = {A note on the multiplicative gamma process},
journal = {Statistics \& Probability Letters},
volume = {122},
pages = {198-204},
year = {2017},
issn = {0167-7152},
doi = {https://doi.org/10.1016/j.spl.2016.11.014},
url = {https://www.sciencedirect.com/science/article/pii/S016771521630253X},
author = {Daniele Durante},
}

@article{koenker1978regression,
  title={Regression quantiles},
  author={Koenker, Roger and Bassett Jr, Gilbert},
  journal={Econometrica: journal of the Econometric Society},
  pages={33--50},
  year={1978},
  publisher={JSTOR}
}

\end{document}